\documentclass[12pt]{article}
\usepackage{amssymb}
\usepackage{graphicx}
\usepackage{epsf}
\usepackage{lscape}
\usepackage{epstopdf}
\usepackage[cp866]{inputenc}
\usepackage[T2A]{fontenc}
\usepackage[english]{babel}
\usepackage{amsmath}
\begin{document}

\title{\large{\bf  Dispersion (asymptotic) theory of charged particle transfer reactions at low energies   and    nuclear astrophysics: I. the   ``non-dramatic'' case   }}
\author{    R. Yarmukhamedov,$^{\rm{1}}$\thanks{Corresponding author, E-mail:
rakhim@inp.uz}\, and K. I. Tursunmakhatov$^{\rm{2}}$   }

 \maketitle{\it
$^{\rm{1}}$Institute of Nuclear Physics,  100214 Tashkent, Uzbekistan\\
$^{\rm{2}}$  Gulistan State University,  120100 Gulistan, Uzbekistan}

\bigskip

 \begin{abstract}
 { \small
A new  dispersion (asymptotic) theory is proposed for the peripheral sub- and above-barrier charged particle transfer $A(x,y)B$  reaction in the  three-body ($A$, $a$ and $y$) model where $ x= y +a$ and $B=A+a$, and $ a$  is a transferred particle.  It is based on the combination of   the dispersion theory and the conventional DWBA method.
   The explicit forms have been derived for the exact  three-body   pole  amplitude  and  differential cross section in which    the contribution of the three-body ($A$, $a$ and $y$) Coulomb  dynamics of the  transfer mechanism  in  the peripheral partial amplitudes, corresponding to   partial waves  with  $l_i>>1$,   is taken into account correctly.
 For the specific peripheral proton and triton transfer reactions,   the   comparative analysis  of    the peripheral partial amplitudes at $l_i>>1$ , which correspond  to    the one-step pole  and   exact  three-body  pole amplitudes  as well as  those of   the  ``post''-approximation and the post form of the conventional DWBA,  is performed with each other. 
It shows the absolute inapplicability of  the  ``post''-approximation
usually applied   for getting an information about specific asymptotic normalization coefficients  being astrophysical interest.
 }
\end{abstract}

\newpage

 \vspace{1.0cm}
\begin{center}
 {\bf I. INTRODUCTION}
  \end{center}

\bigskip

In the  last two decades,  a number of    methods of  analysis of  experimental data for different nuclear processes   were proposed  to obtain  information on  the ``indirect determined'' (``experimental'') values of  the specific asymptotic normalization coefficients (or respective nuclear vertex constants (NVC)) with the aim of their application to   nuclear astrophysics (see, for example, Refs. [1--6] and the available references therein).     One of such methods uses the modified DWBA \cite{Ar96,Mukh2}   for peripheral  nuclear transfer reactions  in which   the differential cross section (DCS) is    parametrized in the terms of the asymptotic normalization coefficients. One notes that an   asymptotic normalization coefficient  (ANC), which is proportional to the NVC for the virtual decay $B\,\to\,A\,+\,a$,  determines the amplitude of the tail of the overlap function corresponding  to the wave function of nucleus $B$ in the binary  ($A\,+\,a$) channel (denoted by $A\,+a\,\to\, B$ everywhere below) \cite{Blok77}.  As   the ANC for $ A\,+a\,\to\, B$ determines  the probability of the configuration $A+a$ in nucleus $B$ at distances greater than the radius of nuclear $Aa$ interaction,     the ANC arises   naturally in  expressions for the cross sections of  the peripheral   nuclear reactions between charged particles at low energies, in particular, of the peripheral exchange    $A{\rm{(}}B,\,A{\rm{)}}B$, transfer $A$($x$, $y$)$B$ and  nuclear-astrophysical  $A(a,\,\gamma)B$ reactions.

\begin{figure}
\begin{center}
\includegraphics[height=4cm, width=14cm]{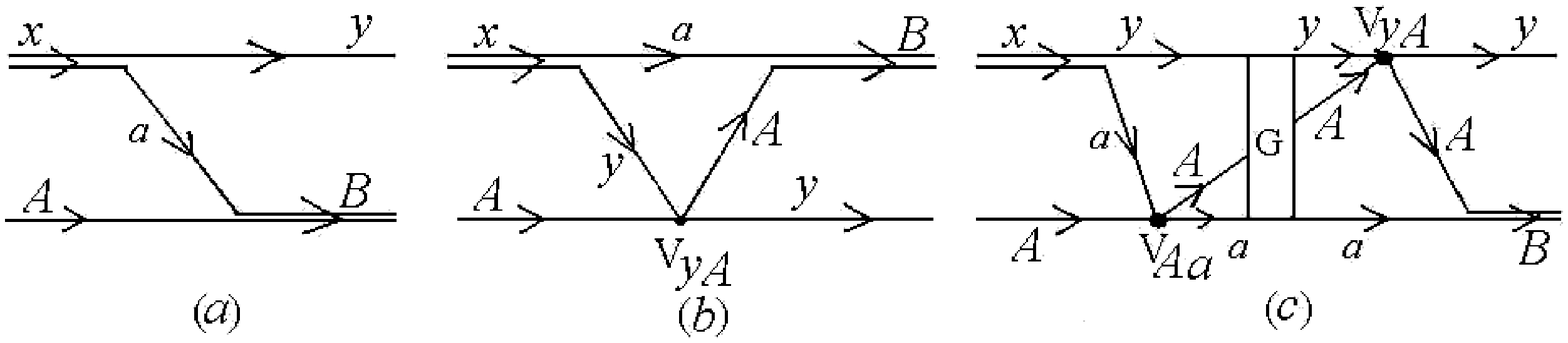}
 \end{center}
\caption{\label{fig1} Diagrams describing transfer of the particle $a$ and taking into account possible subsequent Coulomb-nuclear rescattering of particles ($A$, $a$ and $y$) in the  intermediate state.}
\end{figure}

 In the present work, the peripheral charged particle  transfer reaction
\begin{equation}
\label{subeq1}
x\,+\,A\,\longrightarrow \,y\,+\,B
\end{equation}
is considered in the framework  of the three-body ($A$, $a$ and $y$) model,   where $x$=($y\,+\,a$) is a projectile,  $B$=($A\,+\,a$) and  $a$ is a transferred particle.  The main  idea of consideration        is based on   the following two  assumptions: i) the   peripheral  reaction (\ref{subeq1}) is  governed by the singularity  of the  reaction  amplitude   at $\cos\theta=\,\xi>$ 1, where $\xi$ is the  nearest to physical (-1 $\le\cos\theta\le$ 1) region  singularity generated by the pole mechanism
(Fig. \ref{fig1}$a$) \cite{Shapiro}   and $\theta$ is the center-of-mass scattering angle;  ii)  the dominant role played by this nearest  singularity  is the result of the peripheral nature of the considered reaction   at least in the angular range of  the main peak of the angular distribution \cite{DDM1973}. Consequently,  it is necessary to know  the  behavior of the  reaction  amplitude at the  nearest singularity $\xi$ \cite{Av86,Mukh10}, which in turn  defines the behavior of the true  peripheral partial amplitudes   at  $l_i\,\gtrsim L_0>>$ 1
($L_0\sim k_iR^{{\rm{ch}}}_i$ with  $R_i^{{\rm{ch}}}\,\gtrsim\, R_N$) \cite{Popov1964}  giving the dominant contribution to the reaction amplitude  at least in the angular range of the main peak of the angular distribution \cite{DDM1973,KMYa1988}, where   $l_i$, $k_i$,  $R^{ch}_i$ and  $R_N$   are a  partial  wave,  a  number wave (or a relative momentum),  a channel radius  and  the radius of the nuclear interaction       of  the colliding nuclei, respectively.

 In practice,  the ``post''-approximation  and  the post form   of  the modified DWBA   \cite{Ar96,Mukh2}    are used for the analysis  of the specific peripheral proton transfer reactions. They are restricted  by the zero- and first-order terms of the  perturbation theory   over  the optical Coulomb polarization potential  $\Delta V_f^C$ (or $\Delta V_i^C$) in   the transition operator,  respectively, which   are   sandwiched by the initial and final state wave functions  in  the matrix element of the reaction (\ref{subeq1}). At this, it is assumed  that the contribution of the first-order term over $\Delta V_f^C$ (or $\Delta V_i^C$)
 to   the matrix element is  small \cite{Mukh2}. But, it was shown in Refs. \cite{Yarm2013,Mukh10,Yar97,Igam072}  that, when the residual nuclei  $B$ are formed in
weakly bound states being  astrophysical interest, this assumption is not guaranteed for the peripheral
charged particle  transfer reactions and, so,
the extracted ``experimental''  ANC values   may not have
 the necessary accuracy for their  astrophysical application (see, for example, \cite{Igam072} and Table 1 in  \cite{Yarm2013}). In this case, in
the transition operator an inclusion of all other   orders (the  second and higher orders) of the power expansion in a series over $\Delta V_f^C$ (or $\Delta V_i^C$) is required  for the DWBA cross section calculations since they  strongly change the power of the peripheral partial amplitudes at $l_i\,>>$ 1 \cite{Mukh10,Igam072}.

For these reasons,    it is of great   interest to derive the expressions  for the amplitude and the  differential cross section (DCS) of the peripheral  reaction (\ref{subeq1}) within the so-called hybrid theory: the  DWBA approach and the dispersion peripheral   model \cite{DDM1973,Av86}. The main advantage of the hybrid theory  as compared to the modified DWBA  used in \cite{Ar96,Mukh2} is that, first,  it   allows one  to derive  the expression for the part of  the reaction amplitude having   the contribution only from  the nearest singularity $\xi$ in which  the influence  of   the three-body Coulomb dynamics of the  transfer mechanism on the peripheral partial amplitudes at $l_i>>$ 1    is taken into account in a correct manner within the dispersion theory. Second, it accounts for the distortion effects in the initial and final states within  the DWBA approach, which is  more accurate than  as it was done     in \cite{DMYa1978} in   the dispersion peripheral model \cite{DDM1973}. They allow one to treat the important issue: to what extent does a  correct taking into account of   the  three-body Coulomb effects in the initial, intermediate and final states  of the peripheral reaction (\ref{subeq1}), firstly,   influences the spectroscopic information deduced from the analysis of the experimental
 DCS's and,  secondly,   improves  the accuracy of the modified DWBA analysis  used for obtaining  the ``experimental''  ANC values of astrophysical interest.   Besides, the proposed asymptotic  theory  can  also be applied to strong sub-barrier transfer reactions for which the main  contribution  to the reaction amplitude  comes to several  lowest partial waves $l_i$ ($l_i\sim\,k_iR_i^{{\rm{ch}}}$ =0, 1,..., where $k_i\,\to$  0 and $R_i^{{\rm{ch}}}\gtrsim R_N$) and the contribution of   peripheral partial waves  $l_i$ ($l_i\,>>$ 1) is strongly suppressed.

 The similar  theory   was   proposed earlier in  \cite{KMYa1988} for  the  peripheral    neutron transfer reaction induced by the   heavy ions at above-barrier energies,  which was also  implemented successfully for the specific  reactions.  However,
  for   peripheral charged particle transfer reactions   this task  requires  a special consideration. This is connected with  the considerable  complication occurring in  the main mechanisms  of the  reaction (\ref{subeq1}) because of   correct   taking into account  of   the three-body  Coulomb dynamics of the transfer mechanism   \cite{Av86,Mukh10}.

   Below, we use the system of units  $\hbar$=$c$=  1 everywhere,   except where  they   are specially pointed out.

 \vspace{1cm}
 \begin{center}
  {\bf II.  THREE-BODY COULOMB DYNAMICS OF THE TRANSFER MECHANISM    AND THE GENERALIZED    DWBA}
\end{center}
\bigskip

\hspace{0.6cm} We  consider the  reaction (\ref{subeq1}) within the framework of the     three ($A$, $a$ and $y$)     charged  particles.
 Within the framework of the  three-body  Schr\"{o}dinger approach, the amplitude for the reaction  (\ref{subeq1}) is given by  \cite{Gre1966,Austern1964}
\begin{equation}
\label{subeq2}
 M^{{\rm{TB}}}{\rm{(}}E_i,\,cos\theta{\rm{)}}\,=\,\sum_{M_a}\langle\chi^{(-)}_{{\mathbf k}_f}I_{Aa}|V^{{\rm{TB}}}|I_{ay}\chi^{(+)}_{{\mathbf k}_i}\rangle
\end{equation}
and
\begin{equation}
\label{subeqTB}
V^{{\rm{TB}}}\,=\,\bigtriangleup V_f\,+\,\bigtriangleup V_fG\bigtriangleup V_i.
\end{equation}
 Here $\chi^{(+)}_{{\mathbf k}_i} $ and $ \chi^{(-)}_{{\mathbf k}_f}$ are the optical Coulomb--nuclear distorted wave functions in the entrance and exit channels   with the relative momentum  ${\mathbf  k}_i$ and ${\mathbf k}_f$, respectively ($E_i\,=\,k^2_i/2\mu_{Ax}$ and $E_f\,=\,k^2_f/2\mu_{By}$); $I_{Aa}$(${\mathbf r}_{Aa}$)($I_{ay}$(${\mathbf r}_{ay}$))  is  the overlap integral of the  bound-state      $\psi_A$, $\psi_a$ and $\psi_B$ ($\psi_y$, $\psi_a$ and $\psi_x$) wave functions \cite{Ber1965,BSat1965};     $\bigtriangleup V_f\,=\,V_{ay}\,+\,V_{yA}\,-\,V_f$;  $\bigtriangleup V_i\,=\,V_{Aa}\,+\,V_{yA}\,-\,V_i$;
 $G$ = (${\cal E}\, -\, H\,+\,i\cdot{\rm{0}}$)$^{-1}$ is  the operator of the three-body ($A$, $a$ and $y$)  Green's function  and $M_a$ is the spin  projections of the transferred particle $a$, where  $V_{ij}=V^N_{ij}+V^C_{ij}$, $V^N_{ij}$($V^C_{ij}$)  is the   nuclear (Coulomb) interaction  potential  between the centers  of mass of the  particles $i$ and $j$, which does not depend on the coordinates of the constituent nucleus;
$V_i$ and $V_f$ are the optical Coulomb--nuclear potentials in the entrance and  exit states, respectively; $H$ is the Hamiltonian operator for the three-body ($A$, $a$ and $y$) system;
  ${\cal E}\, =\, E_i\, -\, \varepsilon_{ay}\,=\, E_f\,-\,\varepsilon _{Aa}$ in which  $\varepsilon_{ij}$ is the binding energy of the  bound ($ij$)   system  in respect to    the    ($i\,+\,j$)  channel; ${\mathbf r}_{ij}\,=\, {\mathbf r}_{i}\,-\,{\mathbf r}_{j}$, ${\mathbf r}_{i}$ is the radius-vector of the center of mass of  the particle $i$ and $\mu_{ij}$ = $m_im_j/m_{ij}$ is the reduced mass of  the  $i$ and $j$   particles in which $m_{ij}=m_i\,+\,m_j$ and $m_j$ is the mass of the $j$ particle.

 The operator of the three-body Green's function $G$  can be presented as
\begin{equation} \label{subeq3TB}
G\,=\,G_C\,+\,G_CV^NG, \,\,\, G_C=G_{{\rm{0}}}+G_{{\rm{0}}}V^CG_C,
\end{equation}
where $G_C$\,=\,(${\cal E}\, - \,T\, -\, V^C\,+\,i\cdot{\rm{0}}$)$^{-1}$ and $G_{{\rm{0}}}$\,=\,(${\cal E}\, - \,T\,+\,i\cdot{\rm{0}}$)$^{-1}$   are  the operators of the three-body
($A$, $a$  and  $y$)  Coulomb  and free Green's functions, respectively;  $T$ is  the kinetic energy  operator for  the three-body
 ($A$, $a$ and $y$) system;  $V^N\,=\,V^N_{ay}\,+\,V^N_{Aa}\,+\,V^N_{yA}$ and  $V^C\,=\,V^C_{ay}\,+\,V^C_{Aa}\,+\,V^C_{yA}$.

The  overlap function $I_{Aa}{\rm{(}}{\mathbf r}_{Aa}{\rm{)}}$  is
given by \cite{Blok77}
$$ I_{Aa}{\rm{(}}{\mathbf
r}_{Aa}{\rm{)}}\,=\,N_{Aa}^{1/2}\langle\psi_A{\rm{(}}\boldsymbol{\zeta}_A{\rm{)}}\psi_a{\rm{(}}\boldsymbol{\zeta}_a{\rm{)}}|\psi_B{\rm{(}}\boldsymbol{\zeta}_A,\boldsymbol{\zeta}_a;{\mathbf r}_{Aa}{\rm{)}}\rangle
$$
\begin{equation} \label{subeq3TBOI}
=\,\sum_{l_B\mu_Bj_B\nu_B}C_{j_B\nu_BJ_AM_A}^{J_BM_B}C_{l_B\mu_B J_aM_a}^{j_B
 \nu_B}i^{l_B}Y_{l_B\mu_B}{\rm{(}}\hat{{\mathbf r}}_{Aa}{\rm{)}}I_{Aa;\,l_Bj_B}{\rm{(}}r_{Aa}{\rm{)}}.
  \end{equation}
Here $J_j(M_j)$ is the spin
(its projection) of the  particle $j$; $\hat{{\mathbf r}}_{Aa}={\mathbf r}_{Aa}/r_{Aa}$, $j_B$ and $\nu_B$
($l_B$ and $\mu_B$) are the total (orbital) angular momentum and its projection   of the particle $a$ in the nucleus $B$[=\,($A+a$)], respectively;
$C_{a\alpha\, b\beta }^{c\gamma}$ is the Clebsch-Gordan coefficient, and $N_{Aa}$
 is the factor taking into account the nucleons' identity \cite{Blok77}, which
is absorbed in the radial overlap function $I_{Aa;l_Bj_B}{\rm{(}}r_{Aa}{\rm{)}}$ being  not normalized to unity \cite{Ber1965}.   In the matrix element (\ref{subeq3TBOI}), the integration is taken over all the internal relative coordinates $\boldsymbol{\zeta}_A$ and $\boldsymbol{\zeta}_a$ for the $A$ and $a$ nuclei.

 The asymptotic
behavior   of $I_{Aa;l_Bj_B}{\rm{(}}r_{Aa}{\rm{)}}$  at  $r_{Aa}>r^{{\rm{(}}N{\rm{)}}}_{Aa}$  is given by the relation
\begin{equation}
 I_{Aa;l_Bj_B}{\rm{(}}r_{Aa}{\rm{)}}\,\simeq\, C_{Aa;l_B\,j_B}\frac{W_{-\eta_B;\,l_B+1/2}{\rm{(}}2\kappa_{Aa}
r_{Aa}{\rm{)}}}{r_{Aa}}, \label{subeq4SFOI} \end{equation}
 where
$W_{\alpha;\beta}{\rm{(}}r_{Aa}{\rm{)}}$ is the
Whittaker function, $\eta_B\,=\,z_Az_ae^2\mu_{Aa}/\kappa _{Aa}$ is
the Coulomb parameter for the
$B$ [=($A\,+\,a$)] bound state, $\kappa _{Aa}\,=\,\sqrt{2\mu_{Aa}\varepsilon
_{Aa}}$,
$r_{ij}^{{\rm{(}}N{\rm{)}}}$ is the nuclear interaction
radius between $i$ and $j$ particles in the  bound  ($i\,+\,j$) state
and $C_{Aa;\,l_Bj_B}$ is the ANC  for  $A\, + \,a\,\to\, B$,  which is related to the NVC ($G_{Aa;\,l_Bj_B}$) for the virtual decay $B\,\to\,A\, + \,a$  as \cite{Blok77}
\begin{equation}
G_{Aa;l_Bj_B}=-i^{l_B+\eta_B}\frac{\sqrt{\pi}}{\mu_{Aa}}C_{Aa;l_Bj_B}\cdot
\label{subeq4CG}
\end{equation}
 Eqs. (\ref{subeq3TBOI})--(\ref{subeq4CG})  and the expression for the matrix element $M_{Aa}{\rm{(}}{\mathbf{q}}_{Aa}{\rm{)}}$ for  the virtual decay $B\,\to\,A\,+\,a$, which is  given by Eq. (B1) in Appendix B,  hold for the  matrix element   $M_{ay}{\rm{(}}{\mathbf{q}}_{ay}{\rm{)}}$ of the virtual decay $x\,\to\,y\,+\,a$ and the overlap function $I_{ay}{\rm{(}}{\mathbf r}_{ay}{\rm{)}}$.

The first ($V_{ay}$) and second ($V_{yA}$) terms,  entering the first term of the right-hand side (r.h.s.) of (\ref{subeqTB}),   correspond to the mechanisms described by the pole and triangle diagrams in Figs. \ref{fig1}$a$ and  \ref{fig1}$b$, respectively, where  the Coulomb-nuclear core-core ($A\,+\,y\,\longrightarrow\,A\,+\,y$) scattering in the  four-ray  vertex  of the  triangle diagram of Fig. \ref{fig1}$b$ is taken  in the Born approximation. The $\bigtriangleup V_fG\bigtriangleup V_i$ term in the r.h.s. of (\ref{subeqTB}) corresponds to more complex mechanisms than the pole and triangle ones. This term  is described by a sum of nine diagrams obtained from the basic diagrams presented in   Figs. \ref{fig1}$a$ and  \ref{fig1}$b$,  which take into account  all possible subsequent mutual Coulomb-nuclear rescattering of the particles $A$, $a$ and $y$ in the intermediate state. One of the nine diagrams corresponding to the term $ V_{yA}GV_{Aa}$ is plotted in  Fig. \ref{fig1}$c$, where  the Coulomb-nuclear ($y\,+\,A\,\longrightarrow\,y\,+\,A$ and $A\,+\,a\,\longrightarrow\,A\,+\,a$) scatterings in the four-ray vertices, including in all four-ray vertices for the  others of   eight diagrams,   are taken in  the  Born approximation. This term corresponds to  the mechanism of subsequent Coulomb-nuclear rescattering of  the $y$ and $a$ particles, virtually emitted by the projectile $x$,  on the target $A$  in the intermediate state. In particular,     for the nucleon  ($N$) transfer $A$($d$, $N$)$B$ reaction, this mechanism corresponds     to     that    of      the  subsequent  rescatterings of  the proton ($p$) and neutron ($n$),   virtually emitted  by  the deuteron in the field of the $A$ target, in which the transferred particle  is either $p$ or $n$, where    $B$\,=\,$A$\,+\,$N$.

If the reaction (\ref{subeq1}) is peripheral, then its dominant mechanism, at least in the angular range of the main peak of the angular distribution, corresponds to the pole diagram  in Fig. \ref{fig1}$a$ \cite{DDM1973,KMYa1988}. The amplitude of this diagram has the  singularity at $\cos\theta\,=\,\xi$, which is the nearest one to the physical (-1 $\le\,\cos\theta\,\le$ 1) region \cite{Shapiro,DDM1973} and is given by the expression
\begin{equation}
\label{subeqSB19}
 \xi\,=\,\frac{k_{i{\rm{1}}}^2\,+\,k_f^2\,+\,\kappa_{ay}^2}{2k_{i{\rm{1}}}k_f}\,=\,\frac{k_i^2\,+\,k_{f{\rm{1}}}^2\,+\,\kappa_{Aa}^2}{2k_{i }k_{f{\rm{1}}}},
\end{equation}
 where  $k_{i{\rm{1}}}\,=\,{\rm{(}}m_{y}/m_x{\rm{)}}k_i$ and $k_{f{\rm{1}}}\,=\,{\rm{(}}m_A/m_B{\rm{)}}k_f$.
  However, if nuclear interactions   in   the second ($V_{yA}$)  and the third  ($V_f$) terms of the first $\bigtriangleup V_f$ term of   the  r.h.s. of (\ref{subeqTB})    as well as in the  $\bigtriangleup V_fG\bigtriangleup V_i$  one   are ignored  by  the corresponding  replacement
\begin{equation}
\label{subeqSB19}
 V_{yA}\,\longrightarrow\,V_{yA}^C,\,\, V_f\,\longrightarrow\,V_f^C,\,\,
 \bigtriangleup V_fG\bigtriangleup V_i\, \longrightarrow\,\bigtriangleup V_f^CG_C\bigtriangleup V_i^C,
\end{equation}  where $\bigtriangleup V_f^C\,=\,V_{ay}^C\,+\,V_{yA}^C\,-\,V_f^C$ and  $\bigtriangleup V_i^C\,=\,V_{Aa}^C\,+\,V_{yA}^C\,-\,V_i^C$, then we can separate the part of the amplitude (\ref{subeq2}), denoted by $M^{{\rm{TBDW}}}{\rm{(}}E_i,\,cos\theta{\rm{)}}$ below, which has the singularity at $\cos\theta=\,\xi$ (the type of  branch point).  The remainder of the $ M^{{\rm{TB}}}{\rm{(}}E_i,\,cos\theta{\rm{)}}$ amplitude is given by  the sum of an infinite series of the diagrams of the type in Figs. \ref{fig1}$b$ and  \ref{fig1}$c$.  They  contain  all possible  nuclear rescattering of  the particles $A$, $a$ and $y$ from each other  in the intermediate state.  Therefore, the corresponding  amplitudes of these diagrams  have  singularities ($\zeta_i$), which are located farther away  from the left ($\cos\theta$= -1) and right ($\cos\theta$= 1) boundary of the physical (-1$\le\cos\theta\le$1) region than the singularity $\xi$ ($\mid\zeta_i\mid>\xi$) \cite{Shapiro,BDP1963}. Consequently,   their contribution to the amplitude $M^{{\rm{TB}}}{\rm{(}}E_i,\,cos\theta{\rm{)}}$ in the angular range of the main peak of the angular distribution can be ignored \cite{DDM1973}.
  In this approximation,  the amplitude $ M^{{\rm{TB}}}{\rm{(}}E_i,\,cos\theta{\rm{)}}$ can be reduced to  the form
\begin{equation}
 M^{{\rm{TB}}}{\rm{(}}E_i,\,cos\theta{\rm{)}}\,\approx\, M^{{\rm{TBDW}}}{\rm{(}}E_i,\,cos\theta{\rm{)}}\,=\,M^{{\rm{DW}}}_{{\rm{post}}}{\rm{(}}E_i,\,cos\theta{\rm{)}}\,+\, \bigtriangleup M^{{\rm{TBDW}}}{\rm{(}}E_i,\,cos\theta{\rm{)}}.
\label{subeqSB5}
\end{equation}
Here
 \begin{equation}
M^{{\rm{DW}}}_{{\rm{post}}}{\rm{(}}E_i,\,cos\theta{\rm{)}}\,=\,\sum_{M_a}\langle\chi^{(-)}_{{\mathbf k}_f}I_{Aa}|V_{ay}\,+\,V_{yA}^C\,-\,V_f^C |I_{ay}\chi^{(+)}_{{\mathbf k}_i}\rangle
 \label{subeqSB6}
\end{equation}
and
 \begin{equation}
 \bigtriangleup M^{{\rm{TBDW}}}{\rm{(}}E_i,\,cos\theta{\rm{)}} \,=
  \,\sum_{M_a}\langle\chi^{(-)}_{{\mathbf k}_f}I_{Aa}|\bigtriangleup V_f^CG_C\bigtriangleup V_i^C|I_{ay}\chi^{(+)}_{{\mathbf k}_i}\rangle\cdot
\label{subeqSB7}
\end{equation}
 In Eqs (\ref{subeqSB5})--(\ref{subeqSB7}), the contribution of   the three-body ($A$, $a$ and $y$) Coulomb dynamics of  the transfer mechanism  in the intermediate state  involves    all orders of the perturbation theory over the optical  Coulomb polarization potentials $\bigtriangleup V^C_{f,i}$, whereas  the Coulomb-nuclear distortions ($V_i$ and $V_f$) in the entrance and exit channels are  taken into account within the framework of  the optical model.  The amplitude  $M^{{\rm{TBDW}}}{\rm{(}}E_i,\,cos\theta{\rm{)}}$ can be considered as generalization of the post form of  the DWBA amplitude ($M^{{\rm{DW}}}_{{\rm{post}}}{\rm{(}}E_i,\,cos\theta{\rm{)}}$) \cite{LOLA} in which the three-body Coulomb dynamics of the main transfer mechanism are taken into account  in a correct manner.  The pole-approximation of the DWBA   amplitude (denoted by    $ M^{{\rm{DW}}}_{{\rm{pole}}}{\rm{(}}E_i,\,cos\theta{\rm{)}} $ below)    corresponds to  the simplest mechanism described by the diagram  in Fig. \ref{fig1}$a$. Its amplitude     can be  obtained from  Eq. (\ref{subeqSB6})  if the   $V_{yA}^C\,-\,V_f^C$ term in the transition operator is ignored.
  One notes that  the amplitude $ M^{{\rm{TBDW}}}{\rm{(}}E_i,\,cos\theta{\rm{)}}\,$ passes to the amplitude of the     so-called    ``post''-approximation of the  DWBA \cite{Austern1964}   if all the terms of $\bigtriangleup V_{f,i}^C$ contained in the transition operators of Eqs. (\ref{subeqSB6})  and (\ref{subeqSB7})   are ignored.

\begin{center}
\vspace{1cm}
  {\bf III.   DISPERSION APPROACH AND   DWBA}
  \end{center}
  \bigskip

  The amplitudes given by  Eqs. (\ref{subeqSB6}) and (\ref{subeqSB7})   defines   the behavior  both of the  amplitude $ M^{{\rm{TB}}}{\rm{(}}E_i,\,cos\theta{\rm{)}}$ at $\cos\theta$ = $\xi$ \cite{Mukh10}  and   of    the   corresponding   peripheral partial amplitudes at $l_i\,>>$ 1 \cite{Popov1964}.  Besides,    owing to   the presence of nuclear distortions in the entrance and exit states,  these amplitudes have also the singularities  located farther from the physical (-1$\le\cos\theta\le$1) region than  $\xi$.   Therefore, according  to  \cite{Mukh10},  the behavior of the $M^{{\rm{DW}}}_{{\rm{pole}}}{\rm{(}}E_i,\,cos\theta{\rm{)}} $ and  $M^{{\rm{DW}}}_{\rm{post}}{\rm{(}}E_i,\,cos\theta{\rm{)}}$ amplitudes   near  $\cos\theta\,=\,\xi$, denoted by  $M^{{\rm(}s{\rm{)\,DW}}}_{{\rm{pole}}}{\rm{(}}E_i,\,cos\theta{\rm{)}}$  and $M^{{\rm(}s{\rm{),DW}}}_{{\rm{post}}}{\rm{(}}E_i,\,cos\theta{\rm{)}}$ below, respectively,  can be defined from Eq. (\ref{subeqSB6}) as  the Coulomb-nuclear distortions in the entrance and exit states  are substituted  by purely Coulomb ones.  The singular $M^{{\rm(}s{\rm{)\,DW}}}_{{\rm{pole}}}{\rm{(}}E_i,\,cos\theta{\rm{)}}$ and  $M^{{\rm(}s{\rm{)\,DW}}}_{{\rm{post}}}{\rm{(}}E_i,\,cos\theta{\rm{)}}$ amplitudes near at $\cos\theta=\,\xi$  can be presented in the form:
  \begin{equation}
 M^{{\rm{DW}}}_{{\rm{pole}}}{\rm{(}}E_i,\,cos\theta{\rm{)}}\approx M^{{\rm(}s{\rm{)}}\,{\rm{DW}}}_{{\rm{pole}}}{\rm{(}}E_i,\,cos\theta{\rm{)}}= N^{{\rm{DW}}}_{{\rm{pole }}}{\tilde{M}}^{{\rm(}s{\rm{)}}}_{{\rm{pole}}}{\rm{(}}E_i,\,cos\theta{\rm{)}}
\label{subeqpole}
\end{equation} and
    \begin{equation}
 M^{{\rm{DW}}}_{{\rm{post}}}{\rm{(}}E_i,\,cos\theta{\rm{)}}\approx M^{{\rm(}s{\rm{)}}\,{\rm{DW}}}_{{\rm{post}}}{\rm{(}}E_i,\,cos\theta{\rm{)}}\,=\,{\cal{R}}^{{\rm{DW}}}_{{\rm{post}}}M^{{\rm(}s{\rm{)}}\,{\rm{DW}}}_{{\rm{pole}}}{\rm{(}}E_i,\,cos\theta{\rm{)}},
 \label{subeqSB8}
\end{equation}
\begin{equation}{\cal{R}}^{{\rm{DW}}}_{{\rm{post}}}\,=\,N^{{\rm{DW}}}_{{\rm{post}}}/N^{{\rm{DW}}}_{{\rm{pole }}},
\label{subeqSB8a}
\end{equation}
where the explicit forms of   ${\tilde{M}}^{{\rm(}s{\rm{)}}}_{{\rm{pole}}}{\rm{(}}E_i,\,cos\theta{\rm{)}}$ and of the corresponding  peripheral partial amplitudes at $l_i>>$ 1 (${\tilde{M}}^{{\rm(}as{\rm{)}}}_{l_i;\,{\rm{pole}}}{\rm{(}}E_i{\rm{)}}$) are     given   by Eqs. (A1) and (A2) of Appendix A.  The   peripheral partial amplitudes at $l_i>>$ 1 corresponding to the  $M^{{\rm{DW}}}_{{\rm{pole}}}{\rm{(}}E_i,\,cos\theta{\rm{)}}$ and $M^{{\rm{DW}}}_{{\rm{post}}}{\rm{(}}E_i,\,cos\theta{\rm{)}}$ amplitudes are given in Eqs. (A3) and (A4) of Appendix A, respectively.
      In (\ref{subeqSB8a}),    $N^{{\rm{DW}}}_{{\rm{pole}}}$ and $N^{{\rm{DW}}}_{{\rm{post}}}$ are the Coulomb  renormalized  factors (CRF's) for  the $M^{{\rm(}s{\rm{)}}\,{\rm{DW}}}_{{\rm{pole}}}{\rm{(}}E_i,\,cos\theta{\rm{)}}$     and    $M^{{\rm(}s{\rm{)}}\,{\rm{DW}}}_{{\rm{post}}}{\rm{(}}E_i,\,cos\theta{\rm{)}}$ amplitudes, respectively.    One notes that the CRF's above  are  complex numbers and depend on the energy $E_i$, the binding energies $\varepsilon_{ay}$ and $\varepsilon _{Aa}$ as well as  the Coulomb  ($\eta_x$, $\eta_B$, $\eta_i$ and $\eta_f$) parameters, where  $\eta_i$ and $\eta_f$ are the
 Coulomb parameters in the entrance and exit channels, respectively.
 Below, for the sake of simplicity of the inscription, in   the $N^{{\rm{DW}}}_{{\rm{pole}}}$ and $N^{{\rm{DW}}}_{{\rm{post}}}$  the dependences mentioned above   will not   be pointed out explicitly, except  only the dependence  on   $E_i$. This point is also related to the $N^{{\rm{TBDM}}}$  and $N^{{\rm{TBDW}}}$ CRF's, which are given by Eq. (\ref{subeqSB10}) below  and Eq. (A38) in Appendix.
   The explicit forms of the   $N^{{\rm{DW}}}_{{\rm{pole}}} $ and $N^{{\rm{DW}}}_{{\rm{post}}}$ CRF's  are presented  in  \cite{Mukh10} by  the expressions  of    (14) and (26), respectively, which contain the integrals with  the cumbersome   integrand. Nevertheless, the approximated analytical   forms for the CRF's  can be   derived   and they are presented in Appendix A (see Eqs. (A5) - (A26) there).

The accuracy of the    $M^{{\rm(}s{\rm{)}\,DW}}_{{\rm{post}}}{\rm{(}}E_i,\,cos\theta{\rm{)}}$ amplitude can be defined  by the extent of proximity of the CRF's $N^{\rm{DW}}_{{\rm{pole}}}$,  $N^{{\rm{DW}}}_{{\rm{post}}}$ and  $N^{\rm{TBDM}}$ each other. The $N^{\rm{TBDM}}$ CRF determines the power of leading singular  term
 ($M^{{\rm{(}}s{\rm{)}}\,{\rm{TBDM}}}{\rm{(}}E_i,\,cos\theta{\rm{)}}$) of the
      exact (in the model of three ($A$, $a$ and $y$) charged particles)   $M^{{\rm{TBDM}}}{\rm{(}}E_i,\,cos\theta{\rm{)}}$  amplitude for the pure sub-barrier peripheral reaction (\ref{subeq1}) at $\cos\theta=\xi$, which has the form \cite{Av86}
     \begin{equation}
 M^{{\rm{TBDM}}}{\rm{(}}E_i,\,cos\theta{\rm{)}}\approx M^{{\rm{(}}s{\rm{)}}\,{\rm{TBDM}}}{\rm{(}}E_i,\,cos\theta{\rm{)}}=N^{\rm{TBDM}}{\tilde{M}}^{{\rm{(}}s{\rm{)}}}_{\rm{pole}}{\rm{(}}E_i,\,cos\theta{\rm{)}},
\label{subeqSBDM}
\end{equation}
  The explicit form of the CRF  $N^{TBDM}$  was  obtained in  \cite{Av86} by combination of   the dispersion method  and the three-body Faddeev's equations and is also  given by     the expressions (A27) -- (A31) of Appendix A. Nevertheless, one notes only that the $M^{{\rm{(}}s{\rm{)}}\,{\rm{TBDM}}}{\rm{(}}E_i,\,cos\theta{\rm{)}}$ amplitude includes also all possible subsequent mutual  Coulomb rescattering of  the $A$, $a$ and $y$ particles in the intermediate state. They  are also  described by  infinite series of diagrams constructed on the   basis of the diagrams in Figs. \ref{fig1}$a$ and  \ref{fig1}$b$ in which the four-ray vertexes describing  the Coulomb $Aa$-, $yA$- and $ay$-rescattering  correspond to the  total off-shell Coulomb amplitudes \cite{DM1966} but not their Born approximations that   used in  \cite{Mukh10}.

   As is seen from Eqs. (\ref{subeqpole}), (\ref{subeqSB8}) and (\ref{subeqSBDM}),       the $M^{{\rm{DW}}}_{\rm{pole}}{\rm{(}}E_i,\,cos\theta{\rm{)}}$,
$M^{{\rm{DW}}}_{\rm{post}}{\rm{(}}E_i,\,cos\theta{\rm{)}}$ and  $M^{{\rm{TBDM}}}$ ${\rm{(}}E_i,\,cos\theta{\rm{)}}$
 amplitudes  near  $\cos\theta=\,\xi$  behave identically but they differ from each other only  by the power.  Then,    the behavior of the exact  three-body  $ M^{{\rm{TB}}}{\rm{(}}E_i,\,cos\theta{\rm{)}}$ DWBA amplitude near  the  singularity at $\cos\theta$ = $\xi$,    denoted by $ M^{{\rm{(}}s{\rm{)}}\,{\rm{TBDW}}}{\rm{(}}E_i,\,cos\theta{\rm{)}}$ below, can be presented in   the form as
\begin{equation}
M^{ {\rm{TB}}}{\rm{(}}E_i,\,cos\theta{\rm{)}}\approx M^{{\rm(}s{\rm{)}}\,{\rm{TBDW}}}{\rm{(}}E_i,\,cos\theta{\rm{)}}\,=\,{\cal{R}}^{TBDM}M^{{\rm(}s{\rm{)}}}_{{\rm{pole}}}{\rm{(}}E_i,\,cos\theta{\rm{)}},
 \label{subeqSB9}
\end{equation} where
\begin{equation}
{\cal{R}}^{{\rm{TBDM}}}{\rm{(}}E_i{\rm{)}}\,=\,N^{{\rm{TBDM}}}{\rm{(}}E_i{\rm{)}}/N^{{\rm{DW}}}_{{\rm{pole}}}{\rm{(}}E_i{\rm{)}}.
\label{subeqSB10}
\end{equation} One notes that the expressions  (\ref{subeqSB9}) and (\ref{subeqSB10}) combine  the dispersion method in a correct way by  taking into account  the three-body Coulomb dynamics in the transfer mechanism and   the Coulomb distorted effects in the entrance and exit states, as it  is  done   within of the framework of  the conventional DWBA.
 Besides,   as is seen from   Appendix A, the amplitudes  given by  Eqs. (\ref{subeqSB8}) and (\ref{subeqSB9}) can   define   the  peripheral partial amplitudes for $l_i\,>>$ 1 of the conventional DWBA and the generalized DWBA, respectively, which differ  from each other by their power. Their comparison each other would  make   it possible to test   the accuracy  of  both the pole-approximation and the post form of the conventional  DWBA. Meanwhile,  the various relationships are possible between the CRFs $N^{{\rm{DW}}}_{{\rm{pole}}}$, $N^{{\rm{DW}}}_{{\rm{post}}}$ and $N^{{\rm{TBDM}}}$ and their ratios ${\cal{R}}^{{\rm{DW}}}_{{\rm{post}}}$, ${\cal{R}}^{{\rm{TBDW}}}_{{\rm{post}}}$ and ${\cal{R}}^{{\rm{TBDM}}}$, where ${\cal{R}}^{{\rm{TBDW}}}_{{\rm{post}}}=N^{{\rm{TBDM}}}/N^{{\rm{DW}}}_{{\rm{post }}}$.

 However, as is pointed out  in \cite{Mukh10},   in Eqs. (\ref{subeqSB8a}) and (\ref{subeqSB10})  the most ``dramatic'' situation arises for the calculated CRF's and their ratios above  at the values of    the Coulomb parameters $\eta_x$,  $\eta_B$   or their sum $\eta_{xB}$ ($\eta_{xB}=\eta_x+\,\eta_B$) near to a natural number.  This situation is related  the so-called ``damatic'' case \cite{Mukh10}.
  In Table \ref{table1}, as an  example related to the ``dramatic'' case,
 the results of the  calculations of the  CRF's for first two the specific reactions are presented (see the first--eighth lines). Those reactions
  were considered in Refs. \cite{Azh1999,Azh1999-1,Tab06} within the framework of the post form of DWBA.
  In Table \ref{table1}, for simplicity,  the renormalized CRF's ${\tilde{N}}_{{\rm{pole}}}^{{\rm{DW}}}=\,N^{{\rm{DW}}}_{{\rm{pole}}}/\Gamma$, ${\tilde{N}}_{{\rm{post}}}^{{\rm{DW}}}=\,N^{{\rm{DW}}}_{{\rm{post}}}/\Gamma$ and $\tilde{N}^{{\rm{TBDM}}}=\,N^{{\rm{TBDM}}}/\Gamma$  are presented since all the CRF's  ($N^{{\rm{DW}}}_{{\rm{pole}}}$, $N^{{\rm{DW}}}_{{\rm{pole}}}$ and $N^{{\rm{TBDM}}}$) contain the common multiplier  $\Gamma{\rm{(}}\equiv\Gamma{\rm{(1}}\,-\,\eta_{xB}\, +\,i\eta_{ij}{\rm{)}}{\rm{)}}$,  where  $\Gamma{\rm{(}}\cdot\cdot\cdot{\rm{)}}$  is the Euler's function,  $\eta_{if}=\,\eta_i\,+\,\eta_f$, and     $\eta_i$ and $\eta_f$ are the Coulomb parameters for the entrance and exit channels, respectively. Hence, the ratios of the CFR's presented in the fifth column of  Table \ref{table1} do not depend on the multiplier $\Gamma$.     As is seen from Table \ref{table1}, the values of  the  $\tilde{N}^{{\rm{DW}}}_{{\rm{pole}}}$, $\tilde{N}^{{\rm{DW}}}_{{\rm{post}}}$ and $\tilde{N}^{{\rm{TBDM}}}$ factors   calculated in the present work  for the peripheral proton transfer
  ${\rm {^{10}B(^7Be,\,^8B)^9Be}}$ \cite{Azh1999} and  ${\rm {^{14}N(^7Be,\,^8B)^{13}C}}$  \cite{Azh1999-1,Tab06}    reactions at the projectile energy of 85 MeV differ significantly from each other.
  Besides,   the calculated values of    $\mid {\cal{R}}^{{\rm{DW}}}_{{\rm{post}}}\mid$= $\mid\tilde{N}^{{\rm{DW}}}_{{\rm{post}}}/ \tilde{N}^{{\rm{DW}}}_{{\rm{pole }}}\mid$,  $\mid{\cal{R}}^{{\rm{TBDW}}}_{{\rm{post}}}\mid=\,\mid\tilde{N}^{{\rm{TBDM}}}/\tilde{N}^{{\rm{DW}}}_{{\rm{post }}}\mid$   and  $\mid{\cal{R}}^{{\rm{TBDM}}}\mid=\,\mid\tilde{N}^{{\rm{TBDM}}}/\tilde{N}^{{\rm{DW}}}_{{\rm{pole}}}\mid$, which are also presented in the curly brackets of the last column of Table  \ref{table1},  noticeably differ from each other. One notes that  the  CRF $\tilde{N}^{{\rm{TBDM}}}$ determines the power of the peripheral partial amplitudes at $l_i>>$ 1 of the true three-body $ M^{{\rm{TBDM}}}{\rm{(}}E_i,\,cos\theta{\rm{)}}$ amplitude. Therefore, it is clear that the calculations of the peripheral partial amplitudes at $l_i>>$1, which  are  determined by Eq. (A4) and are dominant in the DWBA amplitude  of
    the   reactions  considered above (at least in the angular range of the main peak of the angular distribution),   cannot be  performed only with  the account   the first order of the perturbation theory in $\Delta V_f^{{\rm{C}}}$ in the amplitude (\ref{subeqSB5}). Hence,
     the expressions  (\ref{subeqSB9}) and  (\ref{subeqSB10}) cannot be used  for the specific peripheral proton transfer reactions considered above.

A   provenance of the main reason of the ``dramatic'' case    is  discussed in detail in Appendix A. Nevertheless, we should only note   the following fact. In that case,  as noted in Appendix A, in the transition operator of  the expressions   (\ref{subeqSB6}) and (\ref{subeqSB7}),  the poor  convergence  occurs for    a series of  the power expansion over $\Delta V_{i,f}^{{\rm{C}}}$.
It is mainly caused owing to  the  presence  of  the vertex Coulomb $F_C$[=$F_C$($\eta_x,\,\eta_B$)] factor  as a multiplier in  the expressions for the $\tilde{N}^{{\rm{DW}}}_{{\rm{pole}}}$ and $\tilde{N}^{{\rm{DW}}}_{{\rm{post}}}$   CRF's derived  within   the conventional DWBA (see Appendix A).  As is shown in Appendix, the $F_C$ factor, which    is  defined   by Eq. (A17) of Appendix A,   enters  implicitly     the $\tilde{N}^{{\rm{DW}}}_{{\rm{pole}}}$ and $\tilde{N}^{{\rm{DW}}}_{{\rm{post}}}$   CRF's  presented approximately in the forms  of Eqs. (A25) and (A25) of Appendix.    In the ``dramatic'' case,  as it is shown by the calculations performed by us,  the value  of the  $F_C$  factor is not    sufficiently close to unity. It happen when the values of the Coulomb parameters $\eta_{x}$, $\eta_{B}$ or their sum ($\eta_{xB}=\eta_{x}+\eta_{B}$) being in the vicinity of a natural number \cite{Mukh10}. It mainly is  one of the main reasons of initiation of this difference observed   between  the $\tilde{N}^{{\rm{DW}}}_{{\rm{pole}}}$,  $\tilde{N}^{{\rm{DW}}}_{{\rm{post}}}$  and  $\tilde{N}^{{\rm{TBDM}}}$ CRF's for the peripheral proton transfer reactions \cite{Azh1999,Azh1999-1,Tab06}. For example, as is seen from Table \ref{table1}, the calculated values of the vertex Coulomb  $F_C$ factor,   are equal to 0.695  ($\eta_{xB}$= 1.823) for the ${\rm {^{10}B(^7Be,\,^8B)^9Be}}$ reaction  and   to 0.366 ($\eta_{xB}$= 1.921) for the ${\rm {^{14}N(^7Be,\,^8B)^{13}C}}$ one, i.e., they differ noticeably from unity.   Perhaps, that is one of the possible reasons why the ANC value for ${\rm{^7Be}}\,+\,p\to{\rm{^8B}}$ recommended in  \cite{Azh1999-1,Tab06} is underestimated as a comparison with that of Refs. \cite{IgYa2008,TYAS2016}, which  leads in turn to  the underestimated astrophysical $S$ factor for the direct radiative capture ${\rm{^7Be(}}p,\,\gamma {\rm{)^8B}}$ reaction at solar energies (see  \cite{IgYa2008,TYAS2016}).
Therefore,  in the ``dramatic'' case,    the next terms ($\bigtriangleup V_f^CG_C\bigtriangleup V_i^C$)    of the transition operator in  the series in $\bigtriangleup V_{f,\,i}^C$ should  directly be  added to the  $ M_{{\rm{post}}}^{{\rm{DW}}}{\rm{(}}E_i,\,cos\theta{\rm{)}}$ (or
$M ^{{\rm{TBDW}}}{\rm{(}}E_i,\,cos\theta{\rm{)}}$) amplitude defined by Eq. (\ref{subeqSB5}). This assertion is suggested by the fact that  the ``dramatic'' case does not occur both  for the peripheral neutron transfer reactions considered in Ref.  \cite{KMYa1988}, where  $\Delta V_f^{C}=V_{yA}^C$ and $F_C$=1 ($\eta_x$=0 and $\eta_B$=0),  and  for  the $A{\rm{(}}d,\,n{\rm{)}}B$ reaction considered in Ref.  \cite{TYAS2016}, where    $\Delta V_f^{C}$=0 and  $F_C$= 1 ($\eta_x$=0 and $\eta_B\ne$0). Besides, as shown in \cite{Mukh10}, the ``dramatic'' case  does not arise    for peripheral charged-particle transfer reactions  as  the values of the vertex Coulomb  $F_C$ factor, calculated at $\eta_x\ne$0
 and $\eta_B\ne$0,  close to 1. The latter occurs    when the values of    the Coulomb parameters $\eta_x$,  $\eta_B$   or their sum $\eta_{xB}$  are not in the vicinity of   a natural number.   This case in \cite{Mukh10}   is called by the ``non-dramatic'' case \cite{Mukh10}. As is seen from Table \ref{table1},  the  peripheral  transfer ${\rm {^9Be(^{10}B,^9Be)^{10}B}}$,   ${\rm{^{16}O(^3He}}, d{\rm {)^{17}F}}$ and ${\rm{^{19}F(}}p,\,\alpha {\rm {)^{16}O}}$  reactions  are related to the ``non-dramatic'' case. Therefore, below those reactions  will be considered    by us  in which the residual  ${\rm{^{10}B}}$ nucleus  is formed in    the ground ($E^*$=0.0; $J^{\pi}$=3$^+$) state, the  first
    ($E^*$= 0.718 MeV; $J^{\pi}$=1$^+$), second ($E^*$= 1.740 MeV; $J^{\pi}$=0$^+$) and third ($E^*$= 2.154 MeV; $J^{\pi}$=1$^+$) excited states (denoted by ${\rm{^{10}B_0}}$, ${\rm{^{10}B_1}}$, ${\rm{^{10}B_2}}$ and ${\rm{^{10}B_3}}$, respectively, below) \cite{Mukh2}, and  the residual  ${\rm {^{17}F}}$ nucleus is formed in  the ground (0.0; $J^{\pi}$=$\frac{{\rm{5}}}{{\rm{2}}}^+$)   and   first ($E^*$=0.495 MeV; $\frac{{\rm{1}}}{{\rm{2}}}^+$) excited states (denoted by ${\rm {^{17}F_0}}$ and ${\rm {^{17}F_1}}$, respectively,  below). While, for the  ${\rm{^{19}F(}}p,\,\alpha {\rm {)^{16}O}}$  reaction [32-34], the residual nucleus is formed in the ground state.

    In   the ninth -- fifty sixth lines   of Table \ref{table1}, the results of the  calculations of the  CRF's and  their ratios   are presented in Table \ref{table1}  for the reactions mentioned above.
   As is seen from  Table  \ref{table1}, for    the  peripheral  reactions related  to the ``non-dramatic'' case  the values the  $F_C$  factor  become   sufficiently close to unity and, consequently,
 the  difference between the values of the CRF's  and their  ratios mentioned above      is  significantly less than between those calculated  for    the ``dramatic''  ${\rm {^{10}B(^7Be,\,^8B)^9Be}}$  and  ${\rm {^{14}N(^7Be,\,^8B)^{13}C}}$    reactions for which    the calculated values of the $F_C$ factor differ considerably  from unity, as noted above.
  This shows the absolute inapplicability of the  ``post''-approximation of the conventional DWBA   used in \cite{Ar96} for  the ${\rm{^{16}O(^3He}}, d{\rm {)^{17}F}}$ DWBA analysis.

It follows from here that  the expressions (\ref{subeqSB8}), (\ref{subeqSB8a}),  (\ref{subeqSB9}) and (\ref{subeqSB10}) can  be used for   the  peripheral  transfer  reactions (\ref{subeq1}), which is   related only  to the ``non-dramatic'' case, including  the  specific peripheral  proton and triton reactions listed   in   Table \ref{table1}.

       For this aim, below we will first show how to obtain the singular part of the pole  $M^{{\rm{DW}}}_{{\rm{pole}}}{\rm{(}}E_i,\,cos\theta{\rm{)}}$  DWBA amplitude corresponding  to   the one-step pole transfer mechanism, which is described by the pole diagram of Fig. \ref{fig1}$a$,   by separating  the contribution  from the nearest singularity $\xi$ to it. Then, from the expression derived for this amplitude,   we obtain the generalized DWBA amplitude valid only for  the ``non-dramatic'' case where the contribution of  the three-body ($A$, $a$ and $y$) Coulomb dynamics of the main transfer mechanism to    the   peripheral partial amplitudes for  $l_i\,>>$ 1    are taken into account  in a correct manner.

\vspace{1cm}
\begin{center}
  {\bf IV.   DISTORTED-WAVE POLE APPROXIMATION}
  \end{center}
\bigskip

 The pole-approximation of the  DWBA  amplitude  can  be obtained  from  Eq.
   (\ref{subeqSB6}). As a result, it has  the form as
 \begin{equation}
  M^{{\rm{DW}}}_{{\rm{pole}}}{\rm{(}}E_i,\,cos\theta{\rm{)}}\,=\,\int d{\mathbf{r}}_id{\mathbf{r}}_f \chi^{(-)^*}_{{\mathbf k}_f}({\mathbf{r}}_f)
I^*_{Aa}({\mathbf{r}}_{Aa}) V_{ay}({\mathbf{r}}_{ay}) I_{ay}({\mathbf{r}}_{ay}) \chi^{(+)}_{{\mathbf{ k}}_i}({\mathbf{r}}_i).
 \label{subeqSB13}
\end{equation}Here ${\mathbf{r}}_i\,\equiv\,{\mathbf{r}}_{xA} $,  ${\mathbf{r}}_f\,\equiv\,{\mathbf{r}}_{yB} $  and
 $$
{\mathbf{r}}_{ay}\,=\,\bar{a}{\mathbf{r}_i}\,-\,\bar{b} {\mathbf{r}_f},
$$
 \begin{equation}
\label{subeqSB14}
 {\mathbf{r}}_{Aa}\,=\,-\,\bar{c}{\mathbf{r}_i}\,+\,\bar{d}{\mathbf{r}_f},
  \end{equation}where $\bar {a}$= $\mu_{Ax}/m_a$, $\bar{b}$= $\mu_{Ax}/\mu_{Aa}$, $\bar{c}$= $\mu_{By}/\mu_{ay}$   and $\bar{d}$= $\mu_{By}/m_a$.
  To obtain the explicit  singular behavior of  $M^{{\rm{DW}}}_{{\rm{pole}}}{\rm{(}}E_i,\,cos\theta{\rm{)}}$ at $\cos\theta=\,\xi$, the integral (\ref{subeqSB13}) should be rewritten  in the momentum representation making use  of Eq. (B1) from  Appendix B and the Fourier integrals for the distorted optical  wave functions in the entrance and exit channels. It takes  the form
\begin{equation}
  M^{{\rm{DW}}}_{{\rm{pole}}}{\rm{(}}E_i,\,cos\theta{\rm{)}}\,=\,\int\frac{ d{\mathbf{k}}}{(2\pi)^3}\frac{d{\mathbf{k}}^{\,\prime}}{(2\pi)^3} \chi^{(+)}_{{\mathbf k}_f}({\mathbf{k}}^{\,\prime})
{\cal{ M}}^{{\rm{DW}}}_{{\rm{pole}}}{\rm{(}}{\mathbf{k}}^{\,\prime},\,{\mathbf{k}}{\rm{)}} \chi^{(+)}_{{\mathbf{ k}}_i}({\mathbf{k}}),
 \label{subeqSB15}
\end{equation}
$$
 {\cal{ M}}^{{\rm{DW}}}_{{\rm{pole}}}{\rm{(}}{\mathbf{k}}^{\,\prime},\,{\mathbf{k}}{\rm{)}}
 \,= \,\sum_{M_a}\langle {\mathbf{k}}^{\,\prime},\,I_{Aa}({\mathbf{q}}_{Aa})|V_{ay}({\mathbf{q}}_{ay})|
I_{ay}  ({\mathbf{q}}_{ay}),\,{\mathbf{k}}\rangle
$$
\begin{equation}
=\,-\,\sum_{M_a}\frac{M_{ay}{\rm{(}}{\mathbf{q}}_{ay}{\rm{)}} M_{Aa}^*{\rm{(}}{\mathbf{q}}_{Aa}{\rm{)}}}{\frac{q_{Aa}^{{\rm{2}}}}{{\rm{2}}\mu_{Aa}}\,+\,\varepsilon_{Aa}}
 \label{subeqSB16}
\end{equation}
 Here ${\cal{ M}}^{{\rm{DW}}}_{{\rm{pole}}}{\rm{(}}{\mathbf{k}}^{\,\prime},\,{\mathbf{k}}{\rm{)}}$ is the off-shell   of the Born (pole) amplitude;     $\chi^{(+)}_{{\mathbf{ k}}_i}({\mathbf{k}})$ and
   $ \chi^{(+)}_{{\mathbf{ k}}_i}({\mathbf{k}})$ are Fourier components of the  distorted wave functions  in the  entrance and exit channels, respectively; $I_{ay}({\mathbf{q}}_{ay})$ and  $I_{Aa}({\mathbf{q}}_{Aa})$ as well as $V_{ay}({\mathbf{q}}_{ay})$    are the  same
    for    the overlap  functions  of the Coulomb-nuclear wave functions  for the bound ($y$ + $a$) and ($A$ + $a$) states as well as for  the Coulomb-nuclear $V_{ay}({\mathbf{r}}_{ay}) $ potential,  respectively;       ${\mathbf{q}}_{ay}\,=\,{\mathbf{k}}_1\,-\,{\mathbf{k}}^{\,\prime}$ and   ${\mathbf{q}}_{Aa}\,=\,-\,{\mathbf{k}}\,+\,{\mathbf{k}}^{\,\prime}_{{\rm{1}}}$, where  ${\mathbf{k}}_1\,=\,{\rm{(}}m_y/m_x{\rm{)}}{\mathbf{k}}$ and ${\mathbf{k}}^{\,\prime}_{{\rm{1}}}\,=\,{\rm{(}}m_A/m_B{\rm{)}}{\mathbf{k}}^{\,\prime}$. In Eq. (\ref{subeqSB16}), $M_{ay}{\rm{(}}{\mathbf{q}}_{ay}{\rm{)}}$   is  the vertex matrix element (or so-called the vertex function) for the virtual decay $x\to\,y\,+\,a$. Its explicit form is similar  to that  for the virtual decay  $B\to\,A\,+\,a$ given by  Eq. (B1)   in Appendix B.

  Using Eq. (B1) from Appendix B and the  corresponding expression  for  $M_{ay}{\rm{(}}{\mathbf{q}}_{ay}{\rm{)}}$,   the  ${\cal{ M}}^{{\rm{DW}}}_{{\rm{pole}}}{\rm{(}}{\mathbf{k}}^{\,\prime},\,{\mathbf{k}}{\rm{)}} $ amplitude can be presented in  the form
$$
 {\cal{ M}}^{{\rm{DW}}}_{{\rm{pole}}}{\rm{(}}{\mathbf{k}}^{\,\prime},\,{\mathbf{k}}{\rm{)}}
 \,=\, \sum_{\alpha_B\alpha_xM_a}C{\rm{(}}\alpha_B\alpha_x;\,{\rm{(}}J,\,M{\rm{)}}_{x,\,A,\,y,\,B};\,J_aM_a{\rm{)}}
 {\tilde{{\cal{ M}}}}^{{\rm{DW}}}_{{\rm{pole}};\,\alpha_B\alpha_x}{\rm{(}}{\mathbf{k}}^{\,\prime},\,{\mathbf{k}}{\rm{)}},
$$
\begin{equation}
{\tilde{{\cal{ M}}}}^{{\rm{DW}}}_{{\rm{pole}};\,\alpha_B\alpha_x}{\rm{(}}{\mathbf{k}}^{\,\prime},\,{\mathbf{k}}{\rm{)}}\,=\,I^{^*}_{Aa;\,\alpha_B}{\rm{(}}{\mathbf{q}}_{Aa}{\rm{)}}W_{ay;\,\alpha_x}{\rm{(}}{\mathbf{q}}_{ay}{\rm{)}}.
 \label{subeqSB17}
\end{equation} Here
$$
C{\rm{(}}\alpha_B\alpha_x;\,{\rm{(}}J,\,M{\rm{)}}_{x,\,A,\,y,\,B};\,J_aM_a{\rm{)}}\,=\,C_{j_x\nu_xJ_yM_y}^{J_xM_x}C_{l_x\mu_xJ_aM_a}^{j_x\nu_x}C_{j_B\nu_BJ_AM_A}^{J_BM_B}
C_{l_B\nu_BJ_aM_a}^{j_B\nu_B}
$$  and
\begin{equation}
I_{Aa;\,\alpha_B}{\rm{(}}{\mathbf{q}}_{Aa}{\rm{)}}=\,-\,{\rm{2}}\mu_{Aa}\frac{W_{Aa;\,\alpha_B}{\rm{(}}{\bf q}_{ya}{\rm{)}}}{q_{Aa}^{{\rm{2}}}+\kappa_{Aa}^{{\rm{2}}}},
\label{subeqSB17a}
\end{equation}  where $\alpha_{\lambda}\,=\,{\rm{(}}l_{\lambda},\,\mu_{\lambda},\,j_{\lambda},\,\nu_{\lambda}{\rm{)}}$;  $\lambda\,=\,x,\,B$; ($J,\,M$) is the set of $J_{\lambda}$ and $M_{\lambda}$ ($\lambda\,=\,x,\,A,\,y,\,B$),   and
$$
W_{Aa;\,\alpha_B}{\rm{(}}{\bf q}_{Aa}{\rm{)}}=\,\sqrt{{\rm{4}}\pi}G_{Aa;\,l_Bj_B}{\rm{(}}q_{Aa}{\rm{)}}Y_{l_B\mu_B}(\hat{{\mathbf q}}_{Aa}{\rm{)}},
$$
\begin{equation}
W_{ay;\,\alpha_x}{\rm{(}}{\mathbf{q}}_{ay}{\rm{)}}=\,\sqrt{{\rm{4}}\pi}G_{ay;\,l_xj_x}{\rm{(}}q_{ay}{\rm{)}}Y_{l_x\mu_x}(\hat{{\mathbf q}}_{ay}{\rm{)}}
\label{subeqSB17b}
\end{equation} are     the reduced  vertex functions for the virtual decays  $B\,\to\,A\,+\,a$  and $x\,\to\,\,y\,+\,a$, respectively.

\begin{figure}
\begin{center}
\includegraphics[height=4cm, width=14cm]{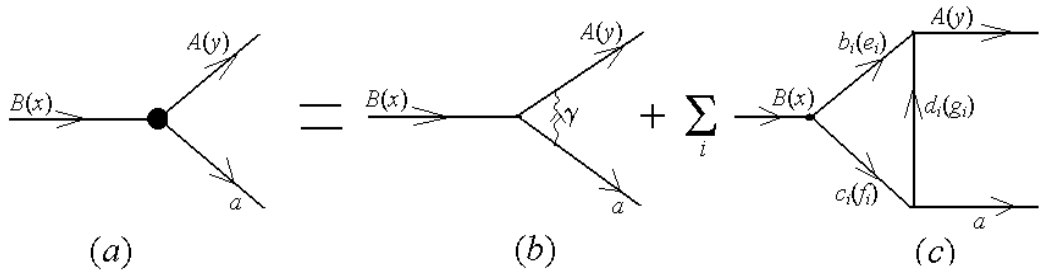}
 \end{center}
\caption{\label{fig2} Diagrams describing   the matrix element for  the virtual decay $B\,\to\,A\,+\,a$ ($x\,\to\,y\,+\,a$).}
\end{figure}

  In the presence of the long-range Coulomb interactions between particles of $A$,  $a$  and $y$, the reduced  vertex functions    can be described by the sum of the nonrelativistic diagrams plotted in Fig. \ref{fig2}. The diagram  in Fig. \ref{fig2}$b$ corresponds  to the Coulomb part of the   vertex function, which  has a branch point singularity at  $q^{{\rm{2}}}_{Aa}+\kappa^{{\rm{2}}}_{Aa}$ =0 ($q^{{\rm{2}}}_{ay}\,+\,\kappa^{{\rm{2}}}_{ay}$ = 0) and   generates    the singularity $\xi$ of the $  M^{{\rm{DW}}}_{{\rm{pole}}}{\rm{(}}E_i,\,cos\theta{\rm{)}}$ amplitude  at  $k=k_i$ and $k^{\prime}=k_f$. The sum in Fig. \ref{fig2}$c$ involves more complicated diagrams  and this part  of the vertex function  corresponds to the   Coulomb-nuclear vertex function, which   is regular at the point $q_{Aa}\,=\,i\kappa_{Aa}$  ($q_{ay}\,=\,i\kappa_{ay}$).  Then, the vertex functions  $W_{Aa;\,\alpha_B}{\rm{(}}{\bf q}_{Aa}{\rm{)}}$  and $W_{ay;\,\alpha_x}{\rm{(}}{\mathbf{q}}_{ay}{\rm{)}} $ can be presented in the forms \cite{DDh1971}
  $$
  W_{Aa;\,\alpha_B}{\rm{(}}{\mathbf q}_{Aa}{\rm{)}}=
 W^{{\rm{(C)}}}_{ Aa;\,\alpha_B}{\rm{(}}{\mathbf q}_{Aa}{\rm{)}}\,+\,
W^{{\rm{(CN)}}}_{Aa;\,\alpha_B}{\rm{(}}{\mathbf q}_{Aa}{\rm{)}},\,\,\,
  $$
\begin{equation}
W_{ay;\,\alpha_x}{\rm{(}}{\bf q}_{ay}{\rm{)}}=\,
W^{{\rm{(C)}}}_{ay;\,\alpha_x}{\rm{(}}{\mathbf q}_{ay}{\rm{)}}\,+\,
W^{{\rm{(CN)}}}_{ya;\,\alpha_x}{\rm{(}}{\mathbf q}_{ay}{\rm{)}}.
\label{subeqSB17WBx}
\end{equation} Here, the $W^{{\rm{(C)}}}_{Aa;\,\alpha_B}$ and $W^{{\rm{(CN)}}}_{Aa;\,\alpha_B}$
($W^{{\rm{(C)}}}_{ay;\,\alpha_x}$ and $W^{{\rm{(CN)}}}_{ay;\,\alpha_x}$) functions are the pure Coulomb and Coulomb-nuclear parts of the vertex functions, respectively.
   All terms of the sum in  Fig. \ref{fig2}$c$  have   dynamic singularities, which are  generated     by internuclear interactions  responsible for    the so-called  dynamic recoil effects \cite{Austern1964,LOLA}.  These singularities are   located   at the points  $q_{Aa}\,=\,i\lambda_i\kappa_i$ and $q_{ay}\,=\,i\bar{\lambda}_i\bar{\kappa}_i$  \cite{BDP1963,Blokh2013}, where $\lambda_i\,=\,m_A/m_{b_i}$,  $\kappa_i=\,  \kappa_{b_ic_i}\,+\,\kappa_{b_id_i}$,  $\bar{\lambda}_i\,=\,m_y/m_{e_i}$ and $\bar{\kappa}_i=\, \kappa_{e_if_i}\,+\,\kappa_{e_ig_i} $. At  $k=k_i$ and $k^{\prime}=k_f$, they   generate   other singularities $\xi_i$ and $\bar{\xi}_ i $ of the $  M^{{\rm{DW}}}_{{\rm{pole}}}{\rm{(}}E_i,\,cos\theta{\rm{)}}$ amplitude, which are determined by
$$
\xi_i=\,\frac{{\rm{(}}k_im_{b_i}/m_A{\rm{)}}^{{\rm{2}}}\,+\,{\rm{(}}k_fm_{b_i}/m_B{\rm{)}}^{{\rm{2}}}\,+\,\kappa_i^{{\rm{2}}}}{{\rm{2}}k_ik_fm_{b_i}^{{\rm{2}}}/m_Am_B}
$$ and
$$
\bar{\xi}_i=\,\frac{{\rm{(}}k_im_{e_i}/m_x{\rm{)}}^{{\rm{2}}}\,+\,{\rm{(}}k_fm_{e_i}/m_y{\rm{)}}^{{\rm{2}}}\,+\,\bar{\kappa}_i^{{\rm{2}}}}{{\rm{2}}k_ik_fm_{e_i}^{{\rm{2}}}/m_xm_y}.
$$
As a rule, they   are   located  farther from the physical (-1 $\le\,\cos\theta\,\le$ 1) region than $\xi$ ($\xi_i\,>\,\xi$ and $\bar{\xi}_i\,>\,\xi$) \cite{BDP1963,DDh1971}. For illustration, the  positions of these singularities ($\xi$, $\xi_i$ and $\bar{\xi}_i$), $\kappa$, $\kappa_i$ and $\bar{\kappa}_i$ calculated  for the  specific  peripheral reactions    are presented
 in Table \ref{table2}.  There, the positions of only several  singularities $\xi_i$ ($\bar{\xi}_i$), which are the  closest to the singularity $\xi$, are presented.   As can be  seen from Table \ref{table2}, the singularities $\xi_i$ and $\bar{\xi}_i$    are located farther from the physical (-1$\le\,\cos\theta\,\le$ 1) region  than the singularity $\xi$.  Besides, in the diagram in Fig. \ref{fig2}$c$, the particle $d_i$($g_i$) can be the neutral $\pi^{{\rm{0}}}$ pion meson. In this case, the positions of the singularities  are
  located at the point $q_{Aa}\,=\,i{\rm{(}}\kappa_{Aa}\,+\,\lambda_{\pi^{{\rm{0}}}}^{{\rm{-1}}}{\rm{)}}$     ($q_{ay}\,=\,i{\rm{(}}\kappa_{ay}\,+\, \lambda_{\pi^{{\rm{0}}}}^{{\rm{-1}}}{\rm{)}}$), where $\lambda_{\pi^0}=\hbar/m_{\pi^{\rm{0}}}c$ is the Compton wave-length of the particle $\pi^{{\rm{0}}}$   equal to 1.414 fm ($ \lambda_{\pi^{{\rm{0}}}}^{{\rm{-1}}}$= 0.707 fm$^{{\rm{-1}}}$). Therefore, the corresponding singularity   $\xi_i$ ($\bar{\xi}_i$) is also located farther from the physical region on the $\cos\theta$-plane  than the singularity $\xi$.

     For the surface reaction (\ref{subeq1}), the contribution of the interior nuclear range to the $M^{{\rm{DW}}}_{{\rm{pole}}}{\rm{(}}E_i,\,cos\theta{\rm{)}} $ amplitude, which is generated  by the singularities of the $W^{{\rm{(CN)}}}_{Aa;\,\alpha_B}$ and  $W^{{\rm{(CN)}}}_{ay;\,\alpha_x}$ functions,  can be ignored at least in the angular range of the main peak
      of the angular distribution \cite{DDM1973,KMYa1988}.   Therefore, in Eqs. (\ref{subeqSB17}) and (\ref{subeqSB17a}),     the vertex functions  given by  Eq. (\ref{subeqSB17b})  can be replaced by their singular behavior of the  corresponding  Coulomb parts  in the vicinity of    nearest singularities to the physical points $q_{ay}^{{\rm{2}}}$=0 and $q_{Aa}^{{\rm{2}}}$=0   (the branch points). These singularities   are  located  at the points  $q_{ay}^{{\rm{2}}}$=-$\kappa_{ay}^{{\rm{2}}}$ ($q_{ay}\,=\,i\kappa_{ay}$)  and  $q_{Aa}^{{\rm{2}}}$=-$\kappa_{Aa}^{{\rm{2}}}$ ($q_{Aa}\,=i\kappa_{Aa}$)   on the $q_{ay}^{{\rm{2}}}$-and $q_{Aa}^{{\rm{2}}}$-planes, respectively.  In the vicinities of these singularity points,   the   vertex functions above    behave as \cite{DDh1971}
$$
 W^{{\rm{(C)}}}_{\beta\gamma;\,\alpha_{\alpha}}{\rm{(}}{\mathbf q}_{\beta\gamma}{\rm{)}}
  \,\simeq\, W^{{\rm{(C}};\,s{\rm{)}}}_{\beta\gamma;\,\alpha_{\alpha}}{\rm{(}}{\mathbf q}_{\beta\gamma}{\rm{)}}
    =\,\sqrt{{\rm{4}}\pi}\Gamma{\rm{(}}1\,-\,\eta_{\alpha}{\rm{)}}
$$
 \begin{equation}
\label{subeqSB18}
\times\Big(\frac{q_{\beta\gamma}}{i\kappa_{\beta\gamma}}\Big)^{l_{\alpha}}
\Big(\frac{{\mathbf{q}}^{\rm{2}}_{\beta\gamma}
\,+\,\kappa_{\beta\gamma}^{\rm{2}}}
{{\rm{4}}i\kappa_{\beta\gamma}^{\rm{2}}}\Big)^{\eta_{\alpha}}
G_{\beta\gamma;\,\,l_{\alpha}j_{\alpha}}
{\rm{(}}i\kappa_{\beta\gamma}{\rm{)}}Y_{l_{\alpha}\nu_{\alpha}}{\rm{(}}\hat {\bf q}_{\beta\gamma}{\rm{)}}
\end{equation} for  $q_{\beta\gamma}\,\to\,
i\kappa_{\beta\gamma}$,  where $G_{\beta\gamma;\,\,l_{\alpha}j_{\alpha}}{\rm{(}}i\kappa_{\beta\gamma}{\rm{) (}}\equiv G_{\beta\gamma;\,\,l_{\alpha}j_{\alpha}}$) is the NVC for the virtual decay $\alpha\,\to\,\beta +\gamma$($\gamma$ = $a$;    $\alpha$ = $x$   and $\beta$ = $y$ for the virtual decay $x\,\to\,y\,+\,a$, and    $\alpha$ = $B$   and $\beta$ = $A$ for the virtual decay   $B\,\to\,A\,+\,a$).

As is seen from  Eqs. (\ref{subeqSB17}), (\ref{subeqSB17a}) and  (\ref{subeqSB18}), the off-shell Born amplitude ${\cal{ M}}^{{\rm{DW}}}_{{\rm{pole}}}{\rm{(}}{\mathbf{k}}^{\,\prime},\,{\mathbf{k}}{\rm{)}}$ (\ref{subeqSB17}) at ${\mathbf{k}} $ = ${\mathbf{k}}_i$ and  ${\mathbf{k}}^{\,\prime}$ = ${\mathbf{k}}_f$ has the  nearest dynamic singularity at $\cos\theta$ = $\xi$.
 Then, the   ${\tilde{{\cal{ M}}}}^{{\rm{DW}}}_{{\rm{pole}};\,\alpha_B\alpha_x}{\rm{(}}{\mathbf{k}}^{\,\prime},\,{\mathbf{k}}{\rm{)}}$ amplitude  in the approximation (\ref{subeqSB18}) takes the form
 \begin{equation}
\label{subeqSB19}
{\tilde{{\cal{ M}}}}^{{\rm{DW}}}_{{\rm{pole}};\,\alpha_B\alpha_x}{\rm{(}}{\mathbf{k}}^{\,\prime},\,{\mathbf{k}}{\rm{)}}\,\approx\,{\tilde{{\cal{ M}}}}^{{\rm{(}}s{\rm{)}};\,{\rm{DW}}}_{{\rm{pole}};\,\alpha_B\alpha_x}{\rm{(}}{\mathbf{k}}^{\,\prime},\,{\mathbf{k}}{\rm{)}}\,=\,I^{^*(s)}_{Aa;\,\alpha_B}{\rm{(}}{\mathbf{q}}_{Aa}{\rm{)}}W_{ay;\,\alpha_x}^{(s)}{\rm{(}}{\mathbf{q}}_{ay}{\rm{)}},
\end{equation} where
 \begin{equation} \label{subeqSB20x}
  W_{ay;\,\alpha_x}^{{\rm{(}}s{\rm{)}}}{\rm{(}}{\bf q}_{ay}{\rm{)}}=\,\sqrt{{\rm{4}}\pi}G_{ay;\,l_xj_x} \Gamma{\rm{(}}1-\eta_{ay}{\rm{)}}\Big(\frac{q_{ay}}{i\kappa_{ay}}\Big)^{l_x}\Big(\frac{q_{ay}^{{\rm{2}}}+\kappa_{ay}^{{\rm{2}}}}{{\rm{4}}i\kappa^{{\rm{2}}}_{ay}}\Big)^{\eta_x}Y_{l_x\nu_x}(\hat {\bf q}_{ay}),
\end{equation}
$$
  I_{Aa;\,\alpha_B}^{*\,{\rm{(}}s{\rm{)}}}{\rm{(}}{\bf q}_{Aa}{\rm{)}}=\,-\sqrt{4\pi}G_{Aa;\,l_Bj_B}  \Gamma{\rm{(}}1\,-\,\eta_B{\rm{)}}
  \left(\frac{q_{Aa}}{i\kappa_{Aa}}\right)^{l_B}
  \left(\frac{q_{Aa}^{{\rm{2}}}\,+\,
  \kappa_{Aa}^{{\rm{2}}}}{{\rm{4}}
  i\kappa^{{\rm{2}}}_{Aa}}\right)
  ^{\eta_B}
$$
 \begin{equation}  \label{subeqSB20}
 \times\frac{{\rm{2}}\mu_{Aa}}{q_{Aa}^{{\rm{2}}}\,+
  \,\kappa_{Aa}^{{\rm{2}}}}Y^*_{l_B\nu_B}{\rm{(}}\hat {\mathbf q}_{Aa}{\rm{)}}.
\end{equation}

 The vertex formfactors $G_{Aa;\,l_Bj_B}{\rm{(}}q_{Aa}{\rm{)}}$ and $G_{ay;\,l_xj_x}{\rm{(}}q_{ay}{\rm{)}}$, defined from by the expressions  (\ref{subeqSB17b}),
 (\ref{subeqSB19})--(\ref{subeqSB20}),  have the kinematic singularities (branch points)  for odd-values of the quantum numbers $l_B$ and $l_x$ \cite{DDM1973}. They arise due to their behaviors as  $G_{Aa;\,l_Bj_B}{\rm{(}}q_{Aa}{\rm{)}}\propto\,q_{Aa}^{l_B}$  at $q_{Aa}\to$ 0 and $G_{ay;\,l_xj_x}{\rm{(}}q_{ay}{\rm{)}}\propto\,q_{ay}^{l_x}$  at $q_{ya}\to$ 0. Nevertheless, as is  seen from Eq. (\ref{subeqSB17WBx}) as well as  from Eqs. (B3) and (B4) of  Appendix B rewtitten over the ${\mathbf q}_{ay}$ and ${\mathbf q}_{Aa}$ variables , the $ W_{Aa;\,\alpha_B}{\rm{(}}{\mathbf q}_{Aa}{\rm{)}}$ ($ I_{Aa;\,\alpha_B}{\rm{(}}{\mathbf q}_{Aa}{\rm{)}}$) and $W_{ay;\,\alpha_x}{\rm{(}}{\bf q}_{ay}{\rm{)}}$ functions as well as  the ${\tilde{{\cal{ M}}}}^{{\rm{DW}}}_{{\rm{pole}};\,\alpha_B\alpha_x}{\rm{(}}{\mathbf{k}}^{\,\prime},\,{\mathbf{k}}{\rm{)}}$ amplitude   do not have these singularities.

Taking into account
Eqs.  (\ref{subeqSB19}) -- (\ref{subeqSB20}), we  now rewrite the  integral  (\ref{subeqSB15})   in the coordinate representation. First, we  consider this presentation for  the Fourier components of the $ W_{ay;\,\alpha_x}^{{\rm{(}}s{\rm{)}}}{\rm{(}}{\bf q}_{ay}{\rm{)}}$ and $I_{Aa;\,\alpha_B}^{*\,{\rm{(}}s{\rm{)}}}{\rm{(}}{\bf q}_{Aa}{\rm{)}}$ functions:
\begin{equation}
\label{subeqSB21}
W^{{\rm{(}}as{\rm{)}}}_{x;\,\alpha_x}{\rm{(}}{\mathbf{r}}_{ay}{\rm{)}}=\int\frac{d{\mathbf{q}}_{ay}}{{\rm{(}}{\rm 2}\pi)^{{\rm{3}}}}e^{i{\mathbf{r}}_{ay}{\mathbf{q}}_{ay}}W^{{\rm{(}}s{\rm{)}}}_{x;\,\alpha_x}{\rm{(}}{\mathbf{q}}_{ay}{\rm{)}}
\end{equation}
and
\begin{equation}
\label{subeqSB22}
I^{{\rm{(}}as{\rm{)}}}_{B;\,\alpha_B}({\mathbf{r}}_{Aa}{\rm{)}}=\int\frac{d{\mathbf{q}}_{Aa}}{{\rm{(}}{\rm 2}\pi)^{{\rm{3}}}}e^{i{\mathbf{r}}_{Aa}{\mathbf{q}}_{Aa}}I^{(s)}_{Aa;\,\alpha_B}({\mathbf{q}}_{Aa})
\end{equation}
Substituting     Eq. (\ref{subeqSB20x})  in Eq. (\ref{subeqSB21}) and Eq.  (\ref{subeqSB20}) in Eq.   (\ref{subeqSB22}),    the integration over the angular variables   can easely   be  performed  making use  of the    expansion
$$
e^{i{\mathbf{q}}{\mathbf{r}}}=
\,{\rm{4}}\pi\sum_{l\nu}i^lj_l{\rm{(}}qr{\rm{)}}
Y^*_{l\nu}{\rm{(}}\hat {\mathbf q}{\rm{)}}Y_{l\nu}{\rm{(}}\hat {\mathbf r}{\rm{)}},
$$ where  $j_l{\rm{(}}z{\rm{)}}$ is a  spherical Bessel function \cite{Abr1970}. The  remaining integrals in $q_{ay}$  and $q_{Aa}$ can be done   by  using  formula 6.565(4)   and  Eq. (91) from   Refs. \cite{GR1980} and  \cite{Blok77}, respectively.  As a result, one obtains
\begin{equation}
\label{subeqSB23}
 W^{{\rm{(}}as{\rm{)}}}_{ay;\,\alpha_x}{\rm{(}}{\mathbf{r}}_{ay}{\rm{)}}= -\frac{\sqrt{{\rm {2}}}\eta_x}{\pi}G_{ay;\,l_xj_x} \left(\frac{\kappa_{ay}}{r_{ay}}\right)^{3/2}\frac{K_{l_x+{\rm{3/2}}+\eta_x}{\rm{(}}\kappa_{ay}r_{ay}{\rm{)}}}{{\rm{(}}2i\kappa_{ay}r_{ay}{\rm{)}}^{\eta_x}}i^{-l_x}Y_{l_x\nu_x}{\rm{(}}\hat{{\mathbf{r}}}_{ay}{\rm{)}}
\end{equation}
 for $r_{ay}\gtrsim R_x $ and
\begin{equation}
\label{subeqSB24}
I^{*{\rm{(}}as{\rm{)}}}_{Aa;\,\alpha_B}{\rm{(}}{\mathbf{r}}_{Aa}{\rm{)}}\,=\,-\frac{\sqrt{{\rm {2}}}}{\pi}
G_{Aa;\,l_Bj_B}\left(\frac{\mu_{Aa}^{{\rm{2}}}\kappa_{Aa}}{r_{Aa}}\right)^{{\rm{1/2}}}\frac{K_{l_B+{\rm{1}}/{\rm{2}}+\eta_B}{\rm{(}}\kappa_{Aa}r_{Aa}{\rm{)}}}{{\rm{(2}}i\kappa_{Aa}r_{Aa}{\rm{)}}^{\eta_B}}i^{-l_B}Y^*_{l_B\nu_B}{\rm{(}}\hat{{\mathbf{r}}}_{Aa}{\rm{)}}
 \end{equation}
for $r_{Aa}\gtrsim R_B$. Here $K_{{\tilde{\nu}}}{\rm{(}}z{\rm{)}}$ is a modified Hankel function \cite{Abr1970} and  $R_C=r_0C^{\rm{1/3}}$ is the radius of $C$ nucleus, where    $C$  is a   mass number of the  $C$ nucleus.   Using formula 9.235 (2)  from \cite{GR1980} and the relation  (\ref{subeq4CG}), the   leading asymptotic  terms of Eqs. (\ref{subeqSB23}) and (\ref{subeqSB24}) can be reduced  to the forms
\begin{equation}
\label{subeqSB23as}
 W^{{\rm{(}}as{\rm{)}}}_{ay;\,\alpha_x}{\rm{(}}{\mathbf{r}}_{ay}{\rm{)}}\,\approx\, V_{ay}^C{\rm{(}}r_{ay}{\rm{)}}I^{{\rm{(}}as{\rm{)}}}_{ay;\,\alpha_x}{\rm{(}}r_{ay}{\rm{)}}Y_{l_x\nu_x}{\rm{(}}\hat{{\mathbf{r}}}_{ay}{\rm{)}},\,
 \end{equation}
  for $r_{ay}\gtrsim R_x $ and
 \begin{equation}
\label{subeqSB24Bas}
I^{^*{\rm{(}}as{\rm{)}}}_{Aa;\,\alpha_B}{\rm{(}}{\mathbf{r}}_{Aa}{\rm{)}}\,\approx\,C_{l_Bj_B}
\frac{\exp\{-\kappa_{Aa}r_{Aa}-\eta_B\ln{\rm{{(2}}}\kappa_{Aa}r_{Aa}{\rm{)}}\}}{r_{Aa}}
Y^*_{l_B\nu_B}{\rm{(}}\hat{{\mathbf{r}}}_{Aa}{\rm{)}},
 \end{equation} for $r_{Aa}\gtrsim R_B$. In   Eq. (\ref{subeqSB23as}),  $V_{ay}^C{\rm{(}}r_{ay}{\rm{)}}=\, Z_aZ_ye^2/r_{ay}$ is the Coulomb interaction potential  between the centers  of mass of particles $y$ and $a$, and
\begin{equation}
\label{subeqSB25xas}
 I^{{\rm{(}}as{\rm{)}}}_{ay;\,\alpha_x}{\rm{(}}r_{ay}{\rm{)}}\,=\,
C_{l_xj_x}
\frac{\exp\{-\kappa_{ay}r_{ay}-\eta_x\ln{\rm{{(2}}}\kappa_{ay}r_{ay}{\rm{)}}\}}{r_{ay} },
 \end{equation} which coincides with the leading term of the asymptotic behavior of the radial component of the overlap function $I_{ay}{\rm{(}}{\mathbf{r}}_{ay}{\rm{)}} \,\approx\,I^{{\rm{(}}as{\rm{)}}}_{ay;\,\alpha_x}{\rm{(}}r_{ay}{\rm{)}} Y_{l_x\nu_x}{\rm{(}}\hat{{\mathbf{r}}}_{ay}{\rm{)}}$  for $r_{ay}> R_x $.

 Following by  \cite{Blokh2013}, it can show that the leading terms of   the asymptotic expressions for
 the radial components of the Coulomb-nuclear parts of the $W_{ay}{\rm{(}}\mathbf{r}_{ay}{\rm{)}}$ and $I_{Aa}{\rm{(}}\mathbf{r}_{Aa}{\rm{)}}$ functions, which are   generated by the singularities of $\xi_i$ and $\bar{\xi}_i$ of the  $W^{{\rm{(CN)}}}_{ay;\,\alpha_x}{\rm{(}}{\mathbf q}_{ay}{\rm{)}}$ and $W^{{\rm{(CN)}}}_{Aa;\,\alpha_B}{\rm{(}}{\mathbf q}_{Aa}{\rm{)}}$  functions, respectively, behave as
 \begin{equation}
\label{subeqSB26xBNCas}
W^{{\rm{(}CN)}}_{l_xj_x}{\rm{(}}r_{ay}{\rm{)}}\approx\,\sum_iW^{{\rm{(CN}};\,as{\rm{)}}}_{l_xj_x;\,i}{\rm{(}}r_{ay}{\rm{)}}, \,\,\,I^{{\rm{(}CN)}}_{l_Bj_B}{\rm{(}}r_{Aa}{\rm{)}}\approx\,\sum_iI^{{\rm{(CN};\,as{\rm)}}}_{l_Bj_B;\,i}{\rm{(}}r_{Aa}{\rm{)}}.
 \end{equation} Here
 \begin{equation}
\label{subeqSB27xCNas}
 W^{{\rm{(CN}};\,as{\rm{)}}}_{l_xj_x;\,i}{\rm{(}}r_{ay}{\rm{)}}=\,\bar{C}^{{\rm{(\,}}i{\rm{\,)}}}_{l_xj_x}
 \frac{\exp\{-[\bar{\kappa}_ir_{ay}
 \,+\,\eta_{e_if_i}\ln{\rm{(2}}\bar{\lambda}_i\kappa_{e_if_i}r_{ay}{\rm{)}}\,+\,\eta_{e_ig_i}\ln{\rm{(2}}\bar{\lambda}_i\kappa_{e_ig_i}r_{ay}{\rm{)}}]\}}{r_{ay}^{{\rm{2}}}},
  \end{equation}
  \begin{equation}
\label{subeqSB27BCNas}
 I^{{\rm{(CN}};\,as{\rm{)}}}_{l_Bj_B;\,i}{\rm{(}}r_{Aa}{\rm{)}}=\,C^{{\rm{(\,}}i{\rm{\,)}}}_{l_Bj_B}
 \frac{\exp\{-[\kappa_ir_{Aa}
 \,+\,\eta_{b_ic_i}\ln{\rm{(2}}\lambda_i\kappa_{b_ic_i}r_{Aa}{\rm{)}}\,+\,\eta_{b_id_i}\ln{\rm{(2}}\lambda_i\kappa_{b_id_i}r_{Aa}{\rm{)}}]\}}{r_{Aa}^{{\rm{2}}}},
  \end{equation} where $\eta_{\alpha\beta}$ is the Coulomb parameter for the bound ($\alpha\,+\,\beta$) system in the tri-ray vertex of the diagram in Fig. \ref{fig2}$c$.
   Explicit expressions for $\bar{C}^{{\rm{(\,}}i{\rm{\,)}}}_{l_xj_x}$ and $C^{{\rm{(\,}}i{\rm{\,)}}}_{l_Bj_B}$ can be  obtained from Eqs. (B.4) and (B.5) of \cite{Blokh2013}, which are expressed in the terms of the product of the corresponding ANC's for the tri-rays vertices of the diagrams in  Fig. \ref{fig2}$c$. As is seen from Eqs.  (\ref{subeqSB26xBNCas}) --  (\ref{subeqSB27BCNas}), if $\kappa_i\,>\,\kappa_{Aa}$ and
  $\bar{\kappa}_i\,>\,\kappa_{ya}$ , then the  asymptotic  terms  given by the expressions   (\ref{subeqSB27xCNas}) and (\ref{subeqSB27BCNas})  decrease more rapidly with increasing $r_{ay}$ and $r_{Aa}$, respectively,
  than those of (\ref{subeqSB23as}) and (\ref{subeqSB24Bas}). See Table \ref{table2}, where $\kappa_i\,>\,\kappa_{Aa}$ and
  $\bar{\kappa}_i\,>\,\kappa_{ya}$  for all the considered reactions.
 Therefore,   the use of the pole approximation is reasonable in calculations of the leading terms of  the peripheral partial wave amplitudes at $l_i\,>>$ 1   determined correctly by only the nearest singularity $\xi$, which is  in turn equivalent to the replacements of  $V_{ay}({\mathbf{r}}_{ay}) I_{ay}({\mathbf{r}}_{ay})$ and  $ I^{^*}_{Aa;\,\alpha_B}{\rm{(}}{\mathbf{r}}_{Aa}{\rm{)}}
$   by $ W^{{\rm{(}}as{\rm{)}}}_{ay;\,\alpha_x}{\rm{(}}{\mathbf{r}}_{ay}{\rm{)}}$ and $I^{^*{\rm{(}}as{\rm{)}}}_{Aa;\,\alpha_B}{\rm{(}}{\mathbf{r}}_{Aa}{\rm{)}}$ in the integrand function of Eq. (\ref{subeqSB13}), respectively. These peripheral partial wave amplitudes  indeed
   give the dominant contribution to the  $M^{{\rm{DW}}}_{{\rm{pole}}}{\rm{(}}E_i,\,cos\theta{\rm{)}}$ at least in the angular range of the main peak of the angular distribution \cite{DDM1973}.
  
In this case,
    the  $M^{{\rm{DW}}}_{{\rm{pole}}}{\rm{(}}E_i,\,cos\theta{\rm{)}}$ amplitude in the coordinate representation can be reduced  to the form as
$$
M^{{\rm{DW}}}_{{\rm{pole}}}{\rm{(}}E_i,\,cos\theta{\rm{)}}\simeq\, M^{{\rm(}s{\rm{)}}\,{\rm{DW}}}_{{\rm{pole}}}{\rm{(}}E_i,\,cos\theta{\rm{)}}=\,\sum_{\alpha_B\alpha_xM_a}
C{\rm{(}}\alpha_B\,\alpha_x;\,{\rm{(}}J,\,M{\rm{)}}_{x,\,A,\,y,\,B};\,J_aM_a{\rm{)}}
$$
\begin{equation}
\label{subeqSB25}
\times{\tilde{M}}^{{\rm{DW}}}_{{\rm{pole};\,\alpha_B\alpha_x}}{\rm{(}}E_i,\,cos\theta{\rm{)}},
\end{equation}where
\begin{equation}
\label{subeqSB26}
{\tilde{M}}^{{\rm{DW}}}_{{\rm{pole};\,\alpha_B\alpha_x}}{\rm{(}}E_i,\,cos\theta{\rm{)}}=
\,\int d{\mathbf{r}}_i d{\mathbf{r}}_f
\Psi^{*{\rm{(}}-{\rm{)}}}_{{\mathbf{k}_f}}{\rm {(}}{\mathbf{r}}_f{\rm{)}}
I^{^*{\rm{(}}as{\rm{)}}}_{Aa;\,\alpha}({\mathbf{r}}_{Aa}{\rm{)}}
W^{{\rm{(}}as{\rm{)}}}_{ay;\,\alpha_x}{\rm{(}}{\mathbf{r}}_{ay}{\rm{)}}
 \Psi^{{\rm{(}}+{\rm{)}}}_{{\mathbf{k}_i}}{\rm {(}}{\mathbf{r}_i}{\rm{)}}.
\end{equation}

One notes that the expressions for   $W^{{\rm{(}}as{\rm{)}}}_{ay;\,\alpha_x}{\rm{(}}{\mathbf{r}}_{ay}{\rm{)}}$, given by Eqs.  (\ref{subeqSB23}) and  (\ref{subeqSB23as}),  is  valid for   $\eta_x>$ 0. For  $\eta_x$=0,   the  Fourier component of the $W^{{\rm{(}}as{\rm{)}}}_{x;\,\alpha_x}{\rm{(}}{\mathbf{r}}_{ay}{\rm{)}}$  function   in (\ref{subeqSB21}) is given only  by   the kinematic function $q_{ay}^{l_x}$ for $l_x\,>$ 0 and, so, the  Fourier integral becomes singular \cite{KMYa1988}.  In this case, for $\eta_x$ = 0 one obtains
\begin{equation}
\label{subeqSB26xas}
 W^{{\rm{(}}as{\rm{)}}}_{ay;\,\alpha_x}{\rm{(}}{\mathbf{r}}_{ay}{\rm{)}}=\,- \frac{C_{l_xj_x}}{{\rm{2}}\mu_{ay}}{\hat{l}}_x!! {\rm{(}}\kappa_{ay}r_{ay}{\rm{)}}^{-l_x}\delta{\rm{(}}r_{ay}{\rm{)}}r_{ay}^{-{\rm{2}}} Y_{l_x\nu_x}{\rm{(}}\hat{{\mathbf{r}}}_{ay}{\rm{)}},
 \end{equation}where $r_{ay}$ is given by Eq. (\ref{subeqSB14}) and  ${\hat{l}}_x$= ${\rm{2}}l_x\,+\,{\rm{1}}$. This expression   corresponds to    the vertex function for the virtual decay $x\to y+a$ \cite{KMYa1988}  calculated in the well-known zero-range approximation. Therefore, the expression (\ref{subeqSB26xas}) can be    applied jointly   with Eq. (\ref{subeqSB24}) for  the $  M^{{\rm{DW}}}_{{\rm{pole}}}{\rm{(}}E_i,\,cos\theta{\rm{)}}$  amplitude of the peripheral  $A$($d$, $n$)$B$ reaction for example.

We now expand the $M^{{\rm{DW}}}_{{\rm{post;\, pole}}}{\rm{(}}E_i,\,cos\theta{\rm{)}}$ amplitude in  partial waves.  To this end, in (\ref{subeqSB26})  we use the  partial-waves expansions (B3) and (B4) from  Appendix B   and  the expansion
$$
\frac{K_{l_{ay}\,+\,{\rm{3/2}}\,+\,\eta_x}{\rm{(}}\kappa_{ay}r_{ay}{\rm{)}}}
{r_{ay}^{l_{ay}\,+\,\eta_x\,+\,{\rm{3/2}}}}
\frac{K_{l_{Aa}\,+\,{\rm{1/2}}\,+\,\eta_B}{\rm{(}}\kappa_{Aa}r_{Aa}{\rm{)}}}
{r_{Aa}^{l_{Aa}\,+\,\eta_B\,+\,{\rm{1/2}}}}
$$
\begin{equation}
\label{subeqSB27}
=\,{\rm{4}}\pi\sum_{l\mu_l}{\cal{A}}_l{\rm{(}}r_i,\,r_f{\rm{)}}Y_{l\mu_l}{\rm{(}}\hat{{\mathbf r}}_i{\rm{)}}Y^*_{l\mu_l}{\rm{(}}\hat{{\mathbf r}}_f{\rm{)}}.
\end{equation}Here
\begin{equation}
\label{subeqSB27}
{\cal{A}}_l{\rm{(}}r_i,\,r_f{\rm{)}}=\,\frac{\rm{1}}{\rm{2}}\int_{-{\rm{1}}}^{{\rm{1}}}\frac{K_{l_{ay}\,+\,{\rm{3/2}}\,+\,\eta_x}{\rm{(}}\kappa_{ay}r_{ay}{\rm{)}}}
{r_{ay}^{l_{ay}\,+\,\eta_x\,+\,{\rm{3/2}}}}
\frac{K_{l_{Aa}\,+\,{\rm{1/2}}\,+\,\eta_B}{\rm{(}}\kappa_{Aa}r_{Aa}{\rm{)}}}
{r_{Aa}^{l_{Aa}\,+\,\eta_B\,+\,{\rm{1/2}}}}P_l{\rm{(}}z{\rm{)}}dz,
\end{equation}where   $r_{ay}$= [($\bar{a}r_i$)$^{\rm{2}}$ + ($\bar{b}r_f$)$^{\rm{2}}$ - 2$\bar{a}\bar{b}r_ir_fz$]$^{\rm{1/2}}$,   $r_{Aa}$= [($\bar{c}r_i$)$^{\rm{2}}$ + ($\bar{d}r_f$)$^{\rm{2}}$ - 2$\bar{c}\bar{d}r_ir_fz$]$^{\rm{1/2}}$ and $z$\,=\,${\rm{(}}{\hat{{\mathbf{r}}}}_i{\hat{{\mathbf{r}}}}_f{\rm{)}}$. The integration over the angular variables ${\hat{\mathbf{r}}}_i $ and ${\hat{\mathbf{r}}}_f$ in Eq. (\ref{subeqSB26}) can easily be done by using Eqs. (B5) and (B6)   of  Appendix B. After some simple, but cumbersome algebra using the corresponding formulae from \cite{Varsh}, one finds  that the pole amplitude $M^{{\rm{DW}}}_{{\rm{ pole}}}{\rm{(}}E_i,\,cos\theta{\rm{)}}$ in the system ${\it z}\Vert{\mathbf k}_i$ has the form
$$
M^{{\rm{DW}}}_{{\rm{pole}}}{\rm{(}}E_i,\,cos\theta{\rm{)}}=-{\rm{8}}\sqrt{\frac{{\rm{2}}}{\pi}}
 \frac{{\rm{1}}}{\mu_{ay}}\frac{{\rm{1}}}{k_ik_f} \sum_{j_x\,\tau_x\,j_B\,\tau_B}\sum_{J\,M
}\sum_{l_x\,l_B} {\rm{(-1)}}^{j_B-J_a+J}i^{l_x+l_B}{\rm{(}}\hat{l}_x\hat{l}_B{\rm{)}}{\rm{(}}\hat{J}\hat{j}_B{\rm{)}}^{{\rm{1/2}}}
 $$
 \begin{equation}
\label{subeqSB32}
\times  C_{ay;\,l_xj_x}C_{Aa;\,l_Bj_B}C_{j_B\tau_BJ_AM_A}^{J_BM_B} C_{j_x\tau_xJ_yM_y}^{J_xM_x}C_{JMj_B\tau_B}^{j_x\tau_x}W{\rm{(}}l_xj_xl_Bj_B;J_aJ{\rm{)}}
 \end{equation}
 $$
 \times\sum_{l_il_f}\,M^{{\rm{pole}}}_{l_xl_BJl_il_f}{\rm{(}}E_i{\rm{)}}C_{l_i\,{\rm{0}}l_f\,M}^{J\,M}Y_{l_fM}{\rm{(}}\theta,{\rm{0)}}
 $$
  where the explicit form of  $M^{{\rm{pole}}}_{l_xl_BJl_il_f}{\rm{(}}E_i{\rm{)}}$ is  given by Eqs.
   (B7) -- (B10) of  Appendix B.

   It should be noted that just neglecting  the dynamic recoil effect mentioned above, which is caused by using the pole approximation in the  matrix elements for the virtual decays $x\,\to\,y\,+\,a$ and
 $B\,\to\,A\,+\,a$, results in the fact that the radial integral (B8) of the $ M^{{\rm{DW}}}_{{\rm{pole}}}{\rm{(}}E_i,\,cos\theta{\rm{)}}$ amplitude, given in Appendix B, does not contain  the $V_{ya}$ and $V_{Aa}$ potentials  in contrast to that of the conventional DWBA with recoil effects \cite{Austern1964,LOLA}. That is the reason why  the $ M^{{\rm{DW}}}_{{\rm{ pole}}}{\rm{(}}E_i,\,cos\theta{\rm{)}}$ amplitude is parametrized  directly in the terms of the  ANCs (or respective the NVCs) but not in  those of the spectroscopic factors, as it occurs for the conventional DWBA \cite{Austern1964,LOLA}.

\vspace{1cm}
  {\bf V.  THREE-PARTICLE COULOMB DYNAMICS OF THE TRANSFER MECHANISM AND THE GENERALIZED DWBA }
  \bigskip

 \hspace{0.6cm}We now  consider  how to       take into account accurately the contribution of the three-body Coulomb dynamics of  the transfer mechanism   to the $M^{{\rm{DW}}}_{{\rm{pole}}}{\rm{(}}E_i,\,cos\theta{\rm{)}}$ and $M^{{\rm{TBDW}}}{\rm{(}}E_i,\,cos\theta{\rm{)}}$ amplitudes by using Eqs. (\ref{subeqSB9}), (\ref{subeqSB10}) and (\ref{subeqSB32})  as well as  Eqs. (B7) and (B8) from  Appendix B.   To this end, we should compare their partial wave amplitudes for $l_i\,>>$ 1 and $l_f\,>>$ 1 (denoted by  $M^{{\rm{TBDW}}}_{l_il_f}{\rm{(}}E_i{\rm{)}}$
  and $M^{{\rm{DW}}}_{{\rm{pole}};\,l_il_f}$ below) with each other, which can be  determined from the corresponding expressions for    the $M^{{\rm(}s{\rm{)}}\,{\rm{TBDW}}}{\rm{(}}E_i,\,cos\theta{\rm{)}}$  and $M^{{\rm(}s{\rm{)}}\,{\rm{DW}}}_{{\rm{pole}}}{\rm{(}}E_i,\,cos\theta{\rm{)}}$ amplitudes.

According to \cite{Popov1964},   from  Eq. (\ref{subeqSB9}) and  (\ref{subeqSB10}), the   peripheral partial amplitudes at $l_i\,>>$ 1  and $l_f\,>>$ 1 can  be presented in the form as
 \begin{equation}
M^{{\rm{TBDW}}}_{l_i\,l_f}{\rm{(}}E_i{\rm{)}}\,=\,{\cal{R}}^{{\rm{TBDM}}}{\rm{(}}E_i{\rm{)}}
M^{{\rm{DW}}}_{{\rm{pole}};\,l_i\,l_f}{\rm{(}}E_i{\rm{)}}.
\label{subeqSB12}
\end{equation}
Here   $M^{{\rm{DW}}}_{{\rm{pole}};\,l_i\,l_f }{\rm{(}}E_i{\rm{)}}$ is the
 peripheral partial amplitude corresponding to  the pole approximation of the   DWBA amplitude.

   The expression  (\ref{subeqSB12})  can be considered as the peripheral partial amplitude of the generalized DWBA in which the contribution of
  the three-body Coulomb dynamics of the main transfer mechanism is correctly taken into account.
For $l_i\,>>$ 1 and $l_f\,>>$ 1 the asymptotics of      the pole approximation ($M^{{\rm{DW}}}_{{\rm{pole}};\,l_i\,l_f}{\rm{(}}E_i{\rm{)}}$) partial amplitudes   of the pole-approximation DWBA amplitude and   the exact three-body ($M^{{\rm{TBDW}}}_{l_i\,l_f}{\rm{(}}E_i{\rm{)}}$) partial  amplitudes of the exact three-body amplitude have the same dependence on $l_i$ and $l_f$. Nevertheless,  they  differ only in their powers.

Therefore, if the main contribution to the    $M^{{\rm{TBDW}}}{\rm{(}}E_i,\,cos\theta{\rm{)}}$ amplitude   comes from the peripheral partial waves with $l_i\,>>$ 1  and $l_f\,>>$ 1, then the expression (\ref{subeqSB12}) makes it possible    to obtain the amplitude of  the generated three-body DWBA. For this aim, in Eq. (\ref{subeqSB32})    the expression $M^{{\rm{pole}}}_{l_xl_BJl_il_f}{\rm{(}}E_i{\rm{)}}$  at fixed values $l_x,\,l_B$ and $J$
  has to be renormalized by the   replacement
\begin{equation}
 M^{{\rm{pole}}}_{l_xl_BJl_il_f}{\rm{(}}E_i{\rm{)}}\,
 \longrightarrow\,
M^{{\rm{TBDW}}}_{l_xl_BJl_il_f}{\rm{(}}E_i{\rm{)}}=
\,{\cal{N}}^{{\rm{TBDM}}}_{l_il_f}{\rm{(}}E_i{\rm{)}}M^{{\rm{pole}}}_{l_xl_BJl_il_f}{\rm{(}}E_i{\rm{)}}.
\label{subeqTBPWA1}
\end{equation}  Here
 \begin{equation}
{\cal{N}}^{{\rm{TBDM}}}_{l_il_f}{\rm{(}}E_i{\rm{)}}=\,\begin{cases}
{\rm{1}}, &{\rm{ for}} \, \,\, l_i,<\,L_{{\rm{0}}} \,\,{\rm{and}}\,\, l_f\,<\,L_{{\rm{0}}};\cr
 {\cal{R}}^{{\rm{TBDM}}}{\rm{(}}E_i{\rm{)}} \,\,\, &{\rm{for}}\,\, l_i\,\ge\,L_{{\rm{0}}}, \,\, l_f\,\ge\,L_{{\rm{0}}},\cr
\end{cases}
\label{subeqTBPWA2}
\end{equation} where $L_{{\rm{0}}}\sim k_iR_i^{{\rm{ch}}}$  (or $\sim k_fR_f^{{\rm{ch}}}$). In this case, the expression for
the amplitude of  the generalized  three-body  DWBA, $M^{{\rm{TBDW}}}{\rm{(}}E_i,\,cos\theta{\rm{)}}$, is given by
$$
M^{{\rm{TB}}}{\rm{(}}E_i,\,cos\theta{\rm{)}}\approx  M^{{\rm{(}}s{\rm{)}}{\rm{TBDW}}}{\rm{(}}E_i,\,\cos\theta)=\,-{\rm{8}}\sqrt{\frac{{\rm{2}}}{\pi}}
 \frac{{\rm{1}}}{\mu_{ay}}\frac{{\rm{1}}}{k_ik_f}\sum_{j_x\,\tau_x\,j_B\,\tau_B}\sum_{J,\,M
}\sum_{l_x\,l_B}
$$
 \begin{equation}
\label{subeqTBDWA2}
\times   C_{ay;\,l_xj_x}C_{Aa;\,l_Bj_B}C_{j_B\tau_BJ_AM_A}^{J_BM_B} C_{j_x\tau_xJ_yM_y}^{J_xM_x}C_{JMj_B\tau_B}^{j_x\tau_x}\sum_{l_il_f}\,
M^{{\rm{TBDW}}}_{l_xl_BJl_il_f}{\rm{(}}E_i{\rm{)}}C_{l_i\,{\rm{0}}l_f\,M}^{J\,M}Y_{l_fM}{\rm{(}}\theta,{\rm{0)}},
 \end{equation} where the expression for $M^{{\rm{TBDW}}}_{l_xl_BJl_il_f}{\rm{(}}E_i {\rm{)}}$ is obtained from Eq. (B7) of Appendix B by the substitution  of the $M^{{\rm{pole}}}_{l_xl_BJl_il_f}{\rm{(}}E_i{\rm{)}}$ by  $M^{{\rm{TBDW}}}_{l_xl_BJl_il_f}{\rm{(}}E_i{\rm{)}} $ defined  by Eq. (\ref{subeqTBPWA1}). The expression (\ref{subeqTBDWA2}) can be considered as a  generalization of  Eqs. (34) and (35) of Ref. \cite{KMYa1988} derived   within the framework of  the dispersion theory   for  the above-barrier peripheral neutron transfer reaction.
  As is seen from Eqs. (\ref{subeqTBPWA1}) -- (\ref{subeqTBDWA2}) as well as from Eqs. (B7) and (B8) of Appendix B, in Eq. (\ref{subeqTBDWA2}), the contribution of non-peripheral partial waves to   the generalized  three-body  DWBA  amplitude   is taken into account in the pole-approximation.

From Eqs.  (\ref{subeqSB32}), (\ref{subeqTBPWA1}) and (\ref{subeqTBPWA2}), we can now  derive  the expression for the differential cross section for the generalized three-body DWBA, which   has   the form     as
$$
\frac{d\sigma}{d\Omega}=\frac{\mu_{Ax}\mu_{By}}{{\rm{(2}}\pi^{{\rm{2}}}{\rm{)}}^{{\rm{2}}}}\frac{k_f}{k_i}\frac{{\rm{1}}}{\hat{J}_A\hat{J}_x}\sum_{M_AM_xM_BM_y}\mid   M^{{\rm{(}}s{\rm{)}}{\rm{TBDW}}}{\rm{(}}E_i,\,\cos\theta)\mid^{\rm{2}} =\,\frac{{\rm{20}}}{\pi^{{\rm{3}}}}\frac{(\hbar c)^{{\rm{2}}}}{E_iE_f}\left(\frac{\hbar}{\mu_{ay}c}\right)^{{\rm{2}}} \frac{k_f}{k_i} \frac{\hat{J}_B}{\hat{J}_A}
 $$
\begin{equation}
\label{FBCS1}
\times \sum_{j_x\,j_B}\sum_{J\,M}\mid
\sum_{l_x\,l_B}\sum_{l_i\,l_f}\,
\exp\{i[\sigma_{l_i}\,+\,\sigma_{l_f}\,+\,\frac{\pi}{{\rm{2}}}{\rm{(}}l_i\,+\,l_f\,+\,l_x\,+\,l_B)]\}
 C_{ay;\,l_xj_x}C_{Aa;\,l_Bj_B}
\end{equation}
$$
\times {\rm{(}}\hat{l}_x\hat{l}_B{\rm{)}}{\rm{(}}\hat{l}_i^2\hat{l}_f{\rm{)}}^{{\rm{1/2}}} W{\rm{(}}l_xj_xl_Bj_B;J_aJ{\rm{)}}C_{l_i\,0l_f\,M}^{J\,M}
M^{{\rm{TBDW}}}_{l_xl_BJl_il_f}{\rm{(}}E_i{\rm{)}}Y_{l_f\,M}{\rm{(}}\theta,\,{\rm{0)}}
\mid^{{\rm{2}}}.
$$
 Herein,  the ANCs $C$'s , $\kappa_{ij}$($k_i$ and $k_f$) and $d\sigma/d\Omega$ are in fm$^{-{\rm{1/2}}}$, fm$^{-{\rm{1}}}$ and mb/sr, respectively, and $E_i$ and $E_f$ are in MeV. One notes that  Eq. (\ref{FBCS1}) and Eq. (B8) given in  Appendix B  contain the   cut-off parameters $R_i^{{\rm{ch}}}$ and  $ R_f^{{\rm{ch}}}$, which are determined by  only the free parameter $r_{{\rm{0}}}$ (see Appendix B).

 The expression  (\ref{FBCS1})  can  also be applied for peripheral   sub-barrier charged particle transfer   reactions  for which the dominant contribution comes to  rather low partial waves with $l_i\sim k_iR_i^{{\rm{ch}}}\sim{\rm{0,\,1,...}}$, which correspond to   $k_i\to$0 and  $R_i^{{\rm{ch}}}\gtrsim R_N$.     Here, it is assumed that the contribution of  the low partial amplitudes to the reaction amplitude   parametrizing via the product of the  ANCs (or NVCs) for $R_i^{{\rm{ch}}}\gtrsim R_N$ can be taken into account in  the pole-approximation of the DWBA.  In this case,  the contribution of the peripheral partial waves with $l_i>>$1 and $l_f>>$1 to the reaction amplitude is strongly suppressed as   $\tau>>$1  in Eqs. (A2) -- (A4) (see Section IV  below). Nevertheless,  the  influence of the   three-body Coulomb  dynamics of the transfer mechanism  on the DCS (\ref{FBCS1}) is mainly  taken into account     via  the interference term  between  the  low  and peripheral  partial  amplitudes  arising from  Eqs. (\ref{subeqTBPWA1}) and (\ref{subeqTBPWA2}). In this connection, one notes that the analogous situation occurs  for the peripheral direct nuclear-astrophysical   ${\rm{A(}}a,\, \gamma{{\rm)B}}$ reaction at extremely low (sub-barrier)  energies for which the radiative capture proceeds also  at the large relative distances of the colliding particles $r_{Aa}\gtrsim R_N$. For this reaction the main contribution in the long-wavelength  approximation comes  to  the partial waves with  $l_i\sim{\rm{0,\,1,...}}$, and the reaction amplitude can also be  expressed in the term of the ANC for ${\rm{A}}\,+\,a\to\, {\rm{B}}$ \cite{Igam07,IgYa2008}.

\bigskip
\vspace{1cm}
  {\bf VI. RESULTS OF  THE   ANALYSIS OT THE PERIPHERAL PARTIAL AMPLITUDES FOR THE SPECIFIC SUB- AND ABOVE-BARRIER REACTIONS }
  \bigskip

  In this section, we present the results of  calculations of the modulus of the partial ampilidues $\mid M^{{\rm{TBDW}}}_{l_xl_BJ\,l_i\,l_f} \mid $  (denoted by $\mid M_{J\,l_i\,l_f}\mid$  for the fixed values of the angular momentums $l_x$ and $l_B$ below) of  the amplitude (\ref{subeqTBDWA2}). The calculation were performed    for the following peripheral   proton and triton transfer  reactions:\\
   (I) ${\rm {^9Be(^{10}B,^9Be)^{10}B_i}}$ ($i$= 0--3)  at the ${\rm{^{10}B}}$ incident  energy $E_{\rm{^{10}B}}$= 100 MeV \cite{Mukh2};\\
     (II)  ${\rm{^{16}O(^3He}}, d{\rm {)^{17}F_i}}$ ($i$= 0 and 1) at  $E_{\rm{^3He }}$= 29.75 MeV \cite{Mukh6};\\
    (III) ${\rm{^{19}F(}}p,\,\alpha {\rm {)^{16}O}}$(g.s.)   at six  sub-barrier proton projectile  energie of $E_p$= 250, 350 and 450 MeV \cite{HL1978,HAS1991} and $E_p$= 327, 387 and 486 MeV \cite{Ivano2015}.  \\
    One notes once more that all they    are   related to  the  ``non-dramatic'' case (see the first column of Table \ref{table1}).

 For the reactions considered above,   the orbital  ($l_B$ and $l_x$)  angular momentums of the transfer (proton or triton) particle  are  taken   equal to  $l_{{\rm{^{10}B}}_i}$= 1 ($i$=0--3),  $l_{{\rm{^{17}F_0}}}$= 2 and $l_{{\rm{^{17}F_1}}}$= 0,  and $l_{{\rm{^3He}}}$=$l_{\alpha}$= 0.
  Since the energy of incident ${\rm{^3He}}$  in the reaction (II) is moderate, the contribution of the   $D$-state of  the ${\rm{^3He}}$ nucleus  in the vertex  ${\rm{^3He}}\to d\,+p$ is  neglectable small \cite{Blok77}.
    Calculations were performed       the optical potentials in the initial and final  states, which were taken from Refs. \cite{Mukh2,Mukh6} (the sets 1 and 2) and \cite{HAS1991} for the standard values of  the parameter  $r_{{\rm 0}}$ ($r_{{\rm 0}}$= 1.25 fm).

\begin{figure}
\begin{center}
\includegraphics[height=10cm, width=12cm]{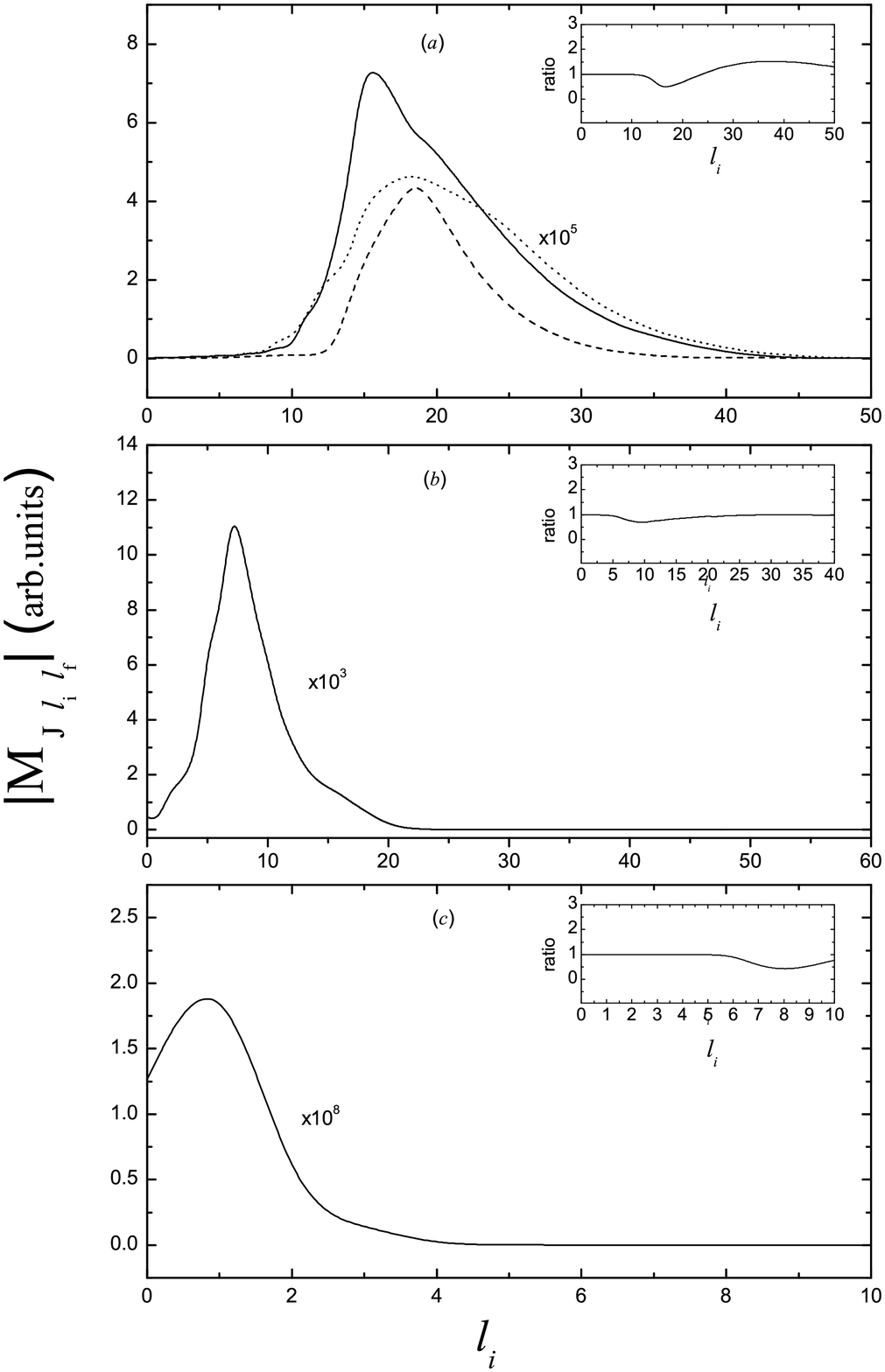}
 \end{center}
\caption{\label{fig3} The $l_i$ dependence of the modulus of the partial wave  amplitudes ($\mid M_{J\,l_i\,l_f}\mid\equiv\mid M^{{\rm{TBDW}}}_{l_xl_BJ\,l_i\,l_f}\mid$) for the ${\rm {^9Be(^{10}B,^9Be)^{10}B_0}}$ ($a$),
   $ {\rm{^{16}O(^3He}}, d{\rm {)^{17}F_0}}$ ($b$) and  ${\rm{^{19}F(}}p,\,\alpha {\rm {)^{16}O}}$ ($c$)     reactions  at projectile energies of  $E_{\rm{^{10}B}}$= 100 MeV,    $E_{\rm{^3He }}$= 29.75 MeV and  $E_p$=250 keV, respectively,   for which $l_{\alpha}=l_{{\rm{^3He}}}$= 0, $l_{{\rm{^{10}B}}}$= 1 and  $l_{{\rm{^{17}F_0}}}$= 2 at different fixed values $J$.  Here $l_i$ and  $l_f$ are  the relative orbital momenta in the entrance and exits channels of the considered reaction, respectively, and $J$ is the transferred angular momentum.  In ($a$),  the solid line is for $J$= 0 and $l_f=l_i $, the dashed line is for  $J$= 1 and $l_f=l_i\,+{\rm{1}} $  and  the dotted line is for  $J$= 2 and $l_f=l_i\,+{\rm{2}} $.  In ($b$),  the solid   line is  for  $J$= 2 ($l_f=l_i\,+{\rm{2}} $).  In ($c$), the solid line is for $J$= 0 ($l_f=l_i $).   The inserts are the  ratio of the   $\mid M_{J\,l_i\,l_f} \mid $   calculated with taking into account of the renormalized  Coulomb ${\cal{N}}^{{\rm{TBDM}}}_{l_il_f}{\rm{(}}E_i{\rm{)}}$ factor     to    that calculated with  ${\cal{N}}^{{\rm{TBDM}}}_{l_il_f}{\rm{(}}E_i{\rm{)}}$= 1 in the peripheral partial amplitudes (see Eqs. (\ref{subeqTBPWA1}) and (\ref{subeqTBPWA2})).    }
 \end{figure}

    In order      to estimate   the influence of the three-body Coulomb dynamics of  the transfer mechanism  on the peripheral partial amplitudes at $l_i>>$ 1 and $l_f>>$ 1, we have     analyzed  only the contribution of the  different  partial wave  amplitudes to  the  amplitude (\ref{subeqTBDWA2}).
      Fig. \ref{fig3} shows  the $l_i$ dependence of the modulus of the  partial amplitudes   ($\mid M_{J\,l_i\,l_f}\mid$)   for the fixed values of $l_x$ and $l_B$ above.  As is seen from Fig. \ref{fig3}$a$, the contribution to the amplitude of  the  ${\rm {^9Be(^{10}B,\,^9Be)^{10}B_0}}$  reaction  from lower partial amplitudes with $l_i<$ 14  is strongly suppressed  due to the strong absorption in the entrance and exit channels. Nevertheless,  for the transferred angular momentum  $J$= 0     the contributions of the three-body Coulomb effects   to the peripheral partial  $\mid M_{J\,l_i\,l_f}\mid$ amplitudes       change from   55\%  to 7\%  for    the   ${\rm {^9Be(^{10}B,\,^9Be)^{10}B_0}}$  reaction at  $ l_i\ge$ 16   (see the inset
    in Fig. \ref{fig3}).    It should be noted that the orbital angular momenta $l_i$ for this reaction are $l_i\sim k_iR_i^{{\rm{ch}}}\approx$ 16  for the channel radius $R_i^{{\rm{ch}}}\approx$ 5.3.   The same situation  occurs for the reaction populating the exited states of
${\rm {^{10}B_i}}$($i$= 1--3)  mentioned above.    Besides, the analogous contribution  is found to be  about 20--30  for the  ${\rm{^{16}O(^3He}}, \,d{\rm {)^{17}F_0}}$ reaction for which  $l_i\sim k_iR_i^{{\rm{ch}}}\approx$ 8   for the channel radius $R_i^{{\rm{ch}}}\approx$ 5 fm (see the inset in Fig. \ref{fig3}$b$).  For the  ${\rm{^{16}O(^3He}}, \,d{\rm {)^{17}F_1}}$ reaction  the influence   of the three-body Coulomb effects on  the peripheral partial amplitudes is extremely larger   as compared with that for  the  ${\rm{^{16}O(^3He}}, \,d{\rm {)^{17}F_0}}$ reaction (see Table ref{table1}). For
 example, the  ratio of the partial  $\mid M_{J\,l_i\,l_f} \mid $ amplitudes,    calculated with taking into account of the Coulomb renormalized  ${\cal{N}}^{{\rm{TBDM}}}_{l_il_f}{\rm{(}}E_i{\rm{)}}$ factor  (see Eqs. (\ref{subeqTBPWA1}) and (\ref{subeqTBPWA2}))     to    that calculated without taking into account of  this factor  (${\cal{N}}^{{\rm{TBDM}}}_{l_il_f}{\rm{(}}E_i{\rm{)}}$= 1)  in the peripheral partial amplitudes, changes about from  1.3x10$^{{\rm -7}}$ to   2.2x10$^{{\rm -7}}$ for $l_i\ge$ 13.
   This is the result of the significant difference between  the ratio $R^{{\rm TBDM}}$ calculated for  the ground and first exited states of the residual  ${\rm{^{17}F}}$ nucleus (see Table \ref{table1}).   In  Fig.  \ref{fig3}$c$, as an illustration,  the same $l_i$ dependence  is displayed     for the sub-barrier ${\rm{^{19}F(}}p,\,\alpha {\rm {)^{16}O}}$ reaction  at the energy  $E_p$=  0.250 MeV for which $l_i\sim k_iR_i^{{\rm{ch}}}\approx$ 1 corresponding to the channel radius $R_i^{{\rm{ch}}}\approx$ 5 fm. As is seen from Fig. \ref{fig3}$c$, the contribution of the peripheral partial waves to the reaction amplitude is suppressed strongly, whereas the main contribution to the amplitude comes to the low partial waves in the vicinity of $l_i\sim$ 1.    The analogous  dependence occurs for
    other considered incident proton energies.  This result is apparently not accidental and can be explained as: for rather low sub-barrier energies ($k_i\to$0),   the position of the nearest singularity $\xi$  moves away  from the right boundary ($\cos\theta$=1) of the physical (-1$\le\cos\theta\le$1) region ($\xi>>$1    and $\tau>>$1). Therefore, as is seen from the fourth column of    Table \ref{table2},       due to a presence  of the factor $\exp{\rm{(}}-l_i\ln\tau{\rm{)}}/\sqrt{\xi^{{\rm{2}}}-{\rm{1}}}$ in Eqs. (A2) -- (A4),  the calculated values of the  peripheral partial amplitudes for $l_i>>$1 become extremely smaller at sub-barrier energies than  those at above-barrier energies for which the position of the singularity $\xi$ is located  rather close to  the  aforementioned   boundary (see the fourth line of Table \ref{table2} and Fig. \ref{fig3}$c$).

     It follows  from here that the influence of the three-body Coulomb effects in the initial, intermediate and final states of  the considered above-barrier  reactions  on  the peripheral partial amplitudes of the reaction  amplitude   can not be ignored even for the ``non-dramatic'' case.  One notes that this influence is ignored  in the calculations  of
     the ``post''-approximation and  the ``post'' form of the    DWBA performed in  \cite{Ar96} and \cite{Mukh2}, respectively. In this connection,  it should be  noted    that this assertion is related also  to    the calculations of    the  dispersion peripheral  model    for  the peripheral proton transfer reactions performed in  \cite{DDh1971} with   taking into account only  the   mechanism described  by the pole  diagram in  Fig. \ref{fig1}$a$.

    The results of the  analysis of   the experimental  differential cross sections \cite{Mukh2,Mukh6,HAS1991,Ivano2015} performed using Eq. (\ref{FBCS1}) and of the ANC values derived  for ${\rm{^9Be}} \,+p\to\, {\rm{^{10}B_i}}$ ($i$= 0--3), ${\rm{^{16}O}}\,+p\to\,{\rm {^{17}F_i}}$ ($i$= 0 and 1)  and ${\rm{^{16}O}} \,+\,t\to \,{\rm{^{19}F}}$(g.s.)    and their comparison with those of the conventional  DWBA obtained by other authors in Refs. \cite{Mukh2,Mukh6,HAS1991}    will be reported in the next paper.  Besides, there, the results of application of the ANC values above for the  nuclear-astrophysical ${\rm{^9Be(}}p,\,\gamma{\rm{)^{10}B}}$, ${\rm{^{16}O(}}p,\,\gamma{\rm{)^{17}F}}$ and ${\rm{^{19}F(}}p,\,\alpha{\rm{)^{16}O}}$ reactions  will also be presented.
 \bigskip

  {\bf VII.  CONCLUSION }
  \bigskip

   \hspace{0.6cm}  Within the   three-body  Schr\"{o}dinger formalism combined with the dispersion theory, a new asymptotic theory   is proposed for the  peripheral  sub- and above-barrier charged-particle transfer  $A$($x$, $y$)$B$ reaction, which is related to  the ``non-dramatic'' case, where  $x$=($y$ + $a$),  $B$=($A$ + $a$) and  $a$ is the transferred particle.  There,   the contribution of the three-body  ($A$, $a$ and $y$)  Coulomb dynamics of the  transfer  mechanism to the main reaction amplitude is taken into account in the correct manner within the framework of the  dispersion theory.  While,   an influence of  the  Coulomb-nulear distortion effects in the entrance and exit channels are kept in mind as it  is done in the conventional  DWBA.  In the asymptotic theory proposed,    the contribution of the three-body Coulomb effects in the initial, intermediate and final states  to the amplitude for the main pole mechanism is taken correctly  into account in all orders over  the Coulomb polarization potential  $V_{i,f}^C$  of the  perturbation theory. Therefore, it can be considered as a  generalization of the ''post''-approximation and  the post form of the conventional   DWBA.

    The explicit forms of  the generalized DWBA amplitude,  the  peripheral partial amplitudes at $l_i>>$ 1 and $l_f>>$1 and   respective the differential cross section   have been obtained. They   are directly expressed in the terms of the  product of  the   ANC's (or respective the NVC's) for  $y$ + $a\to\,x$  and    $A$ + $a\to\,B$ being  adequate to  the physics of the charge particle surface reaction.     In the amplitude derived, the  contributions both of  the rather low partial  waves and of the peripheral partial ones   are taken into account in the pole approximation valid for the channel radius $R_i^{{\rm{ch}}}\gtrsim R_N$.  It   makes it possible to consider simultaneously both the sub-barrier transfer reaction and the above-barrier one.   The calculations  of  the   partial amplitudes  has been perform    for the   specific above- and sub-barrier  peripheral   reactions corresponding to the proton and triton transfer mechanisms, respectively. It is shown quantitatively that  it is necessary to take into account   the three-body Coulomb dynamics  in the main pole transfer mechanism   for calculation   of  the amplitude and the differential cross section where   the  partial amplitudes with $l_i>>$ 1 and $l_f>>$1 provide essential contribution at least in the angular range of the main peak of the angular distribution of the differential cross section.

  \bigskip
  {\bf ACKNOWLEDGEMENT}
   \bigskip

    The authors are deeply grateful to L. D. Blokhintsev for   discussions and constructive suggestions.  This work has been supported in part  by the Academy of Sciences   of the Republic of Uzbekistan.

\bigskip

   {\bf APPENDIX A: Behavior of the pole-approximation  and the ''post'' form of the DWBA amplitudes near $\cos\theta\,\to\,\xi$ and their the peripheral partial amplitudes at $l_i>>$1. The approximate   forms of  the CRFs and the ``dramatic'' case}
\bigskip

  Here,  the explicit approximate  forms for the behavior of the pole-approximation  and the post form of the DWBA amplitudes near $\cos\theta\,\to\,\xi$ and the corresponding peripheral partial  amplitudes at $l_i>>$1 are presented. Besides, below,  we will find out the main reason of a provenance of the ``dramatic'' case  for the CRF values calculated   at the values of the Coulomb parameters $\eta_x$,  $\eta_B$ or $\eta_x+\eta_B$ near to a natural number.

 According to Refs. \cite{Av86,Mukh10} and \cite{Popov1964},    the explicit expressions of  the ${\tilde{M}}^{{\rm(}s{\rm{)}}}_{{\rm{pole}}}{\rm{(}}E_i,\,cos\theta{\rm{)}}$, which determine the behaviors of the pole $M^{{\rm{DW}}}_{{\rm{pole}}}{\rm{(}}E_i,\,cos\theta{\rm{)}} $ amplitude near the  singularity at $\cos\theta=\,\xi$,  and  the corresponding   peripheral partial amplitudes for $l_i>>$ 1    have the forms as
 $$
  {\tilde{M}}^{{\rm(}s{\rm{)}}}_{{\rm{pole}}}{\rm{(}}E_i,\,cos\theta{\rm{)}}=\frac{m_a}{k_ik_f}\frac{ G_{Aa} G_{ay}}{{\rm{(}}\xi-\cos\theta{\rm{)}}^{{\rm{1}}-\eta_{xB}+i\eta_{if}}}.
\eqno(A1)
$$ and
   $$
 \tilde{M}_{l_i;\,{\rm{pole}}}^{{\rm{(}}s{\rm{)}}}{\rm{(}}E_i{\rm{)}}
 \approx\sqrt{\pi}\frac{m_a}{k_ik_f}G_{Aa}G_{ay}
 \frac{{\rm{(}}\xi^{{\rm{2}}}-{\rm{1)}}^{{\rm{(}}\eta_{xB}-i\eta_{if}{\rm{)}}/{\rm{2}}}}
 {\Gamma{\rm{(}}{\rm{1}}-\eta_{xB}+i\eta_{if}{\rm{)}}\sqrt{\tau^{{\rm{2}}}-\rm{1}}}
 $$
 $$
 \times\frac{e^{-l_i\,\ln\tau}}{l_i^{{\rm{1/2}}+\eta_{xB}-i\eta_{if}}}
 \eqno(A2)
 $$  for $l_i>>$ 1, respectively, where $\tau=\,\xi\,+\,\sqrt{\xi^2\,-\,{\rm{1}}}$, $\eta_{xB}=\eta_x+\eta_B$ and $\eta_{if}=\eta_i+\eta_f$, and $\eta_i$ ($\eta_f$) is the Coulomb parameter in the entrance (exit) channel. From Eqs. (\ref{subeqpole}) - (\ref{subeqSB8a}), (A1) and (A2) we can obtain  the explicit  forms of the peripheral partial amplitudes for the
   $M^{{\rm(}s{\rm{)}}\,DW}_{{\rm{pole}}}{\rm{(}}E_i,\,cos\theta{\rm{)}}$ and  $M^{{\rm(}s{\rm{)}}\,DW}_{{\rm{post}}}{\rm{(}}E_i,\,cos\theta{\rm{)}}$ amplitudes. They have the forms as
   $$
   M_{l_i;\,{\rm{pole}}}^{{\rm{DW}}}\approx \tilde{N}^{{\rm {DW}}}_{{\rm{pole}}}\tilde{M}_{l_i;\,{\rm{pole}}}^{{\rm{(}}s{\rm{)}}}{\rm{(}}E_i{\rm{)}}\eqno(A3)
   $$and
   $$
    M_{l_i;\,{\rm{post}}}^{{\rm{DW}}}\approx \tilde{N}^{{\rm {DW}}}_{{\rm{post}}}\tilde{M}_{l_i;\,{\rm{pole}}}^{{\rm{(}}s{\rm{)}}}{\rm{(}}E_i{\rm{)}},\eqno(A4)
   $$
   where $\tilde{N}^{{\rm {DW}}}_{{\rm{pole}}}=N^{{\rm {DW}}}_{{\rm{pole}}}/\Gamma{\rm{(}}{\rm{1}}-\eta_{xB}+i\eta_{if}{\rm{)}}$ and $\tilde{N}^{{\rm {DW}}}_{{\rm{post}}}=N^{{\rm {DW}}}_{{\rm{post}}}/\Gamma{\rm{(}}{\rm{1}}-\eta_{xB}+i\eta_{if}{\rm{)}}$.

The explicit forms of the   $N^{{\rm{DW}}}_{{\rm{pole}}} $ and $N^{{\rm{DW}}}_{{\rm{post}}}$ CRF's contain the integrals over the  variable $t$ (0$\le t\le $ 1) with  the cumbersome   integrand functions. As seen from Ref. \cite{Mukh10},   the  dependence of  the integrand functions on  the vertex Coulomb parameters ($\eta_x$ and $\eta_B$) and the Coulomb parameters in the entrance and  exit channels ($\eta_i$ and $\eta_f$)  are presented in  the factorized forms as  $F^{{\rm{(}}j{\rm{)}}}_{\eta_x\,\eta_B}$($t$)$\tilde{F}^{{\rm{(}}j{\rm{)}}}_{\eta_i\,\eta_f}$($t$) for  the  integral corresponding   to  the   $N^{{\rm{DW}}}_{{\rm{pole}}} $  ($j$= 1) and that corresponding  to $N^{{\rm{DW}}}_{{\rm{post}}}$ ($j$=  2). One notes that the $\tilde{F}^{{\rm{(}}j{\rm{)}}}_{\eta_i\,\eta_f}$($t$) functions are regular at the points $t$=0 and 1, whereas the $F^{{\rm{(}}j{\rm{)}}}_{\eta_x\,\eta_B}$($t$) functions have the integrable singularities at these points. In this case,  the
   approximated explicit  forms  for the $N^{{\rm{DW}}}_{{\rm{pole}}} $ and $N^{{\rm{DW}}}_{{\rm{post}}} $ CRFs   can be obtained  from  the expressions     (14) and (26) of Ref. \cite{Mukh10}, since the modulus of  the $\tilde{F}^{{\rm{(}}j{\rm{)}}}_{\eta_i\,\eta_f}$($t$) functions change slower than the  $F^{{\rm{(}}j{\rm{)}}}_{\eta_x\,\eta_B}$($t$) functions within the integration interval. Therefore,
  the regular $\tilde{F}^{{\rm{(}}j{\rm{)}}}_{\eta_i\,\eta_f}$($t$) functions    can be taken out  from  under the integrations in Eqs.  (14) and (26) of \cite{Mukh10}  at   the point $t$=0 being  a  singular point (a branch one) for the other $\tilde{F}^{{\rm{(}}j{\rm{)}}}_{\eta_x\,\eta_B}$($t$)  functions ($j$= 1 and 2).
  As a result,  the expressions  for the CRF's above    can be reduced to the forms
$$
N^{{\rm{DW}}}_{{\rm{pole}}}\approx N^{\rm{(}{\it ap}{\rm{);\,DW}}}_{{\rm{pole}}}=\, \Gamma{\rm{(1}}\,-\eta_{xB}\,+\,i\eta_{if}{\rm{)}}{\rm{(}}\xi^{{\rm{2}}}\,-\,{\rm{1)}}^{i\eta_{if}/{\rm{2}}}\left(\frac{\lambda_B}{\lambda_x}\right)^{\eta_x}{\cal{D}}_{\eta_x\,\eta_B\,\eta_i\,\eta_f}{\rm{(}}k_i,\,k_f{\rm{)}}
I_{{\rm{pole}}}{\rm{(}}\eta_x,\,\eta_B{\rm{)}} \eqno(A5)
$$ and
$$
N^{{\rm{DW}}}_{{\rm{post}}}\approx N^{\rm{(}{\it ap}{\rm{);\,DW}}}_{{\rm{post}}}=
\,N^{\rm{(}{\it ap}{\rm{);\,DW}}}_{{\rm{pole}}}\,+\, N^{\rm{(}{\it ap}{\rm{);\,DW}}}_{\Delta}\,-\, N^{\rm{(}{\it ap}{\rm{);\,DW}}}_f,
\eqno(A6)
$$
 $$
N^{{\rm{DW}}}_{\Delta}\approx N^{\rm{(}{\it ap}{\rm{);\,DW}}}_{\Delta}=\,\mu_{ay}\eta_{yA}{\rm{(2}}E_{yA}/\mu_{yA}{\rm{)}}^{{\rm{1/2}}}\Gamma{\rm{(1}}\,-\eta_{xB}\,+\,i\eta_{if}{\rm{)}}{\rm{(}}\xi^{{\rm{2}}}\,-\,{\rm{1)}}^{i\eta_{if}/{\rm{2}}}\left(\frac{\lambda_B}{\lambda_x}\right)^{\eta_x}
$$
$$
\times{\cal{D}}_{\eta_x\,\eta_B\,\eta_i\,\eta_f}{\rm{(}}k_i,\,k_f{\rm{)}}I_{{\rm{1}}}{\rm{(}}\eta_x,\,\eta_B{\rm{)}}, \eqno(A7)
$$
$$
N^{{\rm{DW}}}_f\approx N^{\rm{(}{\it ap}{\rm{);\,DW}}}_f=\,\mu_{ay}{\rm{(}}\eta_fk_f/\mu_f{\rm{)}}\Gamma{\rm{(1}}\,-\eta_{xB}\,+\,i\eta_{if}{\rm{)}} {\rm{(}}\xi^{{\rm{2}}}\,-\,{\rm{1)}}^{i\eta_{if}/{\rm{2}}}\left(\frac{\lambda_B}{\lambda_x}\right)^{\eta_x}
$$
$$
\times {\cal{D}}_{\eta_x\,\eta_B\,\eta_i\,\eta_f}{\rm{(}}k_i,\,k_f{\rm{)}}I_{{\rm{2}}}{\rm{(}}\eta_x,\,\eta_B{\rm{)}}. \eqno(A8)
$$Herein: $E_{yA}={\rm{(}}m_{Ax}E_i\,+\,m_B\varepsilon_{Aa}\,+\,m_x\varepsilon_{ay}{\rm{)}}/m_{yA}$ is the relative  kinetic energy of the $A$ and $y$ cores in the intermediate state,
$$
{\cal{D}}_{\eta_x\,\eta_B\,\eta_i\,\eta_f}{\rm{(}}k_i,\,k_f{\rm{)}}=\,C{\rm{(}}\eta_i,\,\eta_f{\rm{)}}  \left(\frac{k_{i{\rm{1}}}k_f}{{\rm{2}}i\kappa_{ay}^{{\rm{2}}}}\right)^{\eta_x}
\left(\frac{k_ik_{f{\rm{1}}}}{{\rm{2}}i\kappa_{Aa}^{{\rm{2}}}}\right)^{\eta_B}  \eqno(A9)
$$
$$
I_{{\rm{pole}}}{\rm{(}}\eta_x,\,\eta_B{\rm{)}}=\,
\eta_x\int_0^{{\rm{1}}}dt\frac{{\rm{(1}}\,-\tilde{c}t{\rm{)}}^{\eta_{xB}-{\rm{1}}}}{t^{{\rm{1}}+\eta_x}
{\rm{(1}}-t{\rm{)}}^{\eta_B}},\eqno(A10)
$$
$$
I_j{\rm{(}}\eta_x,\,\eta_B{\rm{)}}=\,
\int_0^{{\rm{1}}}dt\frac{{\rm{(1}}\,-\tilde{c}t{\rm{)}}^{\eta_{xB}-
{\rm{1}}}}{t^{\eta_x}{\rm{(1}}-t{\rm{)}}^{\eta_B}\sqrt{\chi_j{\rm{(}}t{\rm{)}}}}. \eqno(A11)
$$
Herein: $\tilde{c}={\rm{1}}-\lambda_x/\lambda_B<$ 1($\lambda_x=m_y/m_x$ and $\lambda_B=m_A/m_B$) and
$$
\chi_j{\rm{(}}t{\rm{)}}=a_jt^{{\rm{2}}}\,+\,b_jt\,+\,c_j\eqno(A12a)
$$
in which
$$ a_{{\rm{1}}}=m_a{\rm{(}}\kappa_{ay}^{{\rm{2}}}\,-\,\kappa_{Aa}^{{\rm{2}}}{\rm{)}}/m_B\,- \,m_a^{{\rm{2}}}\kappa_{Aa}^{{\rm{2}}}/m_Am_B\,-\,{\rm{(}}m_am_{By}k_i/m_x\sqrt{m_Am_B}{\rm{)}}^{{\rm{2}}},
$$
$$
a_{{\rm{2}}}=m_am_{AB}{\rm{(}}\kappa_{ay}^{{\rm{2}}}\,-\,\kappa_{Aa}^{{\rm{2}}}{\rm{)}}/m_B^{{\rm{2}}}\,-\,{\rm{(}}m_am_{By}k_i/m_x m_B{\rm{)}}^{{\rm{2}}},
$$
$$
 b_{{\rm{1}}}=m_A{\rm{(}}\kappa_{ay}^{{\rm{2}}}\,-\,\kappa_{Aa}^{{\rm{2}}}{\rm{)}}/m_B\,+\, m_a^{{\rm{2}}}\kappa_{Aa}^{{\rm{2}}}/m_Am_B\,+\,{\rm{(}}m_am_{By}\\k_i/m_x\sqrt{m_Am_B}{\rm{)}}^{{\rm{2}}}, \eqno(A13)
 $$
 $$
 b_{{\rm{2}}}=m_A^{{\rm{2}}}{\rm{(}}\kappa_{ay}^{{\rm{2}}}\,-\,\kappa_{Aa}^{{\rm{2}}}{\rm{)}}/m_B^{{\rm{2}}}\,+\,m_am_{AB}\kappa_{Aa}^{{\rm{2}}}/m_B^{{\rm{2}}}\,+\,{\rm{(}}m_am_{By}k_i/m_x m_B{\rm{)}}^{{\rm{2}}},$$
 $$
 c_{{\rm{1}}}=\kappa_{Aa}^{{\rm{2}}},\,\,    c_{{\rm{2}}}=m_A^{{\rm{2}}}\kappa_{Aa}^{{\rm{2}}}/m_B^{{\rm{2}}}
 $$
 and
$$
C{\rm{(}}\eta_i,\,\eta_f{\rm{)}}=\, \exp[-{\rm{(}}\eta_f\,-\,\eta_i{\rm{)}}\tilde{\varphi}\,-\,{\rm{(}}\eta_i\,+\,\eta_f{\rm{)}}
\tilde{\psi}], \eqno(A14)
$$where
$$
\tilde{\varphi}=\,\tan^{{\rm{-1}}}\,\left(\frac{k_i\,-\,k_{f{\rm{1}}}}{\kappa_{Aa}}\right),
\,\,\,\tilde{\varphi}=\,\tan^{{\rm{-1}}}\left(\frac{\kappa_{Aa}}{k_i\,+\,k_{f{\rm{1}}}}\right). \eqno(A15)
$$
By using this case, we note that there are misprints in expression (14) of \cite{Mukh10}. There,
in the right-hand side of the equation for $\tilde{N}\rm{(}\eta_{\alpha},\,\eta_{\beta},\,
 \eta_i,\,\eta_f\rm{)}$,  the factor ${\rm{(}}\lambda_{\beta}/\lambda_{\alpha}{\rm{)}}^{\eta_{\alpha}}$
 ($\equiv{\rm{(}}\lambda_B/\lambda_x{\rm{)}}^{\eta_x}$ in  Eq.(A11)) is omitted and factor $e^{-\pi\eta}$ should be substituted by that of  $e^{-\pi\eta/{\rm{2}}}$($\eta\equiv\eta_{if}$).

 In Eq. (A9), the Coulomb  $C{\rm{(}}\eta_i,\,\eta_f{\rm{)}}$  factor   arises because of the aforesaid approximate taking into account of the Coulomb distorted effects in the entrance and exit channels.  One notes that  this factor coincides with that obtained in
 \cite{AT1971} from the  approximate
 amplitude of the sub-barrier neutron transfer  reaction derived within   the diffraction model. Besides, as is seen from Eqs. (A12$a$) and (A13), $\chi_j$ ($t{\rm{)}}>$0 for 0$\le\,t\le$ 1 ($j$= 1 and 2) and  $\chi_{{\rm{1}}}$(0)= $\kappa_{Aa}^{{\rm{2}}}$,   $\chi_{{\rm{2}}}$(0)= $m_A^{{\rm{2}}}\kappa_{Aa}^{{\rm{2}}}/m_B^{{\rm{2}}}$ and  $\chi_j$(1)= $a_j\,+\,b_j\,+\,c_j$= $\kappa_{ay}^{{\rm{2}}}$ as well as $a_j<$ 0 and $b_j>$0,  since $k_i>\kappa_{ay}$ and $k_i>\kappa_{Aa}$.

We now consider the integrals (A10) and (A11).  Integration in Eq. (A10) can be easily done  by using formula 3.197(4) of \cite{GR1980}. It results in the expression
$$
I_{{\rm{pole}}}{\rm{(}}\eta_x,\,\eta_B{\rm{)}}=- \left(\frac{\lambda_x}{\lambda_B}\right)^{\eta_x}F_C{\rm{(}}\eta_x,\, \eta_B{\rm{)}},\eqno(A16)
$$where
$$
 F_C{\rm{(}}\eta_x,\, \eta_B{\rm{)}}=\frac{\Gamma{{\rm(1}}\,-\,\eta_x{\rm{)}}
 \Gamma{\rm{(1}}\,-\,\eta_B{\rm{)}}}{\Gamma{\rm{(1}}\,-\,\eta_x-\,\,\eta_B{\rm{)}}}\cdot \eqno(A17)
$$
 To take the integral    (A11),  firstly,  the $\chi_j$($t$)  function should be presented to the  form
$$
 \chi_j{\rm{(}}t{\rm{)}}
 =a_j{\rm{(}}t\,-\,t_j^{{\rm{(1)}}}{\rm{)(}}t\,-\,t_j^{{\rm{(2)}}}{\rm{)}}, \eqno(A12b)
 $$
 where $t_j^{{\rm{(}}k{\rm{)}}}$ ($k$= 1 and 2) are the solutions  of the equations $\chi_j{\rm{(}}t{\rm{)}}$=0  for which  $t_j^{\rm{(1)}}>$1 and $t_j^{\rm{(2)}}<$0 ($j$= 1 and 2).  Then,  using Eq. (A12$b$),  the [$\chi_j$($t$)]$^{{\rm{-1/2}}}$ functions can be  expanded in  the binomial series at the points $t=t_{j;\,{\rm{0}}}=b_j/{\rm{2}}|a_j|<$1, which     are the extremum (minimum) points of the functions above. The power expansion  for the [$\chi_j$($t$)]$^{{\rm{-1/2}}}$ functions is reduced to the form as
 $$
 [\chi_j{\rm{(}}t{\rm{)}}]^{{\rm{-1/2}}}=f_{{\rm{0}}}^{{\rm{(}}j{\rm{)}}}+ \sum_{n={\rm{2}}}^{\infty} f_n^{{\rm{(}}j{\rm{)}}} {\rm{(}}t\,-\,t_{j;\,{\rm{0}}}{\rm{)}}^n.
  \eqno(A18)
$$ Herein
 $$
 f_n^{{\rm{(}}j{\rm{)}}}={\rm{(-1)}}^n\frac{{\rm{2}}|a_j|^{n+{\rm{1/2}}}}{D_j^{{\rm{(}}n+{\rm{1)/2}}}} \sum_{k={\rm{0}}}^n{\rm{(-1)}}^k
 \frac{{\rm{(2}}k{\rm{-1)}}!![{\rm{2(}}n-k{\rm{)-1}}]!!}{k!{\rm{(}}n-k{\rm{)}}!},  \eqno(A19)
  $$  $f_{{\rm{0}}}^{{\rm{(}}j{\rm{)}}}=\,[\chi_j{\rm{(}}t_{j;\,{\rm{0}}}{\rm{)}}]^{{\rm{-1/2}}}=\,{\rm{(}}|a_j|/D_j{\rm{)}}^{{\rm{1/2}}}$ and $[\chi^{\prime}_j{\rm{(}}t_{j;\,{\rm{0}}}{\rm{)}}]^{{\rm{-1/2}}}$=\,0 in which  the  prime is marked a derivation from   the  $[\chi_j{\rm{(}}t{\rm{)}}]^{{\rm{-1/2}}}$ function, and ${\rm{(-1)}}!!$=\,1.

 Inserting   Eq. (A18) in Eq. (A11)
  and using formulae 3.197(3) and 3.211 from \cite{GR1980} in the obtained expression,  for $I_j{\rm{(}}\eta_x,\,\eta_B{\rm{)}}$  we derive  the following form
$$
I_j{\rm{(}}\eta_x,\,\eta_B{\rm{)}}=\frac{F_C{\rm{(}}\eta_x,\, \eta_B{\rm{)}}}{{\rm{(1}}\,-\,\eta_{xB}{\rm{)}}}\tilde{I}_j{\rm{(}}\eta_x,\,\eta_B{\rm{)}}\eqno(A20)
  $$where
 $$
\tilde{I}_j{\rm{(}}\eta_x,\,\eta_B{\rm{)}}=\, {\rm{2}}\left(\frac{|a_j|}{D_j}\right)^{{\rm{1/2}}}\,+\, \sum_{n={\rm{2}}}^{\infty} \frac{f_n^{{\rm{(}}j{\rm{)}}}}{n!}F_{{\rm{1}}}{\rm{(1}}-\eta_x,-n,{\rm{1}}-\eta_{xB};{\rm{2}}-\eta_{xB};t^{{\rm{-1}}}_{j;\,{\rm{0}}},\tilde{c}{\rm{)}}.
 \eqno(A21)
$$Herein:
$$
 F_{{\rm{1}}}{\rm{(1}}-\eta_x,-n,{\rm{1}}-\eta_{xB};{\rm{2}}-\eta_{xB};t^{{\rm{-1}}}_{j;\,{\rm{0}}},\tilde{c}{\rm{)}}=\,\sum_{q={\rm{0}}}^{n}\frac{n!}{q!(n-q)!}{\rm{(}}-t_{j;\,{\rm{0}}}{\rm{)}}^{-q}
 $$
 $$
 \times\frac{\Gamma{\rm{(1}}\,+q\,-\,\eta_x{\rm{)}}}{\Gamma{\rm{(1}}\,-\,\eta_x{\rm{)}}}\frac{\Gamma{\rm{(2}}\,-\,\eta_{xB}{\rm{)}}}{\Gamma{\rm{(2}}\,+\,q\,-\,\eta_{xB}{\rm{)}}}F{\rm{(}}{\rm{1}}-\eta_{xB},{\rm{1}}\,+q\,-\,\eta_x;{\rm{2}}\,+\,q\,-\,\eta_{xB};\tilde{c}{\rm{)}}   \eqno(A22)
$$is the hypergeometric function of two variables \cite{GR1980}, and $F{\rm{(}}a,b;c;\tilde{x}{\rm{)}}$ is the  known hypergeometric function.

Inserting Eqs. (A16)  and (A17)  in Eq. (A5), and  Eqs. (A7), (A8) and (A20) in Eq. (A6) the $N^{{\rm{DW}}}_{{\rm{pole}}}   $ and $ N^{{\rm{DW}}}_{{\rm{post}}}$ CRF's    can be reduced to the  forms as
$$
N^{{\rm{DW}}}_{{\rm{pole}}}\approx N^{\rm{(}{\it ap}{\rm{);\,DW}}}_{{\rm{pole}}}=\,-\Gamma{\rm{(1}}\,-\eta_{xB}\,+\,i\eta_{if}{\rm{)}}{\rm{(}}\xi^{{\rm{2}}}\,-\,{\rm{1)}}^{i\eta_{if}/{\rm{2}}}F_C{\rm{(}}\eta_x,\, \eta_B{\rm{)}} {\cal{D}}_{\eta_x\,\eta_B\,\eta_i\,\eta_f}{\rm{(}}k_i,\,k_f{\rm{)}}\eqno(A23)
$$ and
$$
 N^{{\rm{DW}}}_{{\rm{post}}}\approx N^{\rm{(}{\it ap}{\rm{);\,DW}}}_{{\rm{post}}}=
 N^{\rm{(}{\it ap}{\rm{);\,DW}}}_{{\rm{pole}}}\{{\rm{1}}\,-
 \frac{\mu_{ay}{\rm{(}}\lambda_B/\lambda_x{\rm{)}}^{\eta_x}}{{\rm{(1}}\,-\,\eta_{xB}{\rm{)}}}
$$
$$
\times\big[\eta_{yA}{\rm{(2}}E_{yA}/\mu_{yA}{\rm{)}}^{{\rm{1/2}}}\tilde{I}_{{\rm{1}}}{\rm{(}}\eta_x,\,\eta_B{\rm{)}} -{\rm{(}}\eta_fk_f/\mu_f{\rm{)}}\tilde{I}_{{\rm{2}}}{\rm{(}}\eta_x,\,\eta_B{\rm{)}}\big]\}. \eqno(A24).
$$

As is seen from Eqs. (A23) and (A24),  the approximate allowance of the Coulomb distortions  in the entrance and exit channels in the expressions for  the  $N^{{\rm{DW}}}_{{\rm{pole}}}$ and $N^{{\rm{DW}}}_{{\rm{post}}}$  CRF's     makes us it possible  to derive their explicit forms in which  the factors, depending both  on the Coulomb $\eta_x$  and $\eta_B$ parameters and on the Coulomb    $\eta_i$  and $\eta_f$ ones, are  separated.
The explicit approximate forms for the  $M^{{\rm(}s{\rm{)}}\,{\rm{DW}}}_{{\rm{pole}}}{\rm{(}}E_i,\,cos\theta{\rm{)}}$ and $ M^{{\rm(}s{\rm{)}}\,{\rm{DW}}}_{{\rm{post}}}{\rm{(}}E_i,\,cos\theta{\rm{)}}$  amplitudes  can be obtained from Eqs.  (\ref{subeqpole}), (\ref{subeqSB8})  and   (\ref{subeqSB8a}) by mean of replacement of the $N^{{\rm{DW}}}_{{\rm{pole}}}$ and $N^{{\rm{DW}}}_{{\rm{post}}}$  CRF's by those given in Eqs. (A23) and (A24), respectively. Similar to Eqs. (A1) and (A2), the amplitudes  above determine the behavior of   peripheral partial amplitudes for $l_i>>$ 1, which have the forms as
$$
M_{{\rm{pole}};\,l_i}^{\rm{DW}} {\rm{(}}E_i{\rm{)}}\approx M_{{\rm{pole}};\,l_i}^{\rm{(}{\it ap}{\rm{);\,DW}}} {\rm{(}}E_i{\rm{)}}=\,\sqrt{\pi}\frac{m_a}{k_ik_f}G_{Aa}G_{ay}
\frac{{\rm{(}}\xi^{{\rm{2}}}-{\rm{1)}}^{\eta_{xB}/{\rm{2}}}}{\sqrt{\tau^{{\rm{2}}}-\rm{1}}}
$$
$$
\times F_C{\rm{(}}\eta_x,\, \eta_B{\rm{)}} {\cal{D}}_{\eta_x\,\eta_B\,\eta_i\,\eta_f}{\rm{(}}k_i,\,k_f{\rm{)}}\frac{e^{-l_i\,\ln\tau}}{l_{l_i}^{{\rm{1/2}}+\eta_{xB}-i\eta_{if}}}
 \eqno(A25)
 $$  and
$$
M_{{\rm{post}};\,l_i}^{\rm{DW}} {\rm{(}}E_i{\rm{)}}\approx M_{{\rm{post}};\,l_i}^{\rm{(}{\it ap}{\rm{);\,DW}}} {\rm{(}}E_i{\rm{)}}= \,{\cal{R}}^{{\rm{(ap);\,DW}}}_{{\rm{post}}}
M_{{\rm{pole}};\,l_i}^{\rm{(}{\it ap}{\rm{);\,DW}}} {\rm{(}}E_i{\rm{)}} \eqno(A26)
$$  for $l_i>>$ 1, where $G_{Aa}\equiv G_{Aa;\,l_Bj_B}$ and  $G_{ay}\equiv G_{ay;\,l_xj_x}$, and $N^{\rm{(}{\it ap}{\rm{);\,DW}}}_{{\rm{post}}}$ is determined by Eq. (A24), and ${\cal{R}}^{{\rm{(ap);\,DW}}}_{{\rm{post}}}=\,N^{\rm{(}{\it ap}{\rm{);\,DW}}}_{{\rm{post}}}/N^{\rm{(}{\it ap}{\rm{);\,DW}}}_{{\rm{pole}}}$.

According to \cite{Av86,Mukh10}, the CRF $N^{\rm{TBDM}}$ for the  $ M^{{\rm{(}}s{\rm{)}}\,{\rm{TBDM}}}{\rm{(}}E_i,\,cos\theta{\rm{)}}$  amplitude, given by Eq. (\ref{subeqSBDM}), can be presented in the form
   $$
N^{\rm{TBDM}}=-{\rm{(}}k_{i{\rm{1}}}k_f/{\rm{2}}i\kappa_{ay}^{{\rm{2}}}{\rm{)}}^{\eta_x}(k_ik_{f1}/{\rm{2}}i\kappa_{Aa}^{{\rm{2}}})^{\eta_B}{\rm{(}}\xi^{{\rm{2}}}-{\rm{1}}{\rm{)}}^{i\eta_{if}/{\rm{2}}}
$$
$$
\times \Gamma{\rm{(}}{\rm{1}}-\eta_{xB}+i\eta_{if}{\rm{)}}F_C{\rm{(}}\eta_x,\,\eta_B{\rm{)}}N{\rm{(}}\eta_x, \,\eta_B,\,\eta_i,\,\eta_f{\rm{)}},
\eqno(A27)
$$
$$
N{\rm{(}}\eta_x, \,\eta_B,\,\eta_i,\,\eta_f{\rm{)}}=
F^{{\rm{-1}}}_C{\rm{(}}\eta_x,\, \eta_B{\rm{)}}\Delta_{\eta_{yA}}{\rm{(}}k_i,\,k_f{\rm{)}}\tilde{\Delta}_{\eta_i\,\eta_f}{\rm{(}}k_i,\,k_f{\rm{)}},
\eqno(A28)
$$
$$
\Delta_{\eta_{yA}}{\rm{(}}k_i,\,k_f{\rm{)}}=\,e^{-{\rm{2}}\eta_{yA}\varphi_{yA}},\,\,\, \tilde{\Delta}_{\eta_i\,\eta_f}{\rm{(}}k_i,\,k_f{\rm{)}}=\,\exp{\rm{(}}-\eta_i\varphi_i\,-\,\eta_f\varphi_f{\rm{)}},
 \eqno(A29)
$$
$$
\varphi_{yA}=\tan^{{\rm{-1}}}\frac{{\rm{(}}m_am_{yA}E_{yA}{\rm{)}}^{{\rm{1/2}}}}
{{\rm{(}}m_xm_A\varepsilon_{ay}{\rm{)}}^{{\rm{1/2}}}\,+\,{\rm{(}}m_Bm_y\varepsilon_{Aa}{\rm{)}}^{{\rm{1/2}}}},\eqno(A30)
$$
$$
\varphi_i=\tan^{{\rm{-1}}}\left(\frac{k_{f{\rm{1}}}^{\rm{2}}-{k_i}^{\rm{2}}}{{\rm{2}}k_i\kappa_{Aa}}\right),\,\,
\varphi_f=\tan^{{\rm{-1}}}\left(\frac{k_{i{\rm{1}}}^{\rm{2}}-{k_f}^{\rm{2}}}
{{\rm{2}}k_{i{\rm{1}}}\kappa_{ay}}\right).\eqno(A31)
$$
The behavior of the   peripheral partial amplitudes  of the  $ M^{{\rm{(}}s{\rm{)}}\,{\rm{TBDM}}}{\rm{(}}E_i,\,cos\theta{\rm{)}}$ amplitude  for $l_i>>$ 1 has the form as  \cite{Av86,Mukh10}
$$
M_{l_i}^{\rm{TBDM}} {\rm{(}}E_i{\rm{)}}\approx \sqrt{\pi}\frac{m_a}{k_ik_f}G_{Aa}G_{ay}\frac{{\rm{(}}\xi^{{\rm{2}}}-{\rm{1)}}^{{\rm{(}}\eta_{xB}-i\eta_{if}{\rm)}/{\rm{2}}}}{\sqrt{\tau^{{\rm{2}}}-\rm{1}}}
\tilde{N}^{\rm{TBDM}}\frac{e^{-l_i\,\ln\tau}}{l_{l_i}^{{\rm{1/2}}+\eta_{xB}-i\eta_{if}}}.
\eqno(A32)
$$

 One notes that, the  three-body CRF  $N{\rm{(}}\eta_x, \,\eta_B,\,\eta_i,\,\eta_f{\rm{)}}$($\equiv N$)  (A28) arises due to  correct taking into account  both  of  the three-body Coulomb dynamics  in the main transfer mechanism  and  of    the Coulomb interactions in the entrance and exit states. As shown in Refs. \cite{Av86} and \cite{Dem1975},    the factors  $F^{{\rm{-1}}}_C{\rm{(}}\eta_x,\, \eta_B{\rm{)}}$  and $\Delta_{\eta_{yA}}{\rm{(}}k_i,\,k_f{\rm{)}}$ in Eq. (A28) by-turn  arise  as a result of taking into account  all possible subsequent mutual   Coulomb interactions of the transferred particle $a$ with the cores $A$ and $y$  and   of the cores $A$ and $y$, respectively, in the main  transfer mechanism.
 Whereas   the factor $\tilde{\Delta}_{\eta_i\,\eta_f}{\rm{(}}k_i,\,k_f{\rm{)}}$ arises due to Coulomb interaction in the initial and final  states.  As is seen from the expressions (A27) and (A28), in Eq. (A27),  the contribution  of the $F_C{\rm{(}}\eta_x,\, \eta_B{\rm{)}}$ factor  to the  $N^{\rm{TBDM}}$ CRF is compensated by an appearance of the factor  $F^{{\rm{-1}}}_C{\rm{(}}\eta_x,\, \eta_B{\rm{)}}$ in the three-body Coulomb factor  $N$. The $F_C{\rm{(}}\eta_x,\, \eta_B{\rm{)}}$ factor  arises because of  the   vertex Coulomb effects in the  three-ray vertexes of  the pole diagram of Fig. \ref{fig1}$a$, which  corresponds  to the pure pole amplitude \cite{Av86,DDh1971}. Besides,  the expression (\ref{subeqSBDM})    coincides with  the behavior of   the pure pole amplitude (Fig. \ref{fig1}$a$) near a  vicinity of the singularity at  $\cos\theta=\,\xi$ when a contribution of the three-body Coulomb effects in the  three-body DWBA amplitude is ignored, i.e.,   the three-body Coulomb factor $N$   should  set equal  to unity  in Eq. (A27).

 We now discuss the main reason of  a provenance of the ``dramatic`` case mentioned above. It  arises because of   taking into account only  the single-Coulomb rescattering  of the transferred particle $a$   with  the cores $A$ and $y$ in the pole-approximation and the ``post'' form of the DWBA  at  values of either $\eta_x$ or $\eta_B$ or $\eta_{xB}$ are in the vicinity of   a natural number. In this case, as it is    seen from Eqs. (A23) and (A24) as well as Eqs.    (A27) and  (A28),   the difference between the CRF's   $  N^{{\rm{DW}}}_{{\rm{pole}}}$,   $  N^{{\rm{DW}}}_{{\rm{post}}}$  and  $N^{{\rm{TBDM}}} $ (or $ {\tilde{N}}^{{\rm{DW}}}_{{\rm{pole}}}$,
 ${\tilde{N}}^{{\rm{DW}}}_{{\rm{post}}}$  and  $\tilde{N}^{{\rm{TBDM}}} $)  becomes  significant. It is due to a presence of the vertex Coulomb  $F_C{\rm{(}}\eta_x,\, \eta_B{\rm{)}}$ and $F_C{\rm{(}}\eta_x,\, \eta_B{\rm{)}}/$(1 - $\eta_{xB}$)  factors in Eqs. (A23) and (A24), respectively, whereas they are absent in the $N^{{\rm{TBDM}}} $ (or ${\tilde{N}}^{{\rm{TBDM}}} $) CRF  in  Eqs. (A27) and (A28).  This means that the power expansion over the Coulomb polarization potential $\Delta V_{i,f}^C$ in the transition operator of  Eqs. (\ref{subeqSB5}) and (\ref{subeqSB6}), which correspond to the zero- and first orders of the perturbation theory over  $\Delta V_{i,f}^C$, has  a poor convergence in the ``dramatic'' case.

 Therefore,  in the ``dramatic'' case,    the next terms ($\bigtriangleup V_f^CG_C\bigtriangleup V_i^C$)    of the transition operator in  the series in $\bigtriangleup V_{f,\,i}^C$ should directly be  taken into account in the  $ M^{{\rm{TBDW}}}{\rm{(}}E_i,\,cos\theta{\rm{)}}$ amplitude. Since  each of the terms of them  has the identical behaviour as that for
 the $M_{{\rm{post}}}^{{\rm(}s{\rm{)}}\,{\rm{DW}}}{\rm{(}}E_i,\,cos\theta{\rm{)}}$
amplitude \cite{Mukh10}, one obtains the expression
$$
 \bigtriangleup M^{{\rm{TBDW}}}{\rm{(}}E_i,\,cos\theta{\rm{)}}\approx\,\bigtriangleup M^{{\rm(}s{\rm{)}}\,{\rm{TBDW}}}{\rm{(}}E_i,\,cos\theta{\rm{)}}=
\,\Delta N^{\rm{TBDW}}\tilde{M}_{{\rm{pole}}}^{{\rm(}s{\rm{)}}}{\rm{(}}E_i,\,cos\theta{\rm{)}},
\eqno(A33)
$$
where  $\Delta N^{\rm{TBDW}}$ is the CRF corresponding to the $\bigtriangleup M^{{\rm{TBDW}}}{\rm{(}}E_i,\,cos\theta{\rm{)}}$ amplitude.  Then,  the expression for  the main singular  term of the $M^{{\rm{TB}}}{\rm{(}}E_i,\,\cos\theta{\rm{)}}$ amplitude near $\cos\theta=\,\xi$ can be  presented in the form
$$
 M^{{\rm{TB}}}{\rm{(}}E_i,\,cos\theta{\rm{)}}\approx\, M^{{\rm(}s{\rm{)}}\,{\rm{TB}}}{\rm{(}}E_i,\,\cos\theta{\rm{)}}=
\,{\cal{R}}^{{\rm{TB}}}{\rm{(}}E_i{\rm{)}}M^{{\rm(}s{\rm{)}}}_{{\rm{pole}}}{\rm{(}}E_i,\,cos\theta{\rm{)}},
 \eqno(A34)
$$  where
 $$
  {\cal{R}}^{{\rm{TB}}}{\rm{(}}E_i{\rm{)}}\,=\,N^{{\rm{TB}}}{\rm{(}}E_i{\rm{)}}/N^{{\rm{DW}}}_{{\rm{pole}}}{\rm{(}}E_i{\rm{)}}
\eqno(A35)
$$ and
 $$
N^{{\rm{TB}}}=\,N^{{\rm{DW}}}_{{\rm{post}}}\,+\,\Delta N^{\rm{TBDW}}. \eqno(A36)
$$
As a result, from  Eqs. (\ref{subeqSBDM}) and (A34), the behavior of the exact  three-body  $ M^{{\rm{TB}}}{\rm{(}}E_i,\,cos\theta{\rm{)}}$ DWBA amplitude near  the  singularity at $\cos\theta$ = $\xi$ is  presented in the form
$$
M^{ {\rm{TB}}}{\rm{(}}E_i,\,cos\theta{\rm{)}}\approx M^{{\rm(}s{\rm{)}}\,{\rm{TB}}}{\rm{(}}E_i,\,cos\theta{\rm{)}}\,=\,{\tilde{{\cal{R}}}}^{TB}{\rm{(}}E_i{\rm{)}}M^{{\rm(}s{\rm{)}}}_{{\rm{pole}}}{\rm{(}}E_i,\,cos\theta{\rm{)}}. \eqno(A37)
$$
 Herein:
$$
{\tilde{{\cal{R}}}}^{{\rm{TB}}}{\rm{(}}E_i{\rm{)}}\,=\,N^{{\rm{TBDM}}}{\rm{(}}E_i{\rm{)}}/N^{{\rm{TB}}}{\rm{(}}E_i{\rm{)}}=\, \tilde{N}^{{\rm{TBDM}}}{\rm{(}}E_i{\rm{)}}/\tilde{N}^{{\rm{TB}}}{\rm{(}}E_i{\rm{)}}, \eqno(A38)
\eqno
$$
 which  is   valid    for the ``dramatic'' case, where $\tilde{N}^{{\rm{TB}}}{\rm{(}}E_i{\rm{)}}=N^{{\rm{TB}}}{\rm{(}}E_i{\rm{)}}/\Gamma{\rm{(}}{\rm{1}}-\eta_{xB}+i\eta_{if}{\rm{)}}$.

  One notes that, in reality,  the expressions (A34)--(A38) are valid  simultaneously both  for the ``dramatic'' case and for the ``non-dramatic'' one.   Therefore, Eqs. (A34)--(A36)
  are  more accurate than the expression  (\ref{subeqSB9}). Consequently, they  may also be used for testing  the accuracy of Eq. (\ref{subeqSB9}).  Hence,   a knowledge of  the explicit form of the $\Delta N^{\rm{TBDW}}$ CRF   is required.  But,     the task of  direct  finding   the explicit form of the $\Delta N^{\rm{TBDW}}$ CRF      is fairly difficult    because of the presence of the three-body Coulomb operator $G_C$ in the transition operator of Eq.  (\ref{subeqSB7})   and, so,   it  requires a special consideration.   At present such work  is in progress within the cycle of works, which are carried by us, on a development of the  asymptotic theory  for the peripheral  reaction (\ref{subeq1}), which must really involve both the ``dramatic'' case and the ``non-dramatic'' one.

         \bigskip

   {\bf APPENDIX B: Formulae and expressions}

  \bigskip

  \hspace{0.6cm} Here we present the necessary formulae and expressions.

  The matrix element   $M_{Aa}{\rm{(}}{\mathbf{q}}_{Aa}{\rm{)}}$ of the virtual decay $B\,\to\,A\,+\,a$ is related to the  overlap function $I_{Aa}{\rm{(}}{\mathbf r}_{Aa}{\rm{)}}$ as \cite{Blok77}
$$
M_{Aa}{\rm{(}}{\mathbf{q}}_{Aa}{\rm{)}}\,=\,N_{Aa}^{{\rm{1/2}}}\int e^{-i{\mathbf{q}}_{Aa}{\mathbf{r}}_{Aa}}V_{Aa}{\rm{(}}{\mathbf{r}}_{Aa}{\rm{)}}I_{Aa}{\rm{(}}{\mathbf{ r}}_{Aa}{\rm{)}}d{\mathbf{ r}}_{Aa}
$$
$$
=-\,N_{Aa}^{{\rm{1/2}}}\Big(\frac{q_{Aa}^{{\rm{2}}}}{{\rm{2}}\mu_{Aa}}\,+\,\varepsilon_{Aa}\Big)
\int e^{-i{\mathbf{q}}_{Aa}{\mathbf{r}}_{Aa}}I_{Aa}{\rm{(}}{\mathbf{ r}}_{Aa}{\rm{)}}d{\mathbf{ r}}_{Aa} \eqno(B1)
$$
$$
=\,\sqrt{{\rm{4}}\pi}\sum_{l_B\mu_Bj_B\nu_B}C_{j_B\nu_BJ_AM_A}^{J_BM_B}C_{l_B\mu_B J_aM_a}^{j_B
 \nu_B}G_{Aa;\,l_Bj_B}{\rm{(}}q_{Aa}{\rm{)}}Y_{l_B\mu_B}(\hat{{\mathbf q}}_{Aa}{\rm{)}},
$$
  where  $G_{Aa;\,l_Bj_B}{\rm{(}}q_{Aa}{\rm{)}}$ is the vertex formfactor  for the virtual decay $B\,\to\,A\,+\,a$,  ${\mathbf{q}_{Aa}}$ is the relative momentum  of the $A$ and $a$ particles  and $G_{Aa;\,l_Bj_B}\equiv G_{Aa;\,l_Bj_B}{\rm{(}}i\kappa_{Aa}{\rm{)}}$, i.e., the NVC coincides with  the vertex formfactor
   $G_{Aa; \,l_Bj_B}{\rm{(}}q_{Aa}{\rm{)}}$ when all the $B$, $a$ and $A$ particles are  on-shell  ($q_{Aa}=\,i\kappa_{Aa}$).
The same relations similar to Eq. (B1)  hold for the  matrix element   $M_{ay}{\rm{(}}{\mathbf{q}}_{ay}{\rm{)}}$ of the virtual decay $x\,\to\,y\,+\,a$ and the overlap function $I_{ay}{\rm{(}}{\mathbf r}_{ay}{\rm{)}}$.

The  partial-waves expansions for the distorted wave functions of  relative motion of the nuclei  in  the initial and exit states   of the reaction under consideration  have the form as \cite{Austern1964}
 $$
 \Psi^{{\rm{(}}+{\rm{)}}}_{{\mathbf{k}_{i}}}{\rm {(}}{\mathbf{r}_{i}}{\rm{)}}=\,\frac{{\rm{4}}\pi}{k_ir_i}\sum_{l_i\mu_i}i^{l_i}e^{i\sigma_{l_i}}\Psi_{l_i}{\rm{(}}k_i;\,r_i{\rm{)}}Y_{l_i\mu_i}{\rm{(}}\hat{{\mathbf r}}_i{\rm{)}}Y^*_{l_i\mu_i}{\rm{(}}\hat{{\mathbf k}}_i{\rm{)}},
 $$
 $$
  \Psi^{*{\rm{(}}-{\rm{)}}}_{{\mathbf{k}_{i}}}{\rm {(}}{\mathbf{r}_{i}}{\rm{)}}=\,\frac{{\rm{4}}\pi}{k_fr_f}\sum_{l_f\mu_f}i^{-\,l_f}e^{i\sigma_{l_f}}\Psi_{l_f}{\rm{(}}k_f;\,r_f{\rm{)}}Y_{l_f\mu_f}{\rm{(}}\hat{{\mathbf r}}_f{\rm{)}}Y^*_{l_f\mu_f}{\rm{(}}\hat{{\mathbf r}}_f{\rm{)}},\eqno(B2)
 $$
where $\Psi_{l}{\rm{(}}k;\,r{\rm{)}}$ is the partial wave functions in the initial state or the final one.

 The expansions of the $r_{ay}^{l_x}Y_{l_x\sigma_x}{\rm{(}}{\hat{\mathbf{r}}}_{ay}{\rm{)}}$ and $r_{Aa}^{l_B}Y_{l_B\sigma_B}^*{\rm{(}}{\hat{\mathbf{r}}}_{Aa}{\rm{)}}$ functions on the bipolar harmonics of the $l_x$ rank  and the $l_B$ one have the forms as
$$
r_{ay}^{l_x}Y_{l_x\sigma_x}{\rm{(}}{\hat{\mathbf{r}}}_{ay}{\rm{)}}=\,\sqrt{{\rm{4}}\pi}\sum_{\lambda_1+\lambda_2=\,l_x}\,\,\sum_{\tilde{ \mu}_{\lambda_1}\tilde{\mu}_{\lambda_2}}
\Big(\frac{\hat{l}_x!}{\hat{\lambda}_1!{\hat{\lambda}}_2!}\Big)^{{\rm{1/2}}}
\Big(\frac{\mu_{Ax}}{m_a}r_i\Big)^{\lambda_1}
\Big(-\frac{\mu_{Ax}}{\mu_{Aa}}r_f\Big)^{\lambda_2}
$$
$$
\times C_{\lambda_1\tilde{\mu}_{\lambda_1}\,\lambda_2\tilde{\mu}_{\lambda_2}}^{l_x\mu_x}Y_{\lambda_1\tilde{\mu}_{\lambda_1}}{\rm{(}}{\hat{\mathbf{r}}}_i{\rm{)}}
Y_{\lambda_2\tilde{\mu}_{\lambda_2}}{\rm{(}}{\hat{\mathbf{r}}}_f{\rm{)}}\eqno(B3)
$$ and
$$
r_{Aa}^{l_B}Y_{l_B\sigma_B}^*{\rm{(}}{\hat{\mathbf{r}}}_{Aa}{\rm{)}}=\,\sqrt{{\rm{4}}\pi}\sum_{\sigma_1+\sigma_2=\,l_B}\,\,\sum_{\tilde{ \mu}_{\sigma_1}\tilde{\mu}_{\sigma_2}}
\Big(\frac{\hat{l}_B!}{\hat{\sigma}_1!{\hat{\sigma}}_2!}\Big)^{{\rm{1/2}}}
\Big(-\frac{\mu_{By}}{\mu_{ay}}r_i\Big)^{\sigma_1}
\Big(\frac{\mu_{By}}{m_a}r_f\Big)^{\sigma_2}
$$
$$
\times C_{\sigma_1\tilde{\mu}_{\sigma_1}\,\sigma_2\tilde{\mu}_{\sigma_2}}^{l_B\mu_B}Y^*_{\sigma_1\tilde{\mu}_{\sigma_1}}{\rm{(}}{\hat{\mathbf{r}}}_i{\rm{)}}Y^*_{\sigma_2\tilde{\mu}_{\sigma_2}}{\rm{(}}{\hat{\mathbf{r}}}_f{\rm{)}}.
\eqno(B4)
$$  Eqs. (B3) and (B4) can be derived from (\ref{subeqSB14}) and
 $$
\int d{\hat{\mathbf{r}}}_iY^*_{\sigma_1\tilde{\mu}_{\sigma_1}}{\rm{(}}{\hat{\mathbf{r}}}_i{\rm{)}}Y_{l\mu_l}{\rm{(}}\hat{{\mathbf r}}_i{\rm{)}}Y_{l_i\mu_{l_i}}{\rm{(}}\hat{{\mathbf r}}_i{\rm{)}}Y_{\lambda_1\tilde{\mu}_{\lambda_1}}{\rm{(}}{\hat{\mathbf{r}}}_i{\rm{)}}=\, {\rm{(-1)}}^{\tilde{\mu}_l}\sum_{I{\tilde{\mu}}_I}
\left(\frac{\hat{l_i}\hat{\lambda}_1\hat{l}\hat{\sigma}_1}{{\rm{(4}}\pi{\rm{)}}^{\rm{2}}\hat{I}\hat{I}}\right)^{{\rm{1/2}}}
$$
$$
\times C_{l_i\,{\rm{0}}\,\lambda_{{\rm{1}}}\,{\rm{0}}}^{I\,{\rm{0}}}
C_{l\,{\rm{0}}\,\sigma_{{\rm{1}}}\,{\rm{0}}}^{I\,{\rm{0}}}
C_{l_i\,\tilde{\mu}_{l_i}\,\lambda_{{\rm{1}}}\,\tilde{\mu}_{\lambda_{{\rm{1}}}}}^{I\,\tilde{\mu}_I}
C_{l\,-\tilde{\mu}_l\,\sigma_{{\rm{1}}}\,\tilde{\mu}_{\sigma_{\rm{1}}}}^{I\,\tilde{\mu}_I}, \eqno(B5)
$$
 $$
\int d{\hat{\mathbf{r}}}_fY^*_{l\mu_l}{\rm{(}}\hat{{\mathbf r}}_f{\rm{)}}Y^*_{\sigma_{\rm{2}}\tilde{\mu}_{\sigma_{\rm{2}}}}{\rm{(}}{\hat{\mathbf{r}}}_f{\rm{)}}Y_{l_f\mu_{l_f}}{\rm{(}}\hat{{\mathbf r}}_f{\rm{)}}Y_{\lambda_{{\rm{2}}}\tilde{\mu}_{\lambda_{{\rm{2}}}}}{\rm{(}}{\hat{\mathbf{r}}}_f{\rm{)}}=\, \sum_{L{\tilde{\mu}}_L}
\left(\frac{\hat{l_f}\hat{\lambda}_{{\rm{2}}}\hat{l}\hat{\sigma}_{{\rm{2}}}}{{\rm{(4}}\pi{\rm{)}}^{\rm{2}}\hat{L}\hat{L}}\right)^{{\rm{1/2}}}
$$
$$
\times C_{l_f\,0\,\lambda_{{\rm{2}}}\,{\rm{0}}}^{L\,{\rm{0}}}C_{l\,{\rm{0}}\,\sigma_{{\rm{2}}}\,{\rm{0}}}^{L\,{\rm{0}}}C_{l_f\,\tilde{\mu}_{l_f}\,\lambda_{{\rm{2}}}\,\tilde{\mu}_{\lambda_{{\rm{2}}}}}^{L\,\tilde{\mu}_L}
C_{l\,\tilde{\mu}_{l}\,\sigma_{{\rm{2}}}\,\tilde{\mu}_{\sigma_{{\rm{2}}}}}^{L\,\tilde{\mu}_L}.
\eqno(B6)
$$
The explicit form of  $M^{{\rm{pole}}}_{l_xl_BJl_il_f}{\rm{(}}E_i{\rm{)}}$ entering Eq. (\ref{subeqSB32}) is given by
 $$
M^{{\rm{pole}}}_{l_xl_BJl_il_f}{\rm{(}}E_i{\rm{)}}=\, e^{i{\rm{(}}\sigma_{l_i}+\sigma_{l_f}{\rm{)}}}
 {\rm{(}}\hat{l}_i^{{\rm{2}}}\hat{l}_f{\rm{)}}^{{\rm{1/2}}}
 $$
 $$
 \times \sum_{\sigma_{{\rm{1}}}+\sigma_{{\rm{2}}}=l_B\,\,\,} \sum_{\lambda_{{\rm{1}}}+\lambda_{{\rm{2}}}=l_x}\sum_{lIL}\hat{l}
\left(
\begin{array}{cc}
2l_{x}\\
2\lambda_1
\end{array}
\right)^{{\rm{1/2}}}
  \left(
\begin{array}{cc}
{\rm{2}}l_{B}\\
{\rm{2}}\sigma_{{\rm{1}}}
\end{array}
\right)^{{\rm{1/2}}}
{\bar{a}}^{\lambda_{{\rm{1}}}}
{\bar{b}}^{\lambda_{{\rm{2}}}}
{\bar{c}}^{\sigma_{{\rm{1}}}}{\bar{d}}^{\sigma_{{\rm{2}}}}C_{l\,{\rm{0}}\sigma_{{\rm{1}}}\,{\rm{0}}}^{I\,{\rm{0}}}C_{l_i\,{\rm{0}}\lambda_1\,{\rm{0}}}^{I\,{\rm{0}}}
\eqno(B7)
$$
$$
\times
C_{l\,{\rm{0}}\,\sigma_{{\rm{ 2}}}\,{\rm{0}}}^{L\,{\rm{0}}}C_{l_f\,{\rm{0}}\lambda_{{\rm{2}}}\,{\rm{0}}}^{L\,{\rm{0}}}W{\rm{(L}}\sigma_{{\rm{2}}}I\sigma_{{\rm{1}}};\,ll_B{\rm{)}}X{\rm{(}}\lambda_{{\rm{1}}}\lambda_{{\rm{2}}}l_x;\,l_il_fJ;\,ILl_B{\rm{)}}{\cal{B}}^{{\rm{pole}}}_{l_xl_Bl_il_f\lambda_{{\rm{1}}}\sigma_{{\rm{1}}}}{\rm{(k}}_i,\,k_f{\rm{)}},
$$
 $$
{\cal{B}}^{{\rm{pole}}}_{l_xl_Bl_il_f\lambda_{\rm{1}}\sigma_{\rm{1}}}{\rm{(}}k_i,\,k_f{\rm{)}}={\rm{(}}\eta_x/4^{\eta_x+\eta_B}{\rm{)}}{\rm{(}}\kappa_{Aa}/{\rm{2}}{\rm{)}}^{l_B}{\rm{(}}\kappa_{ay}/{\rm{2}}{\rm{)}}^{l_x}\kappa_{Aa}\kappa_{ay}^{\rm{3}}
$$
$$
\times \int_{R_i^{{\rm{ch}}}}^{\infty}dr_ir_i^{\lambda_{\rm{1}}\,+\,\sigma_{\rm{1}}\,+\,{\rm{1}}}\Psi_{l_i}{\rm{(r}}_i;k_i{\rm{)}}
\int_{R_f^{{\rm{ch}}}}^{\infty}dr_fr_f^{\lambda_{\rm{2}}\,+\,\sigma_{\rm{2}}\,+\,{\rm{1}}}\Psi_{l_f}{\rm{(r}}_f;\,k_f{\rm{)}}{\tilde{\cal{A}}}_{l_Bl_xl}{\rm{(}}r_i,\,r_f{\rm{)}}, \eqno(B8)
$$
 $$
{\tilde{\cal{A}}}_{l_Bl_xl}{\rm{(}}r_i,\,r_f{\rm{)}}=\,\frac{{\rm{1}}}{{\rm{2}}}\int_{-{\rm{1}}}^{{\rm{1}}}dz P_l{\rm{(}}z{\rm{)}}F_{l_B}{\rm{(}}r_{Aa};\kappa_B,\eta_B-{\rm{1)}}
F_{l_x}{\rm{(}}r_{ay};\kappa_{ay},\eta_x{\rm{)}},\eqno(B9)
$$
 $$
 F_{l}(r;\kappa,\eta) =\frac{\pi^{1/2}}{\Gamma(l+\eta+2)}\int_{1}^{\infty}dte^{-\kappa rt}  (t^2-1)^{l+\eta+1},
 \eqno(B10)
$$
where $W{\rm{(}}l_{{\rm{1}}}j_{{\rm{1}}}l_{{\rm{2}}}j_{{\rm{2}}};j_{{\rm{3}}}j_{{\rm{4}}}{\rm{)}}$  and $X{\rm{(}}\lambda_{{\rm{1}}}\lambda_{{\rm{2}}}l_x;\,l_il_fJ;\,ILl_B{\rm{)}}$ are the standard  Racah and Fano coefficients \cite{Varsh}, respectively; $R_i^{{\rm{ch}}}=\,R_x\,+\,R_A$ and $R_f^{{\rm{ch}}}=\,R_y\,+\,R_B$ are   the cutoff radii in the entrance and exit channels, respectively, which are determined only by the free parameter $r_{{\rm{0}}}$ since $R_C$= $r_{{\rm{0}}}C^{{\rm{1/3}}} $ in which $C$ is a mass number of the nucleus $C$;  $\left(m\atop n\right)$ is the binomial coefficient  and $\hat{j}$= ${\rm{2}}j\,+\,{\rm{1}}$.

\newpage

\newpage

 \newpage

\newpage

\newpage

{\selectlanguage{english}
 \begin{table}[t]
\begin{center}
 \caption{\label{table1}{\small  Reaction $A$($x$, $y$)$B$,   $F_C=F_C{\rm{(}}\eta_x,\, \eta_B{\rm{)}}=\Gamma{{\rm(1}}\,-\,\eta_x{\rm{)}}\Gamma{\rm{(1}}\,-\,\eta_B{\rm{)}}/\Gamma{\rm{(1}}\,-\,\eta_{xB}{\rm{)}}$ ($\eta_{xB}=\,\eta_x\,+\,\eta_B$), incident energy $E_x$,  values of renormalized  CRFs  $ {\tilde{N}}^{{\rm{DW}}}_{{\rm{pole}}}$ and   $ {\tilde{N}}^{{\rm{DW}}}_{{\rm{post}}}$  as well as  $\tilde{N}^{{\rm{TBDM}}} $ corresponding to  the  pole-approximation and  the ''post'' form of DWBA as well as  the exact three-body model, respectively,  and quantities  $ {\cal{R}}^{{\rm{TBDM }}}$=  $N^{{\rm{TBDM}}}/N^{{\rm{DW}}}_{{\rm{pole}}}$=  $\tilde{N}^{{\rm{TBDM}}}/{\tilde{N}}^{{\rm{DW}}}_{{\rm{pole}}}$, $ {\cal{R}}^{{\rm{TBDM}}}_{{\rm{post}}}$= $N^{{\rm{TBDM}}}/N^{{\rm{DW}}}_{{\rm{post}}}$= $\tilde{N}^{{\rm{TBDM}}}/{\tilde{N}}^{{\rm{DW}}}_{{\rm{post}}}$,
 $ {\cal{R}}^{{\rm{DW}}}_{{\rm{post}}}=\,{\cal{R}}^{{\rm{TBDM }}}/ {\cal{R}}^{{\rm{TBDM}}}_{{\rm{post}}}=\,N^{{\rm{DW}}}_{{\rm{post}}}/N^{{\rm{DW}}}_{{\rm{pole}}}$= $\tilde{N}^{{\rm{DW}}}_{{\rm{post}}}/\tilde{N}^{{\rm{DW}}}_{{\rm{pole}}}$.  ${\rm{^{10}B}}_i$  denotes the ground ($i$=0) state and the first--third ($i$=1--3) excited ones of the residual ${\rm{^{10}B}}$ nucleus, while,   ${\rm{^{17}F}}_i$ denotes    the ground ($i$=0) and first ($i$=1) excited states of  the  residual ${\rm{^{17}F}}$ nucleus. Figures in the curly brackets are the modulus of the correspending ratios.  }}
   \vspace{5mm}
{\footnotesize
\begin{tabular}{lllll}
 \hline
 \multicolumn{1}{c}{} &\multicolumn{1}{c}{}&\multicolumn{1}{c}{}&\multicolumn{1}{c}{}&\multicolumn{1}{c}{}\\
\multicolumn{1}{c}{$A$($x$, $y$)$B$;}
 &\multicolumn{1}{c}{\,\,\,  $E_x$, MeV}
  &\multicolumn{1}{c}{$ {\tilde{N}}^{{\rm{DW}}}_{{\rm{pole}}}$ ($ {\tilde{N}}^{{\rm{DW}}}_{{\rm{post}}}$)}
&\multicolumn{1}{c}{\,\,\,$\tilde{N}^{{\rm{TBDM}}} $  }
&\multicolumn{1}{c}{$ {\cal{R}}^{{\rm{TBDM}}}$ ($  {\cal{R}}^{{\rm{TBDM}}}_{{\rm{post}}}$) }\\
\multicolumn{1}{c}{$\eta_B$;\,$\eta_{xB}$ ($F_C$) }
 &\multicolumn{1}{c}{}
  &\multicolumn{1}{c}{ }
&\multicolumn{1}{c}{   }
&\multicolumn{1}{c}{  [${\cal{R}}^{{\rm{DW}}}_{{\rm{post}}}$] }\\
\hline\\
\,\,\,\,\,\,\,\,\,\,\,\,\,\,\,\,\,\,1&\,\,\,\,\,\,\,\,\,\,\,2&\,\,\,\,\,\,\,\,\,\,\,\,\,\,\,\,\,\,\,\,\,\,\,\,\,\,\,\,\,\,\,\,\,\,\,\,3&\,\,\,\,\,\,\,\,\,\,\,\,\,\,\,\,\,\,4&\,\,\,\,\,\,\,\,\,\,\,\,\,\,\,\,\,\,5\\
\hline
 ${\rm {^9Be(^7Be,\,^8B)^{10}Be}}$ &\,\,\,\,84 \cite{Azh1999}&\,\,\,\,\,\,\,\,\,\,\,\,\,\,\,(9.222 -  $i\cdot$33.373)x10$^{{\rm{3}}}$&\,\,\,\,\,\,\,-8.1648x10$^{{\rm{4}}}$&-0.627 - $i\cdot$2.292\\
\,\,\,0.233; 1.823&&\,\,\,\,\,\,\,\,\,\,\,((3.899 - $i\cdot$14.112)x10$^{{\rm{3}}}$)&
&(-0.374 - $i\cdot$1.355)\\
\,\,\,\,\,\,(0.695)   &&&&[1.671 + $i\cdot$4.020x10$^{{\rm{-13}}}$]\\
  &&&&\{1.67(1.46)[2.38]\}\\
 ${\rm {^{14}N(^7Be,\,^8B)^{13}C}}$ &\,\,\,\,85 \cite{Azh1999-1,Tab06}&\,\,\,\,\,\,\,\,\,\,\,\,\,\,\,(-7.865 +  $i\cdot$3.795)x10$^{{\rm{3}}}$&\,\,\,\,\,\,\,-1.6084x10$^{{\rm{4}}}$&16.58 + $i\cdot$8.00 \\
\,\,\,0.331; 1.921    &&\,\,\,\,\,\,\,\,\,\,\,((2.6553 - $i\cdot$1.2813)x10$^{{\rm{4}}}$)&&(1.47 - $i\cdot$0.71)\\
\,\,\,\,\,\,(0.366)   &&&&[11.2 + $i\cdot$2.1x10$^{{\rm{-11}}}$]\\
 &&&&\{18.4(1.63)[11.2]\}\\
   ${\rm {^9Be(^{10}B,^9Be)^{10}B_0}}$ &\,\,\,\,100 \cite{Mukh2}&\,\,\,\,\,\,\,\,\,\,\,\,\,\,\,0.339 - $i\cdot$2.664&\,\,\,\,\,\,\,-4.117&-0.193 - $i\cdot$1.521 \\
\,\,\,0.234; 0.468  &&\,\,\,\,\,\,\,\,\,\,\,(0.515 - $i\cdot$4.053)&&(-0.127 - $i\cdot$1.000)\\
\,\,\,\,\,\,(0.871)   &&&&[1.521 + $i\cdot$1.300x$10^{-15}$]\\
&&&&\{1.533(1.008)[1.521]\}\\
${\rm {^9Be(^{10}B,^9Be)^{10}B_1}}$ &\,\,\,\,100 \cite{Mukh2}&\,\,\,\,\,\,\,\,\,\,\,\,\,\,\,0.215 - $i\cdot$2.833 &\,\,\,\,\,\,\,-4.431&-0.118 - $i\cdot$1.555 \\
\,\,\,0.234; 0.482  &&\,\,\,\,\,\,\,\,\,\,\,(-0.333 - $i\cdot$4.383)&&(-7.639x10$^{{\rm{-2}}}$ - $i\cdot$1.005)\\
\,\,\,\,\,\,(0.852)   &&&&[1.546 - $i\cdot$7.227x$10^{{\rm{-15}}}$]\\
&&&&\{1.559(1.008)[1.546]\}\\
 ${\rm {^9Be(^{10}B,^9Be)^{10}B_2}}$ &\,\,\,\,100 \cite{Mukh2}&\,\,\,\,\,\,\,\,\,\,\,\,\,\,\, 1.658x10$^{{\rm{-2}}}$ - $i\cdot$3.147&\,\,\,\,\,\,\,-5.063&-8.476x10$^{{\rm{-2}}}$ - $i\cdot$1.609 \\
\,\,\,0.234; 0.506  &&\,\,\,\,\,\,\,\,\,\,\,(-2.643 - $i\cdot$5.016)&& (-5.319x10$^{{\rm{-3}}}$ - $i\cdot$1.009)\\
\,\,\,\,\,\,(0.845)   &&&&[1.593 + $i\cdot$1.588x$10^{-14}$]\\
&&&&\{1.609(1.009)[1.593]\}\\
 ${\rm {^9Be(^{10}B,^9Be)^{10}B_3}}$ &\,\,\,\,100 \cite{Mukh2}&\,\,\,\,\,\,\,\,\,\,\,\,\,\,\,-7.489x10$^{{\rm{-2}}}$ - $i\cdot$3.314&\,\,\,\,\,\,\,-5.417&3.691x10$^{{\rm{-2}}}$ - $i\cdot$1.633 \\
\,\,\,0.234; 0.519  &&\,\,\,\,\,\,\,\,\,\,\,(-0.121 - $i\cdot$5.361)&&(2.281x10$^{{\rm{-2}}}$ -  $i\cdot$1.010)\\
\,\,\,\,\,\,(0.836)   &&&&[1.617 + $i\cdot$1.405x$10^{-14}$]\\
&&&&\{1.634(1.010)[1.617]\}\\
   ${\rm{^{16}O(^3He}}, d{\rm {)^{17}F_0}}$&\,\,\, 29.75 \cite{Mukh6}&\,\,\,\,\,\,\,\,\,\,\,\,\,261.48 + $i\cdot$435.04&\,\,\,\,\,\,-590.36 &-0.599 + $i\cdot$0.996 \\
\,\,\, 1.577;   1.632    &&\,\,\,\,\,\,\,\,\,\,\,(279.68 + $i\cdot$465.32)&&(-0560 + $i\cdot$0.932)\\
 \,\,\,\,\,\, (0.983)   &&&&[1.069 - $i\cdot$1.600x$10^{-15}$]\\
 &&&&\{1.162(1.087)[1.069]\}\\
  ${\rm{^{16}O(^3He}}, d{\rm {)^{17}F_1}}$ & &\,\,\,\,\,\,\,\,\, (-2.96 - $i\cdot$4.75)x10$^{{\rm{15}}}$&\,\,\,\,\,-1.33x10$^{{\rm{9}}}$ &(1.26 -  $i\cdot$2.05)x10$^{{\rm{-7}}}$ \\
\,\,\, 3.760; 3.815     &&\,\,\,\,\,\,\,\,\,\,\,((-0.725  - $i\cdot$1.160)x10$^{{\rm{15}}}$)&&((5.14  - $i\cdot$8.26)x10$^{{\rm{-7 }}}$)\\
 \,\,\,\,\,\,(0.887)   &&&&[0.245 - $i\cdot$1.300x$10^{-11}$]\\
  &&&&\{2.406x10$^{{\rm{-7}}}$(9.729x10$^{{\rm{-7}}}$)\\
   &&&&[0.245]\}\\
    \hline\\
   \end{tabular}}
\end{center}
\end{table}}
\newpage

 {\selectlanguage{english}
\begin{table}[t]
\begin{center}
 \caption{{continuation of Table 1}}
\vspace{2mm}
{\footnotesize
\begin{tabular}{lllll}
\hline\\
\,\,\,\,\,\,\,\,\,\,\,\,\,\,\,\,\,\,1&\,\,\,\,\,\,\,\,\,\,\,2&\,\,\,\,\,\,\,\,\,\,\,\,\,\,\,\,\,\,\,\,\,\,\,\,\,\,\,\,\,\,\,\,\,\,\,\,3&\,\,\,\,\,\,\,\,\,\,\,\,\,\,\,\,\,\,4&\,\,\,\,\,\,\,\,\,\,\,\,\,\,\,\,\,\,5\\
\hline
${\rm{^{19}F(}}p,\,\alpha {\rm {)^{16}O}}$(g.s.)&\,\,\,\, 0.250 \cite{HL1978,HAS1991}&\,\,\,\,\,\,\,(-1.360 + $i\cdot$0.453)x10$^{{\rm{-3}}}$&\,\,\,-1.68x10$^{{\rm{-3}}}$&\,\,1.112 + $i\cdot$0.370 \\
\,\,\,0.585; 0.615   &&\,\,\,\,\,\,\,\,\,\,\,((-1.48 + $i\cdot$0.494)x10$^{-3}$)&&(1.023 + $i\cdot$0.340)\\
\,\,\,\,\,\, (0.943)   &&&&[1.088 - $i\cdot$7.150x$10^{-18}$]\\
&&&&\{1.172(1.078)[1.088]\}\\
  &\,\,\,\, 0.350&\,\,\,\,\,\,\,(-3.20 + $i\cdot$1.14 )x10$^{{\rm{-3}}}$&\,\,\,-3.98x10$^{{\rm{-3}}}$&\,\,1.104+ $i\cdot$0.394 \\
   &&\,\,\,\,\,\,\,\,\,\,\,((-3.480 - $i\cdot$1.240)x10$^{-3}$)&&(1.014 + $i\cdot$0.361)\\
     &&&&[1.088 - $i\cdot$6.005x$10^{-18}$]\\
     &&&&\{1.172(1.076)[1.088]\}\\
 &\,\,\,\, 0.450&\,\,\,\,\,\,\,(-5.41 + $i\cdot$2.04 )x10$^{{\rm{-3}}}$&\,\,\,-6.78x10$^{{\rm{-3}}}$&\,\,1.097+ $i\cdot$0.412 \\
   &&\,\,\,\,\,\,\,\,\,\,\,((-5.893  - $i\cdot$2.217)x10$^{-3}$)&&(1.008 + $i\cdot$0.379)\\
    &&&&[1.088]\\
    &&&&\{1.172(1.077)[1.088]\}\\
      &\,\,\,\,  0.327 \cite{Ivano2015}&\,\,\,\,\,\,\,(-2.750 + $i\cdot$0.97)x10$^{{\rm{-3}}}$&\,\,\,-3.42x10$^{{\rm{-3}}}$&\,\,1.110 + $i\cdot$0.390 \\
  &&\,\,\,\,\,\,\,\,\,\,\,((-3.00 + $i\cdot$1.05)x10$^{{\rm{-3}}}$)&&(1.020 + $i\cdot$0.360)\\
    &&&&[1.090]\\
    &&&&\{1.177(1.082)[1.088]\}\\
     &\,\,\,\, 0.387&\,\,\,\,\,\,\,(-3.980 + $i\cdot$1.450 )x10$^{{\rm{-3}}}$&\,\,\,-4.97x10$^{{\rm{-3}}}$&\,\,1.100+ $i\cdot$0.400 \\
   &&\,\,\,\,\,\,\,\,\,\,\,((-4.330 - $i\cdot$1.580)x10$^{{\rm{-3}}}$)&&(1.010 + $i\cdot$0.370)\\
     &&&&[1.090]\\
     &&&&\{1.170(1.076)[1.090]\}\\
      &\,\,\,\, 0.486&\,\,\,\,\,\,\,(-6.30 + $i\cdot$2.41 )x10$^{{\rm{-3}}}$&\,\,\,-7.90x10$^{{\rm{-3}}}$&\,\,1.090+ $i\cdot$0.420 \\
   &&\,\,\,\,\,\,\,\,\,\,\,((-6.850  + $i\cdot$2.62)x10$^{{\rm{-3}}}$)&&(1.010 + $i\cdot$0.380)\\
    &&&&[1.090]\\
    &&&&\{1.168(1.079)[1.090]\}\\
    \hline\\
   \end{tabular}}
\end{center}
\end{table}}

    \newpage

{\selectlanguage{english}
\begin{table}[t]
\begin{center}
 \caption{\label{table2}{\small The specific reactions and the corresponding to them vertices  described by the triangle diagram  Fig. \ref{fig2}$c$,   the positions of  singularities $i\kappa$ and $i\kappa_i$ ($i\bar{\kappa}_i$) in $q_{Aa}$($q_{ay}$) as well as  $\xi$ and $\xi_i$ ($\bar{\xi}_i$) in  the  $\cos\theta$-plane of the reaction amplitude, where $\kappa$ is related either  to the vertex B $\to$ A+$a$ ( $\kappa$=  $\kappa_{Aa}$)  or to the vertex x $\to$ y+$a$ ($\kappa$= $\kappa_{ya}$) }.}

 \vspace{2mm}
{\footnotesize
\begin{tabular}{lllllllll} \hline
\multicolumn{1}{c}{}
&\multicolumn{1}{c}{}
&\multicolumn{1}{c}{ The vertex}
&\multicolumn{1}{c}{}
&\multicolumn{1}{c}{}
&\multicolumn{1}{c}{}
&\multicolumn{1}{c}{}
&\multicolumn{1}{c}{}
&\multicolumn{1}{c}{}\\
\multicolumn{1}{c}{Reaction}
&\multicolumn{1}{c}{$E_x^{{\rm{lab}}}$}
&\multicolumn{1}{c}{$B\,\to\,A\,+\,a$}
&\multicolumn{1}{c}{ $\xi$}
&\multicolumn{1}{c}{$b_i$ }
&\multicolumn{1}{c}{$c_i$}
&\multicolumn{1}{c}{$d_i$}
&\multicolumn{1}{c}{$\kappa_i$($\bar{\kappa}_i$), }
 &\multicolumn{1}{c}{$\xi_i$} \\
\multicolumn{1}{c}{$A$($x$,\,$y$)$B$}
&\multicolumn{1}{c}{MeV}
&\multicolumn{1}{c}{($x\,\to\,y\,+\,a$)}
&\multicolumn{1}{c}{($\kappa$, fm$^{{\rm{-1}}}$)}
&\multicolumn{1}{c}{($e_i$) }
&\multicolumn{1}{c}{($f_i$)}
&\multicolumn{1}{c}{($g_i$)}
&\multicolumn{1}{c}{fm$^{\rm{{-1}}}$}
 &\multicolumn{1}{c}{($\bar{\xi}_i$)} \\
 \hline\\
\multicolumn{1}{c}{${\rm{^9Be}}$(${\rm{^{10}B}}$,\,${\rm{^9Be}}$)${\rm{^{10}B_0}}$}
&\multicolumn{1}{c}{100}
 &\multicolumn{1}{c}{${\rm{^{10}B_0}}\to\,{\rm{^9Be}}\,+\,p$}
&\multicolumn{1}{c}{1.020(0.534)}
&${\rm{^8Be}}$
&$\,\,\,\,d$
&$\,\,\,\,n$
&0.940&1.064\\
 \multicolumn{1}{c}{  }
& \multicolumn{1}{c}{  }
& \multicolumn{1}{c}{  }
 &\multicolumn{1}{c}{}
 &${\rm{^6Li}}$
 &${\rm{^4{He}}}$
 &$\,\,\,\,t$
 &2.024
 &1.479\\
 \multicolumn{1}{c}{  }
& \multicolumn{1}{c}{  }
& \multicolumn{1}{c}{  }
 &\multicolumn{1}{c}{}
 &$\,\,\,\,n$
 &${\rm {^9B}}$
 &${\rm {^8Be}}$
 &0.802
 &4.169\\
     \multicolumn{1}{c}{${\rm{^{16}O}}$(${\rm{^3He}}$,\,$d$)${\rm{^{17}F_0}}$}
&\multicolumn{1}{c}{29.7}
&\multicolumn{1}{c}{${\rm{^{17}F_0}}\to\,{\rm{^{16}O}}\,+\,p$}
&\multicolumn{1}{c}{1.065(0.165)}
&${\rm{^{14}N}}$
&${\rm{^3He}}$
&$\,\,\,\,d$
&2.696
&3.253\\
\multicolumn{1}{c}{  }
& \multicolumn{1}{c}{  }
& \multicolumn{1}{c}{  }
 &\multicolumn{1}{c}{}
 &${\rm{^{13}N}}$
&${\rm{^4He}}$
&$\,\,\,\,t$
&2.645
&3.508\\
\multicolumn{1}{c}{  }
& \multicolumn{1}{c}{  }
& \multicolumn{1}{c}{  }
 &\multicolumn{1}{c}{}
 &$\,\,\,\,p$
 &${\rm{^{16}O}}$
 &${\rm{^{15}N}}$
  &0.905
   &49.551\\
   \multicolumn{1}{c}{}
& \multicolumn{1}{c}{}
 &\multicolumn{1}{c}{(${\rm{^3He}}\to\,d\,+\,p$)}
 & \multicolumn{1}{c}{ 1.065( 0.420)}
 &\,\,\,\,($p$)
 &\,\,\,\,($d$)
 &\,\,\,\,($n$)
 &(0.652)
 &(1.562)\\
   \multicolumn{1}{c}{${\rm{^{19}F}}$($p$,\,$\alpha$)${\rm{^{16}O}}$}
&\multicolumn{1}{c}{0.250}
&\multicolumn{1}{c}{${\rm{^{19}F}}\to\,{\rm{^{16}O}}\,+\,t$}
&\multicolumn{1}{c}{13.648(1.194)}
&${\rm{^{15}N}}$
&${\rm{^4He}}$
&$\,\,\,\,p$
&1.522
&19.720\\
 \multicolumn{1}{c}{ }
&\multicolumn{1}{c}{0.350}
&\multicolumn{1}{c}{ }
&\multicolumn{1}{c}{11.544(1.194)}
 &
& &
&1.522
&16.647\\
 \multicolumn{1}{c}{ }
&\multicolumn{1}{c}{0.450}
&\multicolumn{1}{c}{ }
&\multicolumn{1}{c}{10.190(1.194)}
 &
& &
&1.522
&14.665\\
\hline
 \end{tabular}}
\end{center}
\end{table}}

 \end{document}